# A Tutorial on 5G NR V2X Communications

Mario H. Castañeda Garcia, Alejandro Molina-Galan, Mate Boban, Javier Gozalvez, Baldomero Coll-Perales, Taylan Şahin and Apostolos Kousaridas

*Abstract*— The Third Generation Partnership Project (3GPP) has recently published its Release 16 that includes the first Vehicle-to-Everything (V2X) standard based on the 5G New Radio (NR) air interface. 5G NR V2X introduces advanced functionalities on top of the 5G NR air interface to support connected and automated driving use cases with stringent requirements. This paper presents an in-depth tutorial of the 3GPP Release 16 5G NR V2X standard for V2X communications, with a particular focus on the sidelink, since it is the most significant part of 5G NR V2X. The main part of the paper is an in-depth treatment of the key aspects of 5G NR V2X: the physical layer, the resource allocation, the quality of service management, the enhancements introduced to the Uu interface and the mobility management for V2N (Vehicle to Network) communications, as well as the co-existence mechanisms between 5G NR V2X and LTE V2X. We also review the use cases, the system architecture, and describe the evaluation methodology and simulation assumptions for 5G NR V2X. Finally, we provide an outlook on possible 5G NR V2X enhancements, including those identified within Release 17.

*Index Terms*—5G NR V2X, 5G V2X, 3GPP, Release 16, sidelink, vehicle-to-everything, V2X, 5G, New Radio, 5G NR, LTE V2X, Cellular V2X, C-V2X, connected and automated vehicles, CAV, connected and automated driving.

## I. INTRODUCTION

IN 1999, the US Federal Communications Commission (FCC) allocated 75 MHz of spectrum in the 5.9 GHz band for Intelligent Transportation Services (ITS). The allocation triggered significant research activity around the world to develop and deploy V2X communications over the past two decades (e.g., the CAMP consortium in the US [1], the Car 2 Car Communication Consortium in Europe [2], and countless research projects). V2X communications includes Vehicle-to-vehicle (V2V), Vehicle-to-Network (V2N) or Vehicle-to-Infrastructure (V2I), Vehicle-to-Road Side Unit (V2R), and Vehicle-to-Pedestrian (V2P). The research resulted in a first set of radio standards for V2X completed by 2010. These standards are based on the IEEE 802.11p technology and are referred to as Dedicated Short Range Communications (DSRC) [3]. The development of radio standards was followed by the definition of higher layer standards, message formats, protocols, and applications (e.g., [4] in Europe and [5] in the US).

3GPP Release 12 (Rel. 12) was the first standard to introduce direct Device-to-Device (D2D) communications for proximity services (ProSe) using cellular technologies [6]. This work was used by 3GPP to develop LTE V2X, the first cellular V2X (C-V2X) standards based on the 4G Long Term Evolution (LTE) air interface. LTE V2X was developed under Release 14 (Rel. 14) [7] and was further enhanced in Release 15 (Rel. 15). It is only under Release 16 (Rel. 16) that 3GPP has developed a new cellular V2X standard based on the 5G NR (New Radio) air interface. The precursor to the technical work on Rel. 16 NR V2X was the study item (SI) approved under Rel. 15. This SI developed the evaluation methodology and assumptions for LTE and NR V2X [8] that were necessary to evaluate and compare the various proposals to be included in the 5G NR V2X standard. Next, 3GPP approved a SI and a work item (WI)[1] to develop the first set of 5G NR V2X standards in Rel. 16. Specifically, the SI on radio interface technologies [9] ran until March 2019 followed by a WI [10] that officially concluded in December 2019. This WI resulted in the first set of 5G NR V2X specifications included in the 3GPP technical specifications (TS). Fig. 1 summarizes the timeline of the development of cellular V2X standards under 3GPP with a focus on Radio Access Network (RAN) developments.

The 5G NR standard was developed under Rel. 15 but it did not include sidelink (SL) aspects. SL refers to direct communication between terminal nodes or User Equipments (UEs) without the data going through the network. In NR V2X, UEs are vehicles, Road Side Units (RSUs), or mobile devices that are carried by pedestrians. Rel. 16 is the first to introduce V2X communications, including SL communications, based on the 5G NR air interface. This makes Rel. 16 NR V2X SL the first 5G V2X standard available, and a basis for future enhancements and extensions for V2X and non-V2X SL applications. As noted in [10], the NR V2X SL has been developed to complement and not replace LTE V2X SL communications. The goal of NR V2X SL is to support enhanced V2X (eV2X) use cases related to connected and automated driving [11]. Some of these use cases have requirements that cannot be satisfied by the LTE V2X standard [12].

This paper presents an exhaustive overview of the first standard for 5G NR V2X communications that 3GPP developed in Rel. 16. The main focus of the standard and of this paper is on NR V2X SL communications as the basis to support ubiquitous V2X communications. The paper provides a comprehensive and reference tutorial that introduces the major 3GPP standard developments essential to understand how NR V2X communications operate. To this aim, Section II briefly reviews first the LTE V2X standard designed under Rel. 14 and Rel. 15. This review helps understand and highlight the differences and novelties introduced by NR V2X under Rel. 16. One key difference between LTE V2X and NR V2X is the use cases that each technology should support. To that end, Section III presents the use cases that 5G NR V2X supports following the work done at the 3GPP and the 5G Automotive Association (5GAA) [13]. Section III also includes a review of the main use case requirements and key performance indicators (KPIs). Section IV presents a high level overview of the 5G NR system architecture for V2X communication for SL and uplink/downlink communications. Section V provides an in-depth summary of the NR V2X SL physical layer including its structure, the physical sidelink channels, as well as physical layer sidelink procedures such as HARQ feedback, synchronization, and power control. Section VI describes the resource allocation in NR V2X SL for its two communication modes (mode 1 and mode 2) as well as procedures for supporting congestion control. Section VII describes the framework and mechanisms defined under Rel. 16 to manage quality of service (QoS) in NR V2X communications. Section VIII presents the major enhancements introduced in Rel. 16 to the Uu interface and to the mobility management in order to support V2N communications. NR V2X has been designed to complement LTE V2X. To that end, Section IX explains how the two technologies can co-exist. The publication of Rel. 16 will trigger significant efforts to evaluate the performance and capabilities of NR V2X. Section X presents the evaluation methodology defined in 3GPP that includes new channel models and assumptions for system and link level simulations. Finally, Section XI presents future possible enhancements for NR V2X communications including those already identified as study and work items under Release 17 (Rel. 17).

[1]In 3GPP, SI refers to the more exploratory work that investigates the feasibility and benefit that certain technologies can bring. On the other hand, WI refers to specification work that often builds on top of the work performed in a SI and results in a set of technical specifications (i.e., a standard).

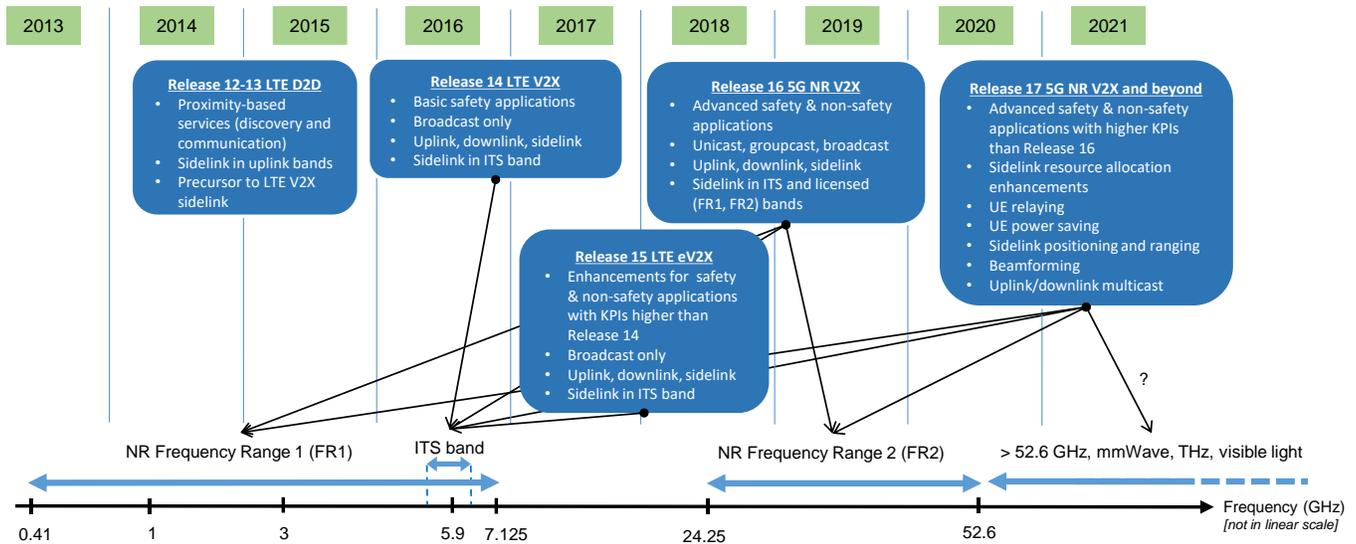

Fig. 1. Progress of 3GPP work on V2X with a focus on RAN.

TABLE I
LIST OF ACRONYMS

| Acronym | Definition | Acronym | Definition |
|---|---|---|---|
| 3GPP | Third Generation Partnership Project | CDM | Code Division Multiplexing |
| 5GAA | 5G Automotive Association | CG | Configured Grant |
| 5GC | 5G Core | CHO | Conditional Handover |
| 5GS | 5G System | CP | Cyclic Prefix |
| 5QI | 5G New Radio Standardized Quality of Service Identifier | CQI | Channel Quality Indicator |
| ACK | Acknowledgement | CR | Channel Occupancy Ratio |
| AF | Application Function | CRC | Cyclic Redundancy Check |
| AGC | Automatic Gain Control | CRlimit | Maximum Channel Occupancy Ratio |
| AMF | Access and Mobility Management Function | CSI-RS | Channel State Information-Reference Signal |
| AS layer | Access Stratum layer | C-V2X | Cellular V2X |
| AS | Application Server | D2D | Device-to-Device (Communications) |
| BS | Base Station | DAPS | Dual Active Protocol Stack |
| BSM | Basic Safety Message | DCC | Decentralized Congestion Control |
| BWP | Bandwidth Part | DCI | Downlink Control Information |
| CAM | Cooperative Awareness Message | DENM | Decentralized Environmental Notification Message |
| CBR | Channel Busy Ratio | DFN | Direct Frame Number |
| CDL | Cluster Delay Line | DG | Dynamic Grant |

TABLE I
LIST OF ACRONYMS

| Acronym | Definition | Acronym | Definition |
|---------|-----------|---------|-----------|
| DL | Downlink | RA | Random Access |
| DMRS | Demodulation Reference Signal | RAN | Radio Access Network |
| DSRC | Dedicated Short Range Communications | RAT | Radio Access Technology |
| eNB | Evolved Node B (LTE Base Station) | RB | Resource Block |
| EPC | Evolved Packet Core | RF | Radio Frequency |
| EPS | Evolved Packet System | RI | Rank Indicator |
| ETSI | European Telecommunications Standardization Institute | RLC | Radio Link Control |
| eV2X | Enhanced Vehicle-to-Everything | RLF | Radio Link Failure |
| EVM | Error Vector Magnitude | RP | Resource Pool |
| FCC | Federal Communications Commission | RRC | Radio Resource Control |
| FDM | Frequency Division Multiplexing | RRI | Resource Reservation Interval |
| FEC | Forward Error Correction | RSRP | Reference Signal Received Power |
| FR1 | Frequency Range 1 | RSRQ | Reference Signal Received Quality |
| FR2 | Frequency Range 2 | RSSI | Received Signal Strength Indicator |
| GLOSA | Green Light Optimal Speed Advisory | RS-SINR | Reference Signal-Signal to Noise and Interference Ratio |
| gNB | Next generation Node B (NR Base Station) | RSU | Road Side Unit |
| gNB-CU | gNB-Control Unit | RV | Redundancy Version |
| gNB-DU | gNB-Distributed Unit | RX UE | Receiving User Equipment |
| GNSS | Global Navigation Satellite System | SA | 3GPP Services and System Aspects |
| HARQ | Hybrid Automatic Repeat Request | SAE | Society of Automotive Engineers |
| HO | Handover | SC-FDMA | Single-Carrier Frequency-Division Multiple Access |
| HPLMN | Home Public Land Mobile Network | SCI | Sidelink Control Information |
| ISD | Inter Site Distance | SC-PTM | Single-Cell Point-To-Multipoint |
| ITS | Intelligent Transportation Systems | SCS | Subcarrier Spacing |
| KPI | Key Performance Indicator | SDAP | Service Data Adaptation Protocol |
| LDPC | Low-Density Parity-Check | SF | Subframe |
| LTE | Long-Term Evolution | SFN | System Frame Number |
| MAC | Medium Access Control | SI | Study Item |
| MAC CE | MAC Control Element | SIB | System Information Block |
| MBMS | Multimedia Broadcast Multicast Services | SID | Study Item Description |
| MBS | Multicast Broadcast Services | SL | Sidelink |
| MBSFN | Multimedia Broadcast Single Frequency Network | SLR | Service Level Requirement |
| MCS | Modulation and Coding Scheme | SLRB | Sidelink Radio Bearer |
| NACK | Negative Acknowledgement | SLSS | Sidelink Synchronization Signal |
| NAS | Non-Access Stratum | SMF | Session Management Function |
| NDI | New Data Indicator | SPS | Semi-Persistent Scheduling |
| NEF | Network Exposure Function | S-PSS | Sidelink Primary Synchronization Signal |
| NF | Network Functions | SR | Scheduling Request |
| NG-RAN | Next-Generation Radio Access Network | SSB | Synchronization Signal Block |
| NR | New Radio | S-SSB | Sidelink Synchronization Signal Block |
| NRF | Network Repository Function | S-SSS | Sidelink Secondary Synchronization Signal |
| NWDAF | Network Data Analytics Function | TAI | Tracking Area Identifier |
| OAM | Operations, Administration and Maintenance | TB | Transport Block |
| OFDM | Orthogonal Frequency Division Multiplexing | TDD | Time Division Duplex |
| P2P | Pedestrian-to-Pedestrian | TDM | Time Division Multiplexing |
| PBCH | Physical Broadcast Channel | TFRP | Time-Frequency Resource Pattern |
| PCF | Policy Control Function | TPC | Transmit power control |
| PDB | Packet Delay Budget | TS | Technical Specification |
| PDCCH | Physical Downlink Control Channel | TTI | Transmission Time Interval |
| PDCP | Packet Data Convergence Protocol | TX UE | Transmitting User Equipment |
| PDSCH | Physical Downlink Shared Channel | UDM | Unified Data Management |
| PFI | PC5 QoS Flow ID | UDR | Unified Data Repository |
| PMI | Precoding Matrix Indicator | UICC | Universal Integrated Circuit Card |
| PPPP | Proximity Service Per-Packet Priority | UL | Uplink |
| PPPR | Proximity Service Per-Packet Reliability | UPF | User Plane Functions |
| PQI | PC5 5QI | URLLC | Ultra-Reliable Low Latency Communication |
| PRB | Physical Resource Block | UTC | Coordinated Universal Time |
| ProSe | Proximity Service | V2I | Vehicle-to-Infrastructure |
| PSBCH | Physical Sidelink Broadcast Channel | V2N | Vehicle-to-Network |
| PSCCH | Physical Sidelink Control Channel | V2P | Vehicle-to-Pedestrian |
| PSFCH | Physical Sidelink Feedback Channel | V2R | Vehicle-to-RSU |
| PSSCH | Physical Sidelink Shared Channel | V2V | Vehicle-to-Vehicle |
| PT-RS | Phase Tracking-Reference Signal | V2X | Vehicle-to-Everything |
| PUCCH | Physical Uplink Control Channel | VPLMN | Visited Public Land Mobile Network |
| PUSCH | Physical Uplink Shared Channel | VRU | Vulnerable Road User |
| QoS | Quality of Service | WI | Work Item |
| R2R | RSU(Road Side Unit)-to-RSU | | |

## II. RELEASE 14/15: LTE V2X

3GPP defined in Rel. 14 (and later refined in Rel. 15) the LTE V2X standard for V2X communications using the LTE air interface [14]. LTE V2X is expected to operate on the 5.9 GHz band reserved in certain markets (e.g., United States, Europe, China) for ITS services. For SL communications, vehicles utilize the so-called PC5 interface, whereas they utilize the Uu interface for V2N. LTE V2X has been designed to support basic cooperative active traffic safety, traffic management, and telematics applications [15]. LTE V2X supports similar services as those supported by DSRC or its European counterpart ITS-G5. These services rely on the broadcast transmission of small awareness messages such as CAMs (Cooperative Awareness Messages) in ITS-G5 [4] or BSMs (Basic Safety Messages) in DSRC [3] to regularly provide basic information such as the location, direction, speed, and acceleration of the transmitting vehicle. LTE V2X defines new PHY (Physical) and MAC (Medium Access Control) layers for V2X and reuses the upper V2X layers and protocols specified by ETSI (European Telecommunications Standardization Institute), IEEE (Institute of Electrical and Electronic Engineers), and SAE (Society of Automotive Engineers).

### A. Physical Layer

LTE V2X uses SC-FDMA (Single-Carrier Frequency-Division Multiple Access) and supports 10 MHz and 20 MHz channels. The channel is divided into 180 kHz Resource Blocks (RBs) that correspond to 12 subcarriers of 15 kHz each. In the time domain, the channel is organized into 1 ms subframes. Fig. 2 illustrates the channelization in LTE V2X. Each subframe has 14 OFDM symbols with normal cyclic prefix. Nine of these symbols are used to transmit data and four of them (3rd, 6th, 9th, and 12th) are used to transmit demodulation reference signals (DMRSs) for channel estimation and combating the Doppler effect at high speeds [17]. The last symbol is used as a guard symbol for timing adjustments and for allowing vehicles to switch between transmission and reception across subframes.

RBs are grouped into sub-channels (Fig. 2). A sub-channel can include RBs only within the same subframe. The number of RBs per sub-channel can vary and is (pre-)configured. (Pre-)configuration refers to a configuration that is: 1) defined by the network and signaled to the UE by the cellular base station (eNB or gNB) when a UE is in network coverage; or 2) predefined in the UE when the UE is out of network coverage. Sub-channels are used to transmit data and control information. The data is organized in Transport Blocks (TBs) (Fig. 2) that are carried in the Physical Sidelink Shared Channel (PSSCH). A TB contains a full packet (e.g., a CAM or a BSM). A TB can occupy one or several sub-channels depending on the size of the packet, the number of RBs per sub-channel, and the utilized Modulation and Coding Scheme (MCS). TBs can be transmitted using QPSK, 16-QAM or 64QAM (introduced in Rel. 15) modulations and turbo coding.

Each TB has an associated Sidelink Control Information (SCI) message (Fig. 2) that is carried in the Physical Sidelink Control Channel (PSCCH) [18]. It is also referred to as Scheduling Assignment (SA). An SCI occupies 2 RBs and includes information such as: an indication of the RBs occupied by the associated TB; the MCS used for the TB; the priority of the message that is being transmitted; an indication of whether it is a first transmission or a blind retransmission of the TB; and the resource reservation interval. A blind retransmission refers to a scheduled retransmission or repetition of the TB (i.e., not based on feedback from the receiver). The resource reservation interval specifies when the vehicle will utilize the reserved sub-channel(s) to transmit its next TB. The SCI includes critical information for the correct reception of the TB. A TB cannot be decoded properly if the associated SCI is not received correctly. A TB and its associated SCI must be transmitted always in the same subframe (Fig. 2). The TB and its associated SCI can be transmitted in adjacent sub-channels. Alternatively, RBs can be divided into two pools. One pool is dedicated to transmit SCIs and the other one to transmit TBs.

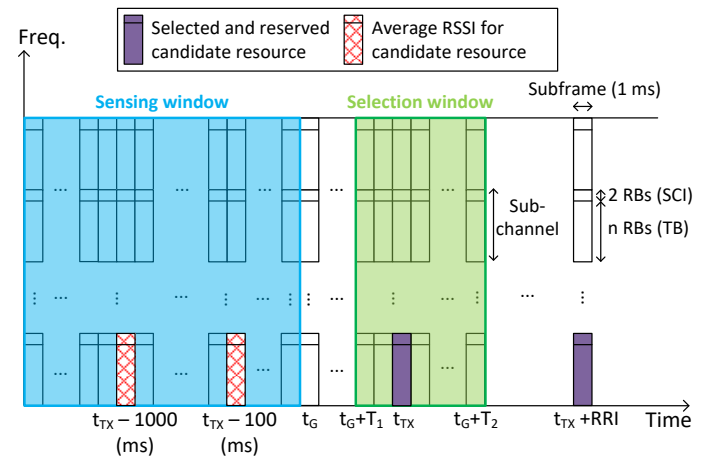

Fig. 2. LTE V2X channelization and illustration of mode 4 sensing-based SPS scheduling (illustrative example when T=100 ms).

### B. Resource Allocation

LTE V2X defines two new resource allocation modes (mode 3 and mode 4) for V2X SL communications [16]. In mode 3, the cellular infrastructure (eNB) manages the V2X SL communications. This includes selecting and configuring the communication resources (sub-channels). Mode 4 can operate without cellular infrastructure support. In this case, vehicles autonomously select, manage and configure the sub-channels. Vehicles utilizing mode 3 need to be in network coverage, while vehicles using mode 4 can operate out of network coverage.

The standard does not specify an algorithm for the selection of sub-channels in mode 3. Instead, it defines two scheduling approaches [14]: dynamic scheduling and Semi-Persistent Scheduling (SPS). With dynamic scheduling, vehicles must request sub-channels from the eNB for each TB. With SPS scheduling, the eNB reserves sub-channels so that a vehicle can transmit several TBs. The eNB can configure the periodicity of the reserved sub-channels [23] using the DCI (Downlink Control Information) transmitted over the Physical Downlink Control Channel (PDCCH). Mode 3 can outperform mode 4 since the scheduling of transmissions is centralized at the eNB ([20], [21]). However, it requires operating in network coverage and introduces cellular uplink (UL) and downlink (DL) signaling overhead. Mode 3 can also encounter challenges at the cell boundaries, in particular when different operators serve neighboring vehicles. 3GPP has defined an architecture to support multi-operator scenarios [22].

Under mode 4, vehicles autonomously select their sub-channels using the sensing-based SPS scheduling scheme specified in Rel. 14/15. A vehicle uses the selected sub-channels for the transmission of its following *Reselection Counter* consecutive TBs. The vehicle announces the reservation of the selected sub-channels for the transmission of the next TB using the *Resource Reservation Interval* (*RRI*) included in the SCI. This is illustrated in Fig. 2 where a vehicle selects sub-channel(s) at subframe $t_{TX}$, and informs neighboring vehicles that it reserves them for its following transmission at subframe $t_{TX} + RRI$. This is done to prevent other vehicles from utilizing the same sub-channels at the same time. The *RRI* can be equal to 0 ms, 20 ms, 50 ms, 100 ms or any multiple of 100 ms up to a maximum value of 1000 ms. A vehicle sets the *RRI* equal to 0 ms to announce neighboring vehicles that it is not reserving the same sub-channels for the next TB. A vehicle can only select *RRIs* values higher than 0 ms from a (pre-)configured list of permitted *RRI* values. This list can contain up to 16 values although currently 3GPP standards only define 12 possible *RRIs* values higher than 0 ms for mode 4 [19]. *Reselection Counter* is randomly chosen between 5 and 15 for a selected *RRI* higher than or equal to 100 ms. It varies between 10 and 30 for a selected *RRI* equal to 50 ms and between 25 and 75 for a *RRI* equal to 20 ms, respectively. *Reselection Counter* is decremented by one after transmitting a TB. A vehicle must select new sub-channels with probability (1-*P*) when *Reselection Counter* is equal to zero, where $P \in [0, 0.8]$. The standard does not specify a fixed value for *P*. New sub-channels must also be selected if a new TB does not fit in the previously reserved sub-channels or if the current reservation cannot satisfy the latency deadline of a new TB. The later occurs if the time until the next reserved sub-channels is higher than the latency deadline of the new TB. Vehicles select new sub-channel(s) using the sensing-based SPS scheduling scheme that operates as follows ([23], [24]):

1) The vehicle first identifies candidate resources within a Selection Window (SW). A candidate resource is a group of adjacent sub-channels within the same subframe where the new SCI+TB to be transmitted fits (see Fig. 2). SW is a time window that includes the subframes in the range $[t_G+T_1, t_G+T_2]$, where $t_G$ is the subframe at which the vehicle wants to select a new candidate resource. $T_1$ is the processing time (in subframes) required by a UE to identify and select candidate resources for transmission. $T_1$ is left to UE implementation but it must be equal to or smaller than 4 subframes. The value of $T_2$ (in number of subframes) is left to UE implementation but must be included within the range $T_{2min} \leq T_2 \leq 100$ if $T_{2min}$ has been (pre-)configured. If $T_{2min}$ has not been (pre-)configured, then the value of $T_2$ (in number of subframes) must be in the range $20 \leq T_2 \leq 100$. $T_{2min}$ depends on the priority of the transmission and its value (in number of subframes) must be within the range $10 \leq T_{2min} \leq 20$ [19]. Additionally, $T_2$ must be set so that $t_G+T_2$ fulfils the maximum latency requirement for transmitting the TB. The maximum latency is equal to 100 ms, 50 ms and 20 ms when vehicles transmit 10 packets per second (pps), 20 pps and 50 pps, respectively.

2) The vehicle then identifies candidate resources from SW that it should exclude. To do so, it senses the transmissions from other vehicles during the last 1000 subframes before $t_G$ (*sensing window* in Fig. 2). If the vehicle was transmitting in any previous subframe $t_N$ in the sensing window, it excludes all candidate resources within any subframe $t_N+q*RRI_i$ located in the selection window where $q$ is an integer value defined in [23]. $RRI_i$ is equal to any value of the *RRI* (in number of subframes in the above expression) included in the (pre-)configured list of permitted *RRIs*. Note that if the list of permitted $RRI_i$ only includes values higher than or equal to 100 ms, then $q$ can take only the value 1. These resources are excluded because the vehicle could not sense the transmissions from other vehicles at the same subframe it was transmitting a TB. We should note that other vehicles could be transmitting with any of the permitted values of $RRI_i$. Additionally, the vehicle should exclude all candidate resources within any subframe $t_F$ of the selection window if it estimates that any of its following $j$ transmissions at $t_F+j*RRI_{TX}$ (with $1 \leq j \leq 10*ReselectionCounter-1$ and $RRI_{TX}$ being the *RRI* selected by the vehicle) can coincide with any subframe $t_N+q*RRI_i$. Again, if the list of permitted $RRI_i$ only includes values higher than or equal to 100 ms, then $q$ can take only the value 1. The vehicle also excludes the candidate resources that it estimates will be used by other vehicles. A candidate resource is estimated to be used by other vehicles, and hence excluded, if the two following conditions are met: 1) the vehicle estimates with an SCI received from another vehicle that this other vehicle will utilize this candidate resource in the current SW or at the same time the vehicle will need it to transmit any of its following 10*ReselectionCounter-1 TBs; 2) the vehicle excludes a candidate resource if its average Reference Signal Received Power (RSRP) measured over the TB associated to the corresponding SCI is higher than a threshold that can be configured. The vehicle checks then if the number of remaining available candidate resources is equal or higher than 20% of all candidate resources within the SW. If this is not the case, the RSRP threshold is increased by 3 dB and the process to identify available candidate resources is repeated iteratively until the number of available candidate resources is at least equal to 20% of the candidate resources within SW. The vehicle creates a list $L_1$ with the identified available candidate resources.

3) The vehicle creates a second list $L_2$ with the candidate resources from $L_1$ experiencing the lowest average RSSI (Received Signal Strength Indicator) in the sensing window. The total number of candidate resources in $L_2$ must be greater than or equal to 20% of all candidate resources in the SW. The RSSI of a candidate resource located at subframe $t_{TX}$ is computed by averaging the RSSI over all the RBs of the candidate resource. The average RSSI is computed over the previous $t_{Tx}$-$T*j$ subframes located in the sensing window, where $j$ is a positive integer (note that $T$ is in number of subframes in the above expression). If the vehicle has selected an $RRI \geq 100$ ms, then $T$=100 ms. When the $RRI$ is set equal to 50 ms then $T$=50 ms and when $RRI$=20 ms then $T$=20 ms.

4) The vehicle randomly chooses one of the candidate resources from $L_2$ to reduce the probability that multiple vehicles select the same candidate resource or candidate resources that partially overlap. The selected candidate resource is used by the vehicle to transmit its new TB, and the vehicle maintains the selection for its next *ReselectionCounter*-1 transmissions.

The operation of LTE V2X mode 4 depends on parameters (e.g., $P$ and RSRP threshold) that can be (pre-)configured. An analysis of their configuration can be found in [20] and [25].

Since LTE V2X mode 4 utilizes a sensing-based scheduling mechanism, it is prone to packet collisions as the network load and channel congestion increase [26]. 3GPP defines two metrics to characterize the channel congestion: Channel Busy Ratio (CBR) and Channel occupancy Ratio (CR). The CBR is defined as the ratio of sub-channels that experience an RSSI higher than a (pre-)configured threshold to the total number of sub-channels in the previous 100 subframes. The CR quantifies the channel occupancy generated by the transmitting vehicle. It is estimated in subframe $n$ as the ratio between the number of sub-channels utilized by the transmitting vehicle in subframes *[n-a, n-1]* and selected by the vehicle for its remaining *Reselection Counter* transmissions in subframes *[n, n+b]* and the total number of sub-channels within *[n-a, n+b]*. $a$ and $b$ must satisfy $a+b+1 = 1000$ with a $\geq 500$. The standard defines up to sixteen CBR ranges. For each range, the standard specifies a $CR_{Limit}$ that cannot be surpassed by a transmitting vehicle and that can take different values as a function of the priority of the transmission. When a vehicle wants to transmit a TB, it measures the CBR and maps it to one of the ranges to get the $CR_{Limit}$. The vehicle also estimates its CR and if it is higher than the $CR_{Limit}$, it adjusts its transmission parameters. For example, the vehicle could drop certain packets (while maintaining the reserved sub-channels) or reduce the transmission power to decrease the CBR measured by other vehicles and maintain its CR. A vehicle could also augment the MCS to utilize less sub-channels to transmit a TB and hence reduce its CR. 3GPP does not specify a particular congestion control mechanism. However, first studies have analyzed the performance achieved with the Decentralized Congestion Control (DCC) technique defined by ETSI [27] or the distributed congestion control mechanism specified in the SAE J2945/1 standard [28].

## III. 5G NR V2X USE CASES

5G NR V2X has been designed to complement LTE V2X. LTE V2X supports basic active safety and traffic management use cases while 5G NR V2X supports advanced use cases and higher automation levels. The 5G NR V2X use cases have been specified by 3GPP Services and System Aspects (SA) Working Group 1 (SA1) and have been further elaborated by the 5GAA Working Group 1 (Use Cases and Technical Requirements). Both 3GPP and 5GAA organize use cases in groups, and a use case can be a member of more than one group. This section provides an overview of the use cases supported by 5G NR V2X along with their main KPIs. Readers are referred to 3GPP [11] and 5GAA [13] documents or papers on the topic [29] for their comprehensive description.

### A. 3GPP use case groups

3GPP technical report (TR) 22.886 [11] and TS 22.186 [12] present a comprehensive description of the NR V2X use cases and requirements, respectively. For each use case, 3GPP further distinguishes different degrees of automation following the SAE automation levels [30] ranging from 0 (no automation) to 5 (full automation). Typically, the higher automation levels of a use case, the more stringent the NR V2X QoS requirements are. The use cases are divided in the following four groups [11]:

1) *Vehicles Platooning*: This group includes use cases for the dynamic formation and management of groups of vehicles in platoons. Vehicles in a platoon exchange data periodically to ensure the correct functioning of the platoon. The inter-vehicle distance between vehicles in a platoon may depend on the available QoS.

2) *Advanced Driving:* This group includes use cases enabling semi-automated or fully-automated driving. In this group, vehicles share data obtained from their local sensors with surrounding vehicles in proximity. In addition, vehicles share their driving intention in order to coordinate their trajectories or maneuvers, thus increasing safety and improving traffic efficiency.

3) *Extended Sensors:* This group enables the exchange of sensor data – either raw or processed – collected through local sensors between vehicles, RSUs, devices of pedestrians, and V2X application servers. The objective is to improve the perception of the environment beyond the perception capabilities of the vehicles' own sensors.

4) *Remote Driving:* This group enables a remote (tele-operated) driver or a V2X application to operate a vehicle. The main use cases are for passengers who cannot drive themselves, for vehicles located in hazardous environments (e.g., construction areas or locations with adverse weather conditions), and for complex situations which automated vehicles are unable to drive safely.

### B. 5GAA use case groups

5GAA combines the use cases and groups defined by 3GPP in Rel. 14 [31] and Rel. 15 [11] with new use cases in Rel. 16, and defines the following groups [13]:

1) *Safety:* This group includes use cases that provide safety for vehicles and other traffic participants. It includes basic safety use cases such as emergency braking, collision warning, etc. [31], along with more advanced use cases requiring higher automation levels (e.g., intersection management).

2) *Vehicle operations management:* This group comprises commercial use cases aimed at improved operation of

TABLE II
REQUIREMENT RANGES FOR THE 3GPP USE CASE GROUPS [12]

| Use case group | Payload (Bytes) | Tx rate (Message/ Sec) | Max end-to-end latency (ms) | Reliability (%) | Data rate (Mbps) | Required communication range (meters) |
|---|---|---|---|---|---|---|
| Vehicles Platooning | 50-6000 | 2-50 | 10-25 | 90-99.99 | <= 65 | 80-350 |
| Advanced Driving | SL: 300-12000 UL: 450 | SL: 10-100 UL: 50 | 10-100 | 90-99.999 | SL: 10-50 UL: 0.25-10 DL: 50 | 360-700 |
| Extended Sensors | 1600 | 10 | 3-100 | 90-99.999 | 10-1000 | 50-1000 |
| Remote Driving | 16000-41700 | 33-200 | 5 | 99.999 | UL: 25 DL: 1 | 1000+ |

Note 1: If not specified otherwise, the requirement applies to all link types (SL, DL, and UL).
Note 2: In case of the Remote Driving use case group, [12] does not specify the values for Payload, Tx rate, and Required communication range. For completeness, we include these missing values based on [127].

vehicles for diverse types of users (e.g., vehicle owners/drivers, transport/delivery companies, etc.). Examples include sensors monitoring, software updates, remote support, etc.

3) *Convenience:* This group includes a diverse set of use cases that provide value and convenience to either the driver or the fleet management company operating the vehicle. Examples include infotainment, assisted navigation, and smart parking.

4) *Autonomous driving:* This group includes the advanced driving, remote driving, and extended sensors groups defined by 3GPP.

5) *Platooning:* This group is the same as the 3GPP vehicles platooning use case group.

6) *Traffic efficiency and environmental friendliness:* This group includes use cases that provide enhanced value to infrastructure or city providers in areas where the vehicles will be operating. Examples include Green Light Optimal Speed Advisory (GLOSA), traffic jam information, routing advice, etc.

7) *Society and community:* This group includes use cases that are of value and interest to the society and public. Examples in this group are Vulnerable Road User (VRU) protection, emergency vehicle approaching, emergency answering points, etc.

### C. Use case requirements

#### 1) 3GPP requirements

3GPP WG SA1 defined in Rel. 14 [32] and Rel. 15 [12] the key requirements for V2X services as follows:

1) *Payload* refers to the amount of data required by a specific service and generated by the application.

2) *Tx rate* is the number of messages per unit time that the transmitter generates and the receiver is expected to receive subject to other relevant requirements (e.g., payload size, latency, communication range, etc.).

3) *Maximum end-to-end latency* is the maximum allowed time between the generation of a message at the transmitter's application and the reception of the message at the receiver's application.

4) *Reliability* is defined as the probability that a transmitted message is correctly received within a specified maximum

end-to-end latency subject to other relevant requirements (e.g., payload size, communication range, etc.).

5) *Data rate* represents the total amount of data that needs to be received by the receiver per unit time. It is related directly to the payload and the *Tx rate* and is measured in bits per second (bps). It is also subject to other requirements (e.g., latency, reliability etc.)

6) *Required communication range* specifies the minimum distance between a transmitter and its intended receiver allowing communication with a targeted payload size, maximum latency, reliability, and data rate.

Table II summarizes the range of values for the above requirements as identified by 3GPP [12] for the four 3GPP use case groups. Each range is defined by the minimum and maximum requirements based on [12]. Use cases typically do not need to meet all of the most stringent requirements simultaneously. For example, platooning might require relatively low latency and high reliability, but the required data rate is moderate (exchange of maneuver information among the platoon members). We note that higher degrees of automation generally lead to more stringent requirements.

#### 2) 5GAA requirements

For each use case, 5GAA defines multiple possible [13] use case scenarios, where the scenarios differ in terms of road configuration, actors involved, service flows, etc. 5GAA complements largely network-centric requirements laid out by 3GPP with the concept of Service Level Requirements (SLRs). SLRs focus more on the automotive-centric requirements (e.g., service level reliability, interoperability, positioning, etc.), and consider system-level aspects (e.g., vehicle density) that need to be supported by the network for a specific use case scenario. SLRs are established for each user story defined within a use case. A user story describes, from the user's perspective, a use case scenario that results in a specific SLR. Key SLRs are defined in Table III.

5GAA has so far not performed an exhaustive analysis of service level requirements for all use cases. Rather, it provided in [13] some examples of use cases, scenarios, and user stories for each use case group. Example of 5GAA use cases and user stories descriptions can be found in [13].

TABLE III
SERVICE LEVEL REQUIREMENTS DEFINED BY 5GAA [13]

| Service level Requirement (SLR) | SLR Unit |
|---|---|
| Range: same as 3GPP definition of required communication range. | [m] |
| Information requested/generated: Information that a service needs or creates. | Quality of information / Information needs |
| Service Level Latency: same as 3GPP definition of maximum end-to-end latency. | [ms] |
| Service Level Reliability: probability that at least one of a set of messages carrying the same information from the transmitter (i.e., including retransmissions on any of the layers) is correctly received within a specified maximum latency, subject to other relevant requirements. | [%] |
| Velocity: maximum speed of the user (vehicle, pedestrian, etc.) that is required by the use case story. | [m/s] |
| Vehicle Density: maximum density of vehicles in an area required by the use case story. | [vehicle/km$^2$] |
| Positioning: maximum allowed error in estimating the location of a user (vehicle, pedestrian, etc.) required by the use case story. | [m] |
| Interoperability/ Regulatory/ Standardization Required (yes/no): requirements by the use case story in terms of: 1) interoperability between users; 2) regulatory action (e.g., in terms of spectrum, safety, etc.); and 3) standardization of the use case story for its functioning. | Not applicable |

## IV. OVERVIEW OF 5G SYSTEM ARCHITECTURE FOR V2X COMMUNICATION OVER PC5 AND Uu

The 5G system architecture supports two operation modes for V2X communication, namely V2X communication over the PC5 reference point or interface and V2X communication over the Uu reference point or interface. The PC5 interface supports SL V2X communications for NR and LTE. V2X communications over Uu for UL and DL transmissions are possible under NR Non-Standalone (NSA) and Standalone (SA) deployments. In Rel. 16, V2X communication over Uu is only supported for unicast communications. However, Rel. 17 includes an ongoing SI [33] to discuss enhancements for multicast and broadcast transmissions in 5G. It should be noted that V2X messages over LTE-Uu can be broadcast in DL via Multimedia Broadcast Multicast Services (MBMS) [34]. Since NR V2X includes both the SL (over the PC5 interface) and UL/DL (over the Uu interface), when referring specifically to either the SL or the UL/DL, in the rest of the paper we name them explicitly (e.g., NR V2X SL).

Fig. 3 shows the high-level view of the 5G System architecture for V2X communication over PC5 and Uu interface. More specifically, Fig. 3 illustrates the roaming architecture with local breakout. Home Public Land Mobile Network (HPLMN) refers to the network that a user is a subscriber to while visited PLMN (VPLMN) is the network to which the UE roams to when leaving the HPLMN. When a UE roams, services provided by the HPLMN are used to provide V2X service related parameters to the VPLMN. The local breakout is a deployment option where the Session Management Function (SMF) for establishment, modification or release of a session and all User Plane Functions (UPFs) that are involved in a Protocol Data Unit (PDU) session (i.e., a logical connection between the UE and network) are under control of the VPLMN. Local breakout is critical to reduce the latency. V2X communication over PC5 supports roaming and inter-PLMN operations. In case of inter-PLMN V2X communication over PC5, the PC5 parameters must be configured consistently among the UEs within a certain geographical area.

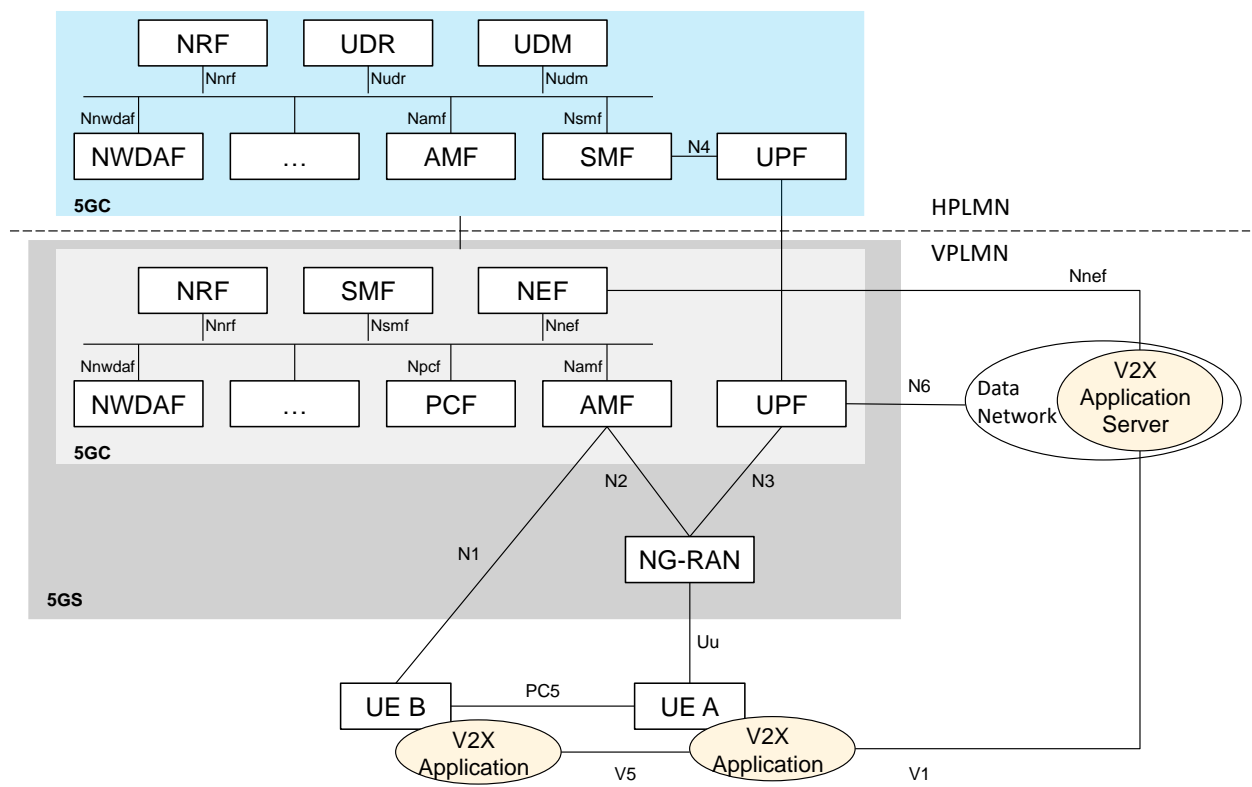

Fig. 3. 5G System architecture for V2X communication over PC5 and Uu – Local breakout scenario (Roaming case).

5G System (5GS) consists of the Next-Generation Radio Access Network (NG-RAN) and the 5G Core network (5GC) domains. The 5GC consists of several Network Functions (NF) such as the Access and Mobility Management Function (AMF), Policy Control Function (PCF), Network Data Analytics Function (NWDAF), Network Repository Function (NRF), Network Exposure Function (NEF), Unified Data Repository (UDR), Unified Data Management (UDM), UPF and SMF among others. It should be noted that in Fig. 3 the NFs within the 5GC Control Plane (that includes all NFs except UPF) use service-based interfaces for their interactions (e.g., Npcf, Nsmf, Nnef, Nnwdaf, etc). In service-based representation, NFs within the Control Plane enable other authorized network functions to access their services. Fig. 3 also shows the application layer interfaces between V2X applications in the UEs (interface V5) and the interface between V2X applications in the UE and in the V2X Application Server (AS) (interface V1). The application layer interfaces can be used to exchange application layer information and configuration parameters useful for the UE to configure its V2X communication. The 5GS architecture specified in [35] does not introduce new NFs to support V2X communication. Instead, existing NFs have been extended with V2X related functionalities. Next, we present a non-exhaustive list of these extensions at the 5GS and the UE.

The UE can report the V2X capability and V2X PC5 capability information to the 5GC (as well as receive V2X parameters from the 5GC) over the N1 reference point and/or from a V2X AS over the V1 reference point (see Fig. 3). The UE includes procedures for V2X communication over PC5 and for the configuration of parameters (e.g., destination Layer-2 IDs, radio resource parameters, V2X AS address information, mapping between V2X service types and V2X frequencies). A UE can receive the parameters for its V2X communications over PC5 and Uu from different sources. In this case, the UE shall consider them in the following priority order: a) provided/updated by the PCF through the AMF using Non-access stratum (NAS) signaling; b) provided/updated by the V2X AS via the V1 reference point; c) (pre-)configured in the Universal integrated circuit card (UICC); and d) pre-configured in the UE. The PCF can provide the UE with authorization and policy parameters for V2X communication over PC5 and Uu, for example, radio parameters when the UE is not served or under coverage by LTE or 5G. The PCF also provides the AMF with necessary parameters that are part of the UE context to configure and manage V2X communications. This includes, for example, parameters related to PC5 QoS flows and profiles. These parameters are retrieved from a UDR. The NRF helps other NFs to discover and select the appropriate PCF taking into account the V2X capabilities.

The 5GS provides NEF services to enable communication between NFs in the PLMN and the V2X AS. In addition, the NEF stores the V2X service parameters in the UDR. The V2X AS includes AF (Application Function) functionalities such as: request 5GC services (e.g., notifications about expected changes of provided QoS, QoS Sustainability Analytics), provision the UE and/or the 5GC with parameters for V2X communications over PC5 and/or Uu (e.g., mapping of V2X service types to V2X frequencies with geographical areas).

The interworking between 5GS V2X and 4G Evolved Packet System (EPS) V2X does not require new interfaces at the architecture level and does not impact existing network function entities in the 4G Evolved Packet Core (EPC) network entities and the 5GC entities. When the UE is in 5GS or EPS, the UE shall use the valid V2X policy and parameters provisioned by the PCF in 5GC or by the V2X Control Function (CF) in EPC for V2X communication. The V2X related parameters for EPS are defined in TS 23.285 [34]. They can be provided by the PCF or by the V2X CF while the V2X policy and parameters for 5GS are provided by the PCF only.

## V. PHYSICAL LAYER DESIGN FOR NR V2X SIDELINK

The physical layer structure for the NR V2X sidelink is based on the Rel. 15 NR Uu design. In addition, the physical layer procedures for the NR V2X SL reuse some of the concepts of Rel. 14 LTE V2X, with the introduction of additional procedures for providing physical layer support for unicast and groupcast transmissions.

### A. Physical Layer Structures for NR V2X sidelink

#### 1) Numerology

Rel. 16 NR V2X sidelink can operate at the same frequencies as Rel. 15 NR Uu [10], [36], i.e., at frequencies within the two following frequency ranges [37], [38]:

- Frequency range 1 (FR1): 410 MHz – 7.125 GHz.
- Frequency range 2 (FR2): 24.25 GHz – 52.6 GHz.

Although both frequency ranges are supported in NR V2X sidelink, the design of NR V2X sidelink has been based mainly on FR1 [10]. For NR V2X sidelink, no specific optimization is performed for FR2 [10], except for addressing phase noise which is more prominent at higher frequencies [39].

Transmissions in NR V2X SL use the orthogonal frequency division multiplexing (OFDM) waveform with a cyclic prefix (CP). The sidelink frame structure is organized in radio frames (also referred simply as frames), each with a duration of 10 ms. A radio frame is divided into 10 subframes, each with a duration of 1 ms. The number of slots per subframe and the subcarrier spacing (SCS) for the OFDM waveform can be flexible for NR V2X. To support diverse requirements and different operating frequencies in FR1 and FR2, a scalable OFDM numerology is considered for NR V2X based on Rel. 15 NR Uu. Each OFDM numerology is defined by an SCS and a CP. NR V2X supports multiples of 15 kHz (i.e., the SCS in LTE V2X) for the SCS of the OFDM waveform [36]. As shown in Table IV, different OFDM numerologies can be obtained with a scalable SCS given by $2^{\mu} \times 15$ kHz, where $\mu$ is an SCS configuration factor. For NR V2X, the SCS configuration factor can be $\mu = 0, 1, 2, 3$ such that the SCS can be equal to 15 kHz, 30 kHz, 60 kHz or 120 KHz. In FR1, 15 kHz, 30 kHz and 60 kHz are supported for the SCS, while 60 kHz and 120 kHz are supported for the SCS in FR2. Supporting higher SCS improves robustness of the OFDM waveform against frequency impairments caused by Doppler effects, carrier frequency offsets and hardware phase

TABLE IV
SUPPORTED NUMEROLOGIES IN NR V2X SIDELINK

| $\mu$ | SCS ($2^{\mu} \times 15$ kHz) | Frequency Range | Cyclic Prefix | Symbols per slot | Slots per subframe ($2^{\mu}$) | Slot duration ($2^{-\mu}$ ms) | Maximum Carrier Bandwidth |
|---|---|---|---|---|---|---|---|
| 0 | 15 kHz | FR1 | Normal | 14 | 1 | 1 ms | 50 MHz |
| 1 | 30 kHz | FR1 | Normal | 14 | 2 | 0.5 ms | 100 MHz |
| 2 | 60 kHz | FR1, FR2 | Normal | 14 | 4 | 0.25 ms | 200 MHz |
| | | | Extended | 12 | | | |
| 3 | 120 kHz | FR2 | Normal | 14 | 8 | 0.125 ms | 400 MHz |

noise [41], which are more prominent in FR2.

In NR V2X SL, the number of slots in a subframe is equal to $2^{\mu}$, i.e., it is determined by the SCS configuration factor $\mu$. As the subframe duration is 1 ms, the slot duration is given by $2^{-\mu}$ ms. Thus, a larger SCS results in a shorter slot duration. As in Rel. 15 NR Uu, two different CP lengths are supported in NR V2X sidelink: a first CP referred to as normal CP and a longer CP referred to as extended CP. A normal CP can be used for all supported SCS while the extended CP is supported only for an SCS of 60 kHz [36]. With the normal CP, increasing the SCS from 15 kHz to 60 kHz leads to a shorter CP (similar to how the slot duration is reduced by a factor of 4). However, this reduced normal CP with an SCS of 60 kHz may not be enough to accommodate for the delay spread in certain deployments and channels. For this purpose, Rel. 15 NR Uu introduced the extended CP for an SCS of 60 KHz, which has a similar duration as the normal CP for an SCS of 15 KHz [41]. Depending on whether a normal CP or an extended CP is used, each slot consists of 14 or 12 OFDM symbols, respectively. The duration of the normal or extended CP as well as the resulting OFDM symbol duration scale with $2^{-\mu}$, i.e., decreasing with increasing SCS. The exact expressions for the CP duration[2] and the resulting OFDM symbol duration is given in [36]. In the following, we refer to OFDM symbols simply as symbols.

As in NR Uu, the maximum bandwidth in NR V2X SL depends on the SCS ([37],[38]), see Table IV. Only one numerology (i.e., one combination of SCS and CP) can be used in a carrier at a time in NR V2X SL. As a general design guideline, the selection of the numerology depends on aspects like the carrier frequency, radio channel conditions (e.g., due to mobility), requirements (e.g., latency) and hardware features (e.g., complexity). For example, a large SCS may be preferred for low latency applications since a larger SCS leads to a shorter slot duration.

The smallest unit of time for scheduling SL transmissions in NR V2X is a slot. This is in contrast to Rel. 15 NR Uu which supports mini-slot scheduling, where a data transmission can be scheduled on only some of the OFDM symbols within a slot[3] [42]. NR V2X does not support mini-slot scheduling in the SL.

*2) Sidelink Bandwidth Parts*

In NR Uu, the maximum carrier bandwidth is 200 MHz for FR1 and 400 MHz in FR2. Although gNBs can support such wide bandwidths, this may not be the case for all UEs, in particular low-end UEs. Furthermore, supporting a very large bandwidth may also imply higher power consumption at the UE, both from the radio frequency (RF) and baseband signal processing perspectives. To support UEs that cannot handle large bandwidths (e.g., due to processing limitations or high power consumption), Rel. 15 NR Uu introduced the concept of bandwidth part (BWP) [43]. A BWP consists of a contiguous portion of bandwidth within the carrier bandwidth where a single numerology is employed. By defining a small BWP, the computational complexity and power consumption of a UE can be reduced. As each BWP can have a different bandwidth and numerology, BWPs enable a more flexible and efficient use of the resources by dividing the carrier bandwidth for multiplexing transmissions with different configurations and requirements.

The concept of BWP has also been adopted for the NR V2X sidelink ([44], [45]) where a SL BWP occupies a contiguous portion of bandwidth within a carrier as depicted in Fig. 4. In a carrier, only one SL BWP[4] is (pre-)configured for all UEs [46]. Sidelink UE transmissions and receptions are contained within the SL BWP and employ the same numerology. Thus, all physical channels, reference signals and synchronization signals in NR V2X sidelink are transmitted within the SL BWP. This also means that in the sidelink a UE is not expected to receive or transmit in a carrier with more than one numerology [45]. The SL BWP is divided into common RBs. A common RB consists of 12 consecutive subcarriers with the same SCS, where the SCS is given by the numerology of the SL BWP.

*3) Resource Pools*

In NR V2X, only certain slots are (pre-)configured to accommodate SL transmissions. Thus, the available sidelink resources consist of slots allocated for sidelink (time resources) and common RBs within a SL BWP (frequency resources). In NR V2X, a subset of the available SL resources is (pre-)configured to be used by several UEs for their SL transmissions. This subset of available SL resources is referred to as a resource pool [45] and is illustrated in Fig. 4. The common resource blocks within a resource pool are referred to as physical resource blocks (PRBs)[5]. A resource pool consists of contiguous PRBs and contiguous or non-contiguous slots that have been (pre-)configured for SL transmissions. A resource pool must be defined within the SL BWP [47].

---

[2]For the extended CP, all symbols have the same CP duration. For the normal CP, the first symbol every 0.5 ms has a slightly longer CP than the CP in the rest of the symbols [40].

[3]Mini-slot scheduling can be beneficial for latency critical transmissions and shorter data transmissions, e.g., ultra-reliable low latency communication (URLLC).

[4]This is in contrast to NR Uu where up to four BWPs can be configured for the downlink and uplink of a UE (although only one BWP is active for the downlink and one BWP is active for the uplink) [36].

[5]The PRBs are indexed within a resource pool while the common RBs are indexed within the SL BWP.

Therefore, a single numerology is used within a resource pool. If a UE has an active UL BWP, the SL BWP must use the same numerology as the UL BWP if they are both included in the same carrier [48]. Otherwise, the SL BWP is deactivated [48].

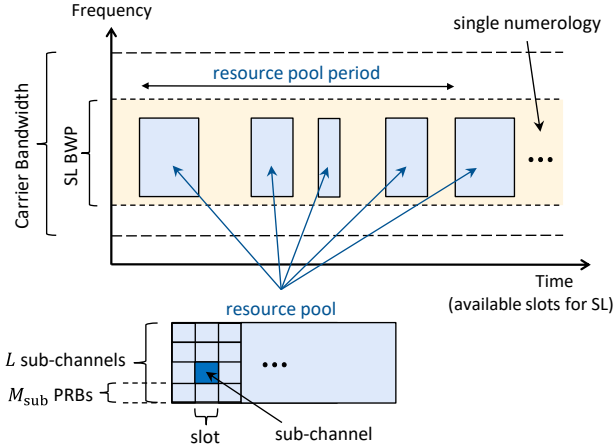

Fig. 4. SL bandwidth part and resource pool for NR V2X sidelink.

In the frequency domain, a resource pool is divided into a (pre-)configured number $L$ of contiguous sub-channels, where a sub-channel consists of a group of consecutive PRBs in a slot [42]. The number $M_{sub}$ of PRBs in a sub-channel corresponds to the sub-channel size, which is (pre-)configured within a resource pool. In NR V2X SL, the sub-channel size $M_{sub}$ can be equal to 10, 12, 15, 20, 25, 50, 75, or 100 PRBs. A sub-channel represents the smallest unit for a sidelink data transmission or reception. A sidelink transmission can use one or multiple sub-channels. In the time domain, the slots that are part of a resource pool are (pre-)configured and occur with a periodicity of 10240 ms [46]. The slots that are part of a resource pool can be (pre-)configured with a bitmap [49]. The length of the bitmap can be equal to 10, 11, 12, …, 160 [46].

At each slot of a resource pool, only a subset of consecutive symbols are (pre-)configured for the sidelink [42], i.e., out of the 14 or 12 symbols per slot for a normal or extended CP, respectively. The subset of SL symbols per slot is indicated with a starting symbol and a number of consecutive symbols, where these two parameters are (pre-)configured per resource pool [55]. The number of consecutive SL symbols[6] can vary between 7 and 14 symbols [42], e.g., depending on the physical channels which are carried within a slot (e.g., see Fig. 5 and Fig. 6).

A resource pool (RP) can be shared by several UEs for their SL transmissions. An RP can be used for all transmission types (i.e., unicast, groupcast, and broadcast). A UE can be (pre-)configured with multiple RPs for transmission (transmit RPs) and with multiple RPs for reception (receive RPs) [45]. A UE can then receive data on resource pools used for SL transmissions by other UEs, while the UE can still transmit on the SL using its transmit resource pools. For the case when UEs in network coverage do not have a stable network connection, exceptional transmit RPs are configured for the UEs [45]. These situations include when a UE is in a transition from idle to connected mode, when a UE experiences a link failure or a handover, or when a UE is changing between different configured transmit RPs [45]. The use of exceptional transmit RPs in such situations aids in improving service continuity [50].

### B. Physical Channels and Signals in NR V2X sidelink

The physical channels specified in NR V2X SL [36]:

1) *Physical Sidelink Control Channel (PSCCH)*: carries control information in the sidelink.
2) *Physical Sidelink Shared Channel (PSSCH)*: carries data payload in the sidelink and additional control information.
3) *Physical Sidelink Broadcast Channel (PSBCH)*: carries information for supporting synchronization in the sidelink. PSBCH is sent within a sidelink synchronization signal block (S-SSB).
4) *Physical Sidelink Feedback Channel (PSFCH)*: carries feedback related to the successful or failed reception of a sidelink transmission.

Furthermore, the following signals (carried in or associated with the previous channels) are specified in NR V2X [36]:

1) *Demodulation reference signal (DMRS)*: used by a receiver for decoding the associated sidelink physical channel, i.e., PSCCH, PSSCH, PSBCH. The DMRS is sent within the associated sidelink physical channel.
2) *Sidelink primary synchronization signal (S-PSS)* and *sidelink secondary synchronization signal (S-SSS)*: used by a receiver to synchronize to the transmitter of these signals. S-PSS and S-SSS are sent within an S-SSB.
3) *Sidelink Channel state information reference signal (SL CSI-RS)*: used for measuring channel state information (CSI) at the receiver that is then fed back to the transmitter. The transmitter can adjust its transmission based on the fed back CSI. SL CSI-RS is sent within the PSSCH region of the slot.
4) *Sidelink Phase-tracking reference signal (SL PT-RS)*: used for mitigating the effect of phase noise (in particular at higher frequencies) resulting from imperfections of the oscillator. SL PT-RS is sent within the PSSCH region of the slot.

Compared to LTE V2X that supports only broadcast transmissions in the sidelink, NR V2X provides physical layer support for unicast, groupcast, and broadcast transmissions in the SL. In the following, we refer to a UE performing a transmission as a TX UE while the intended recipients of a transmission are denoted as RX UEs. In NR V2X SL, data is organized into TBs [42] and each TB is associated with a SCI. A TB is carried in a PSSCH. The SCI indicates the resources used by the PSSCH that carries the associated TB, as well as further information required for decoding the TB. A PSCCH is sent with a PSSCH. The SCI in NR V2X is transmitted in two stages [45] compared to a single one for LTE V2X. The 1st-stage SCI in NR V2X is carried on the PSCCH while the 2nd-stage SCI is carried on the corresponding PSSCH. As discussed in Section V.B.2), the introduction of the 2nd-stage SCI enables a flexible SCI design to support unicast, groupcast, and broadcast transmissions in NR V2X, in contrast to LTE V2X where only broadcast is supported. Splitting the SCI in two stages (1st-stage SCI and 2nd-stage SCI) allows other UEs which are not RX UEs of a transmission to decode only the 1st-stage SCI for channel sensing purposes, i.e., for determining the resources reserved by other transmissions. On the other hand, the 2nd-stage SCI provides additional control information which is required for the RX UE(s) of a transmission. UEs performing channel sensing are referred in the following as sensing UEs.

---

[6]Differently from mini-slot scheduling from NR Uu, the number of symbols configured for a SL transmission in NR V2X is not changed dynamically.

NR V2X introduces the use of feedback in unicast and groupcast communications in order to increase the reliability of SL transmissions. The feedback consists of hybrid automatic repeat request (HARQ) feedback. Two options for sending the SL HARQ feedback are possible for groupcast communication: option 1 and option 2. In option 1, only RX UEs at a specified distance to the TX UE (smaller than the required communication range) should send HARQ feedback. The required communication range (see Section III.C.1)) depends on the service and represents the minimum distance for which the QoS parameters of a service need to be fulfilled [51]. In option 2, all RX UEs send HARQ feedback. The RX UEs send the HARQ feedback on PSFCH in response to a unicast or groupcast transmission carried in a PSSCH. Further details about the HARQ procedure are provided in Section V.C.1). NR V2X also supports CSI reporting in unicast communications. To this aim, a TX UE transmits CSI-RS so that a RX UE can measure the CSI and report it back to the TX UE via CSI reporting carried within a PSSCH.

In NR V2X, a UE can also transmit information for supporting synchronization in the sidelink. If it does, the UE serves as a synchronization reference and is referred to as a SyncRef UE [52]. The synchronization information in NR V2X SL is carried on the S-SSB that consists of the PSBCH, S-PSSS and S-SSS. Nearby UEs that may be out of network or GNSS coverage can receive S-SSB transmissions from a SyncRef UE and synchronize to it. Thus, nearby UEs can then have the same SL timing reference and establish SL communication to/from the SyncRef UE and among nearby UEs. It should be noted that a SyncRef UE is not always a TX UE, i.e., that transmits data.

PSCCH is sent on the same slot as the associated PSSCH, and such a slot is referred to as a PSCCH/PSSCH slot. While PSCCH and PSSCH can be sent in every slot of a resource pool, resources for PSFCH are allocated periodically, with a transmission on PSFCH performed in response to a PSSCH received a few slots before. Furthermore, S-SSBs are transmitted periodically within a SL BWP on slots which are not part of a resource pool. The following sub-sections present in more detail the physical sidelink channels, the S-SSB, the associated signals and the two stage SCI design. A summary is presented afterwards in Table VI.

### 1) Physical Sidelink Control Channel (PSCCH)

The PSCCH carries the 1st-stage SCI that contains control information associated with a PSSCH and the 2nd-stage SCI. For this purpose, SCI format 1-A is used [53]. The 1st-stage SCI indicates the frequency resources (e.g., sub-channels) of the PSSCH carrying the current (re)transmission of a TB, as well as the resource reservation for up to two further retransmissions of the TB. The 1st-stage SCI also informs about the resource reservation period if the UE reserves resources semi-persistently for PSSCH (see Section VI.B). In addition, the 1st-stage SCI includes the priority of the associated PSSCH, as well as the format and size of the 2nd-stage SCI [53]. The 1st-stage SCI also indicates the MCS of the data payload carried in the associated PSSCH. The MCS is determined with an MCS index within an MCS table. In NR V2X SL, a specific MCS table is used by default and additionally up to two further MCS tables can be (pre-)configured per resource pool [54]. The MCS tables are selected among the three MCS tables that are supported for the physical downlink shared channel (PDSCH) in Rel. 15 NR Uu [42]. To indicate the MCS of a PSSCH, the 1st-stage SCI indicates the MCS index as well as the MCS table, if one or two MCS tables are (pre-)configured within the resource pool. For supporting different channel conditions, the DMRS associated with a PSSCH in NR V2X can be carried on different symbols within a PSSCH slot, i.e., with different time patterns. Within a resource pool, multiple time patterns can be (pre-)configured for the PSSCH DMRS, and the 1st-stage SCI indicates which time pattern is used for the associated PSSCH. The 1st-stage SCI also provides the number of ports of the PSSCH DMRS, which can be equal to one or two. This represents the number of layers (i.e., number of data streams) supported in the PSSCH. Thus, by exploiting multiple transmit and receive antennas up to two streams of data can be sent within a PSSCH in NR V2X SL.

The PSCCH is multiplexed in non-overlapping resources with the associated PSSCH in the same slot. The PSSCH is transmitted from the second SL symbol in the slot and starting from the lowest PRB within the sub-channel(s) occupied by the associated PSSCH. The number of symbols for the PSCCH is (pre-)configured per resource pool and can be equal to 2 or 3 symbols. In the frequency domain, the PSCCH occupies a (pre-)configurable number $M_{PSCCH}$ of PRBs per resource pool that can be equal to 10, 12, 15, 20 or 25 PRBs [55]. However, as the PSCCH is to be contained within one sub-channel [56], the number $M_{PSCCH}$ of PRBs for the PSCCH is limited by the number $M_{sub}$ of PRBs in a sub-channel[7], i.e., $M_{PSCCH} < M_{sub}$. The possible number of symbols and PRBs for PSCCH allow different allocations of PSCCH in the time and frequency domain. For instance, for a large sub-channel size (e.g., $M_{sub} = 75$ PRBs), the PSCCH can occupy a large number of PRBs (i.e., $M_{PSCCH} = 25$ PRBs), and hence 2 symbols for the PSCCH may suffice. On the other hand, for a smaller sub-channel size (e.g., $M_{sub} = 15$ PRBs), the PSCCH may only be able to occupy $M_{PSCCH} = 10$ or 12 PRBs and thus, the PSCCH may need to use 3 symbols. Fig. 5 shows examples of a PSCCH/PSSCH slot with a PSCCH of 2 or 3 symbols.

A cyclic redundancy check (CRC) of 24 parity bits is appended to the 1st-stage SCI payload [53] to support error

---

[7]In fact, the sub-channel size $M_{sub}$ should take into account the number $M_{PSCCH}$ of PRBs that a PSCCH may need to occupy, e.g., depending on the

service [57]. For example, if 2 symbols are preferred for PSCCH, e.g., for faster PSCCH decoding, this may impose a minimum number of PRBs for PSCCH.

detection. The $1^{st}$-stage SCI payload with the appended CRC is encoded with the same Polar code as used for the physical downlink control channel (PDCCH) in Rel. 15 NR Uu [53]. The coded bits are then modulated with QPSK and mapped onto the resources for PSCCH [36].

As the number of symbols and number of PRBs for PSCCH are both (pre-)configured, the size of the $1^{st}$-stage SCI is fixed within a resource pool. Since the resource pool can be used for any transmission type [44], the payload size of the $1^{st}$-stage SCI is the same for unicast, groupcast or broadcast transmissions within a resource pool. With $L$ sub-channels within a resource pool, there are $L$ possible locations for a PSCCH in a slot, starting from the second SL symbol in a slot and from the lowest PRB in each sub-channel. To receive a $1^{st}$-stage SCI in NR V2X, a UE needs to check the $L$ possible PSSCH locations at each slot within a resource pool. This is in contrast to LTE V2X where a UE needs to check different pairs of PRBs to determine if they carry a PSCCH [45].

For demodulating the PSCCH, DMRS is transmitted within the PSCCH. The PSCCH DMRS follows the design of the DMRS associated with PDCCH in Rel. 15 NR Uu [44]. The PSCCH DMRS reuses the same pseudo-random sequence used for the PDCCH DMRS in Rel. 15 NR Uu [36], with the PSCCH DMRS sequence initialization based on a (pre-)configured value per resource pool. In addition, every PSCCH symbol contains PSCCH DMRS. The pattern of PSCCH DMRS in the frequency domain also reuses the DMRS frequency pattern employed for the PDCCH in Rel. 15 NR Uu [44].

### 2) Physical Sidelink Shared Channel (PSSCH)

The PSSCH carries the $2^{nd}$-stage SCI and the data payload consisting of a TB [45]. The $2^{nd}$-stage SCI carries information used for decoding PSSCH and for supporting HARQ feedback and CSI reporting [53]. The $2^{nd}$-stage SCI indicates the Layer 1 source ID and destination ID of a transmission that represent identifiers (in the physical layer) of the TX UE and intended recipients (RX UEs) of the TB. The Layer 1 source ID allows an RX UE to know the identity of the TX UE. This is used for determining the PSFCH for HARQ feedback as explained in Section V.C.2). The $2^{nd}$-stage SCI also carries a one-bit new data indicator (NDI) that is used to specify whether the TB sent in the PSSCH corresponds to the transmission of new data or a retransmission. Furthermore, a HARQ process ID is also included in the $2^{nd}$-stage SCI in order to identify a TB. For instance, if the NDI notifies of a retransmission in a PSSCH, an RX UE can determine the TB for which the retransmission corresponds to using the HARQ process ID. The $2^{nd}$-stage SCI also informs about the redundancy version (RV) that depends on the index of a retransmission[8]. The $2^{nd}$-stage SCI also indicates whether HARQ feedback is enabled/disabled for the PSSCH [46]. In NR V2X, two formats are supported for the $2^{nd}$-stage SCI: SCI format 2-A and SCI format 2-B [53].

SCI format 2-A is used when there is no HARQ feedback or for supporting unicast HARQ feedback or groupcast HARQ feedback (option 1 or option 2). Thus, the $2^{nd}$-stage SCI with SCI format 2-A indicates the cast type among broadcast, unicast, and groupcast with HARQ feedback option 1 or option 2 [46]. To request CSI feedback from an RX UE, a one bit CSI request is sent in the $2^{nd}$-stage SCI with SCI format 2-A.

On the other hand, SCI format 2-B is used when there is no HARQ feedback or for supporting groupcast HARQ feedback option 1. For this purpose, the $2^{nd}$-stage SCI with SCI format 2-B also includes the required communication range and the TX UE's zone ID [46]. The zone ID is indicated with 12 bits so a given area can be divided into $2^{12}$ squared regions of equal size[9]. The TX UE's zone ID indicates the zone in which the TX UE is located, i.e., it provides an indication of the TX UE's location [56]. The required communication range is represented with 4 bits using a set of 16 (pre-)configured values that can be selected out of a defined set of possible values[10].

After decoding the $1^{st}$-stage SCI in PSCCH, an RX UE has the necessary information to decode the $2^{nd}$-stage SCI carried in PSSCH. Thus, no blind decoding of the $2^{nd}$-stage SCI is needed [44]. The $2^{nd}$-stage SCI is decoded using PSSCH DMRS. To support error detection, a 24 bits CRC is appended to the $2^{nd}$-stage SCI [46]. Like the $1^{st}$-stage SCI, the $2^{nd}$-stage SCI and the appended CRC is encoded with the same Polar coding as the NR PDCCH. The coded bits are modulated with QPSK [53].

The two-stage SCI in NR V2X reduces the complexity of the SCI decoding, not only for RX UEs, but also for sensing UEs which only need to decode the 1st-stage SCI to know which resources are reserved by a TX UE. This key advantage of the two-stage SCI stems from its design features [57]: (i) the $1^{st}$-stage SCI having a fixed size (independent of transmission type); (ii) the $1^{st}$-stage SCI being carried in PSCCH on a known possible location within a sub-channel; (iii) the $1^{st}$-stage SCI indicating the resources of the $2^{nd}$-stage SCI carried in PSSCH; and (iv) the $2^{nd}$-stage SCI having a varying payload size (depending on transmission type).

The PSSCH also carries the data payload, i.e., the TB. For error detection, A CRC of 24 parity bits is appended to the TB [53]. The TB is encoded using the Low-Density Parity-Check (LDPC) coding employed for PDSCH in Rel. 15 NR Uu [44]. If the TB size is larger than the maximum code block size that can be handled by the LDPC encoder, the TB (including the appended CRC) is divided into code blocks [53]. Each code block is encoded into codewords using the LDPC coding. The codewords of the TB are modulated using QPSK, 16-QAM, 64-QAM or 256-QAM [36]. The MCS is selected from the default MCS table or from the up to two further MCS tables which can be (pre-)configured per resource pool (See Section V.B.1)).

Before being mapped onto a PSSCH, the coded $2^{nd}$-stage SCI and coded TB are multiplexed according to the procedure described in Section 8.2.1 of [53]. Depending on the number of layers (i.e., number of data streams) supported in a PSSCH, the multiplexed $2^{nd}$-stage SCI and TB are mapped to one or two layers and precoded before being mapped to $L_{PSSCH}$ sub-channels of the PSSCH [42]. With $M_{sub}$ PRBs per sub-channel, the number of PRBs spanned by the PSSCH is $M_{PSSCH} = L_{PSSCH} \cdot M_{sub}$ PRBs, starting from the lowest PRB within the sub-channel carrying the corresponding PSSCH.

---

[8]This can be used for soft combining in HARQ with incremental redundancy [58], where each retransmission provides a different RV of the data.

[9]The sides of the zones are configurable per required communication range and resource pool and can be equal to 5, 10, 20, 30, 40 or 50 m [49].

[10]The defined set of possible values include: 20, 50, 80,100, 120, 150, 180, 200, 220, 250, 270, 300, 320, 350, 370, 400, 420, 450, 480, 500, 550, 600, 700, 1000 m [49].

PSSCH can be transmitted from the second SL symbol up to the second to last SL symbol in a slot. As 7 to 14 SL symbols can be (pre-)configured in a slot, PSSCH can be sent in 5 to 12 consecutive SL symbols. The number of PSSCH symbols depends on the number of SL symbols in a slot and whether PSFCH is sent in the slot. In the 2 or 3 SL symbols which carry PSCCH, the PSSCH can be multiplexed in the frequency domain with PSCCH (if the PSCCH does not span the entire $L_{PSSCH}$ sub-channels), resulting in 2 or 3 PSCCH/PSSCH symbols. In the SL symbols without PSCCH, the PSSCH spans all the $L_{PSSCH}$ sub-channels as shown in Fig. 5. The second SL symbol (containing the first PSCCH or PSCCH/PSSCH symbol) is duplicated in the first SL symbol that is used for automatic gain control (AGC[11]) purposes. In addition, the symbol after the last PSSCH symbol is used as a guard symbol. Any remaining SL symbols can be used for PSFCH or for a further guard symbol as discussed in Section V.B.4.

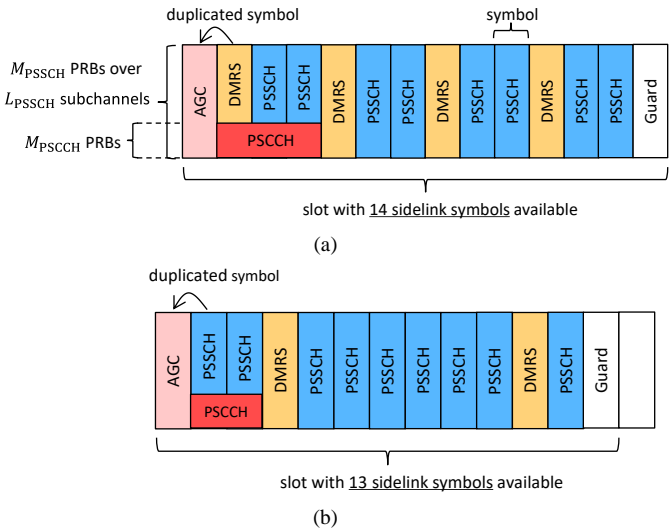

Fig. 5. Example of 2 configurations for a PSCCH/PSSCH slot (for normal CP). (a) Slot with 3 PSCCH, 12 PSSCH including 4 PSSCH DMRS symbols. (b) Slot with 2 PSCCH, 11 PSSCH including 2 PSSCH DMRS symbols.

For demodulating the PSSCH, DMRS is sent within PSSCH. The design of the PSSCH DMRS follows some aspects of the DMRS associated with the physical uplink shared channel (PUSCH) in Rel. 15 NR Uu and others from PDCCH DMRS [44]. The PSSCH DMRS reuses the pseudo-random sequence used for the Rel. 15 PUSCH DMRS [44], with the sequence initialization based on a (pre-)configured value per resource pool. In the frequency domain, the type 1 pattern configuration of PDSCH DMRS is employed for PSSCH DMRS [36].

The PSSCH DMRS can be transmitted in 2, 3, or 4 sidelink symbols at different locations within a slot, i.e., with different time patterns. The different time patterns for the PSSCH DMRS depend on the number of symbols for PSCCH, the number of symbols with PSSCH DMRS and the number of symbols for PSSCH within a slot. The time patterns supported for PSSCH DMRS in NR V2X are listed in Table 8.4.1.1.2-1 of [36]. For a resource pool, one or more time patterns for PSSCH DMRS can be (pre-)configured. In case multiple patterns are (pre-)configured, the DMRS time pattern used in a PSSCH is indicated in the associated 1st-stage SCI.

Fig. 5 depicts two examples of PSCCH/PSSCH slots for different number of PSCCH symbols, number of DMRS and number of PSSCH symbols, which result in different DMRS time patterns[12] based on Table 8.4.1.2.2-1 of [36]. In the figure, a PSSCH symbol with DMRS is shown just as DMRS. For both examples, the first and last sidelink symbol consist of an AGC symbol and a guard symbol, respectively.

### 3) SL CSI-RS and SL PT-RS

In NR V2X, the transmission of SL CSI-RS is supported for unicast transmissions only. The SL CSI-RS is sent in the PSSCH region of the slot. The design of the SL CSI-RS is based on the CSI-RS design of Rel. 15 NR Uu [36]. In addition, the resource mapping of SL CSI-RS in a PRB is based on the CSI-RS resource mapping patterns in NR Uu, which support up to two antenna ports (as in NR V2X SL up to two streams can be supported in a PSSCH). Each PRB within PSSCH uses the same pattern for the SL CSI-RS. SL CSI-RS is not transmitted on symbols containing PSCCH, the 2nd-stage SCI or PSSCH DMRS [46]. The SL CSI-RS configuration includes the resource mapping pattern and the number of antenna ports for SL CSI-RS. The SL CSI-RS configuration is selected by the TX UE and provided to the RX UE via PC5-RRC configuration (see Section VII.C).

The transmission of SL CSI-RS by a TX UE along with a CSI request sent in the 1st-stage SCI triggers the RX UE of a unicast link to feed back a CSI report. The TX UE can configure aperiodic CSI reporting from the RX UE [36]. The RX UE can measure the CSI based on the SL CSI-RS sent by the TX UE as it will be discussed in Section V.C.4. The CSI is fed back over a PSSCH sent from the RX UE to the TX UE. To avoid outdated CSI, the RX UE is expected to feed back the CSI report within a maximum amount of time. This maximum amount of time is referred as the latency bound. The latency bound is determined by the TX UE and signalled to the RX UE via PC5-RRC [54].

The transmission of SL PT-RS is supported for FR2 only. SL PT-RS is sent in the PSSCH region of the slot. The reception of SL PT-RS, allows an RX UE to track phase offsets due to Doppler effects and hardware impairments [59]. The design of SL PT-RS is based on the PT-RS design of Rel. 15 NR Uu [36]. The resource mapping of SL PT-RS is based on the PT-RS resource mapping in NR Uu.

---

[11]The received signal can vary over a wide dynamic range depending on the channel attenuation and interference. AGC is used to adjust the strength of the received signal in order to reduce the quantization error or the clipping of the signal at the analog to digital converter (ADC) [60].

[12]In Table 8.4.1.1.2-1 in [36], the number of PSSCH symbols listed in the first column includes the first sidelink symbol used for AGC as well as the PSSCH DMRS symbols.

### 4) Physical Sidelink Feedback Channel (PSFCH)

In NR V2X, the sole purpose of PSFCH is to carry the HARQ feedback from RX UE(s) to a TX UE. Within a resource pool, resources for PSFCH can be (pre-)configured periodically with a period of 1, 2 or 4 slot(s), i.e., there is a slot with PSFCH every 1, 2 or 4 slot(s) within a resource pool. PSFCH is sent in one symbol among the last SL symbols in a PSCCH/PSSCH slot as shown in the example in Fig. 6. Prior to the PSFCH symbol, one AGC symbol is used consisting of a copy of the PSFCH symbol. The symbol after the PSFCH symbol is used as a guard symbol. The three SL symbols associated with a PSFCH come after the PSSCH symbols as illustrated in Fig. 6. As a result, the number of PSFCH symbols (without the AGC and guard symbol) can be at most 9 symbols when a slot carries PSFCH. The DMRS time pattern of the slot in Fig. 6 results from Table 8.4.1.1.2-1 of [36], based on the given example with 2 PSCCH symbols, 7 PSSCH symbols and 2 PSSCH DMRS symbols.

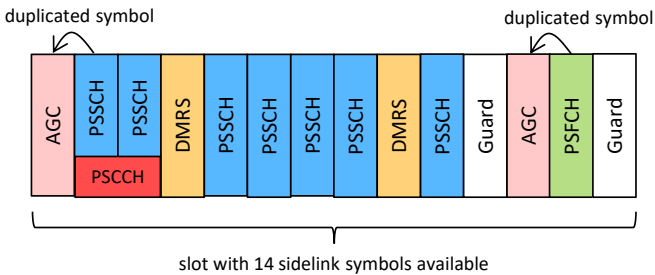

Fig. 6. Example of a PSCCH/PSSCH slot with a PSFCH, 2 PSCCH, 9 PSSCH including 2 PSSCH DMRS symbols (for normal CP).

In a resource pool, the set of PRBs in a PSFCH symbol that are available for PSFCH is indicated using a bitmap. NR V2X supports that all or only a subset[13] of the PRBs in a PSFCH symbol can be used for PSFCH. For a transmission of PSFCH, one PRB in a PSFCH symbol is used that carries a Zadoff-Chu sequence [61] based on the sequences used for the physical uplink control channel (PUCCH) in Rel. 15 NR Uu (PUCCH format 0) [36]. The procedure for selecting a PRB and the sequence in order to send HARQ feedback in a PSFCH is discussed in Section V.C.2).

### 5) Sidelink Synchronization Signal Block (S-SSB): PSBCH, S-PSS and S-SSS

Synchronization information can be transmitted by a SyncRef UE in the sidelink [52], e.g., to expand the synchronization coverage of a synchronization source and to enable that nearby UEs have the same sidelink timing reference. This allows SL communication to/from the SyncRef UE as well as SL communication between nearby UEs. The sidelink synchronization information is carried in an S-SSB that consists of PSBCH, S-PSS and S-SSS. The S-SSB occupies one slot and uses the same numerology as the one (pre-)configured in the SL BWP, i.e., the same numerology as for PSCCH/PSSCH. For a normal CP or extended CP, the PSBCH, S-PSSS and S-SSS are carried in the first 13 or 11 symbols of an S-SSB slot, respectively. An S-SSB slot for a normal CP is depicted in Fig. 7. The last symbol in an S-SSB slot is used as a guard symbol. The S-SSB is not frequency multiplexed with any other sidelink physical channel within the SL BWP, i.e., S-SSBs are not transmitted in the slots of a resource pool. In the frequency domain, the S-SSB spans $M_{S-SSB} = 11$ common RBs within the SL BWP [36]. Since an RB consists of 12 subcarriers, the S-SSB bandwidth spans $11 \times 12 = 132$ subcarriers. The frequency location of an S-SSB is (pre-)configured within a SL BWP [62]. As a result, a UE does not need to perform blind detection in the frequency domain to find an S-SSB [62].

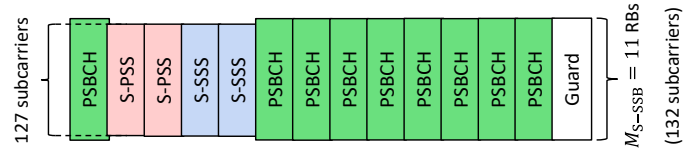

Fig. 7. S-SSB including PSBCH, S-PSS and S-SSS (for normal CP).

The S-PSS and S-SSS are jointly referred to as the sidelink synchronization signal (SLSS). The SLSS is used for time and frequency synchronization. By detecting the SLSS sent by a SyncRef UE, a UE is able to synchronize to the SyncRef UE and estimate the beginning of the frame and carrier frequency offsets. The UE can use the SL timing reference provided by the SyncRef UE for SL transmissions with nearby UEs that are using the same timing reference. Thus, a TX UE does not need to transmit also S-SSBs, i.e., not every TX UE needs to be a SyncRef UE. The S-PSS and the S-SSS consist each of sequences of 127 bits. The S-PSS is generated from maximum length sequences ($m$-sequences) using the same polynomial and initial values as the ones used for generating the $m$-sequences in the primary synchronization signal (PSS) in Rel. 15 NR Uu [36]. In NR Uu, there are three candidate sequences for PSS. However, only two candidate sequences are used for S-PSS[14]. The sidelink-SSS (S-SSS) is generated from Gold sequences that use the same design (i.e., generator polynomials, initial values and cyclic shifts) which is utilized for generating the Gold sequences for the secondary synchronization signal (SSS) in Rel. 15 NR Uu [36]. This results in 336 candidate sequences for S-SSS like for the SSS in NR Uu.

For the transmission of SLSS within an S-SSB, a SyncRef UE selects an S-PSS and S-SSS out of the candidate sequences based on an SLSS ID. The SLSS ID represents an identifier of the SyncRef UE and conveys a priority of the SyncRef UE as in LTE V2X. The procedure for selecting an SLSS ID is discussed in Section V.C.6). Each SLSS ID corresponds to a unique combination of an S-PSS and S-SSS out of the 2 S-PSS and 336 S-SSS candidate sequences, with the following relation SLSS ID $= 336 \times N_{S-PSS} + N_{S-SSS}$, where $N_{S-PSS} = \{0,1\}$ and $N_{S-SSS} = \{0,1, \dots ,335\}$ represent an identifier of the S-PSS and S-SSS among the S-PSS and S-SSS candidate sequences, respectively [36]. Thus, there are $2 \times 336 = 672$ unique SLSS IDs. Once a UE has detected the SLSS and identified the S-PSS and S-SSS transmitted by a SyncRef UE, it can determine the SLSS ID of the SyncRef UE. Thus, the transmission of SLSS within an S-SSB also conveys the SLSS ID of a SyncRef UE.

---

[13] As discussed in Section V.C.2), PRBs in a PSFCH symbol are distributed equally among several sub-channels for the HARQ feedback. This may result in leaving some PRBs unused, depending on the number of PRBs in a symbol.

[14] The two candidate sequences are generated from the same base sequence with two different cyclic shifts, namely, 25 and 65. The concept of cyclic shifts is discussed in Section V.C.2).

The S-PSS and S-SSS are modulated with BPSK such that each sequence occupies 127 subcarriers in a symbol within the S-SSB bandwidth as shown in Fig. 7. The S-PSS and S-SSS do not occupy the entire S-SSB bandwidth of 132 subcarriers, and are allocated from the third subcarrier relative to the start of the S-SSB bandwidth up to the 129th subcarrier. The S-PSS and S-SSS are both repeated in two consecutive symbols in an S-SSB. This provides higher coverage and allows a UE receiving an S-SSB to perform phase tracking between two consecutive symbols carrying the same sequence [45]. Independently of the CP, the S-PSS is sent in the second and third symbol of the S-SSB, while the S-SSS is sent in the fourth and fifth symbol of the S-SSB as shown in Fig. 7. PBSCH is sent in the first symbol and on the eight or six symbols after the S-SSS for a normal or extended CP, respectively. The first symbol of the S-SSB (first PSBCH symbol) is used for AGC. For a given CP, the structure of an S-SSB slot is fixed, in contrast to a PSCCH/PSSCH slot.

The main purpose of the PSBCH is to provide system-wide information and synchronization information that is required by a UE for establishing a sidelink connection. The information carried by the PSBCH includes a one bit indicator whether the SyncRef UE is in coverage of a network or of GNSS [44]. A UE is in coverage of GNSS when GNSS is reliable at the UE. GNSS is reliable if the UE can meet the accuracy requirements specified in [52]. SL transmissions are organized in frames identified by the direct frame number (DFN). Consequently, the PSBCH also indicates the DFN and the slot index as timing information [44]. The DFN enables a UE to synchronize its radio frame transmissions according to the SL timing reference. For a SyncRef UE in network coverage, the DFN can be derived based on the system frame number (SFN), where the SFN provides an indexing of the frames based on the cell timing reference. When a UE is out of network coverage, the DFN can be derived based on the coordinated universal time (UTC) provided by GNSS (i.e., GNSS timing) [63]. For a UE in network coverage but using GNSS as synchronization source for SL communication, the DFN can be derived from the GNSS timing and a timing offset with respect to the cell timing reference [63]. This timing offset enables to align the DFN and SFN timing to achieve a unified SL timing [64]. The time division duplex (TDD) configuration is also carried in the PSBCH [44]. The TDD configuration provides the TDD UL/DL configuration and/or the indication of the slots that can be contained in a resource pool. A UE that is out of coverage of a gNB can receive an S-SSB sent by a SyncRef UE that is in coverage of the gNB. From the PSBCH within the S-SSB, the out of network coverage UE also receives the TDD UL/DL configuration used by the gNB within its cell. This enables the ouf of network coverage UE to be aware of the slots that the gNB uses for the downlink within the cell. This UE can then avoid causing interference to UEs that are receiving a downlink transmission in the cell. To do so, it should avoid using the downlink slots of the cell for its SL transmissions [65]. The payload size of the PSBCH consists of 56 bits, including the 1 bit in coverage indicator, 12 bits for indicating the TDD configuration, 10 bits for the DFN and 7 bits for the slot index. The PSBCH payload also includes 2 reserve bits for future purposes as well 24 bits for a CRC. The PSBCH payload is encoded using the same Polar coding employed for the physical broadcast channel (PBCH) in Rel. 15 NR Uu [53]. The PSBCH occupies 11 RBs of the S-SSB bandwidth as depicted in Fig. 7.

After successfully detecting the SLSS sent by a SyncRef UE, a UE proceeds with decoding the PSBCH. For this purpose, DMRS are transmitted in every PSBCH symbol and on every fourth subcarrier (starting from the first subcarrier), following a similar structure as the DMRS in the PBCH in Rel. 15 NR Uu [36]. The PSBCH DMRS sequence generation is also based on the design of PBCH DMRS for Rel. 15, which resuls in 31-length Gold sequences using QPSK modulation [36]. The SLSS ID is used for the PSBCH DMRS sequence initialization [49].

S-SSBs are sent with a fixed periodicity of 160 ms (i.e., 16 frames). This is in contrast to the SSB structure in Rel. 15 NR Uu, which supports multiple periodicities. Similar to the SSB structure in NR Uu, multiple S-SSBs can be sent within one S-SSB period. Within one S-SSB period, the number of S-SSBs is (pre-)configurable depending on the SCS and the frequency range as given in Table V [44]. For larger SCSs, a larger number of S-SSBs within a period is supported such that sufficient SL coverage can be provided also for larger SCSs. S-SSBs are distributed within an S-SSB period according to two (pre-)configured parameters, namely the offset from the start of the S-SSB period to the first S-SSB and the interval between consecutive S-SSBs (see Fig. 8) [44]. The beginning of an S-SSB period corresponds to the first slot in a frame with the SFN satisfying SFN mod $16 = 0$, where mod is the modulo operator. Each frame within a period of 1024 frames can be identified by the SFN. The SFN allows defining transmission periods which are longer than one frame [41], e.g., as the S-SSB transmissions. When GNSS is used as the synchronization reference, the DFN is used instead of SFN [63].

TABLE V
NUMBER OF S-SSB WITHIN AN S-SSB PERIOD (160 MS)

| Frequency Range | Subcarrier Spacing (SCS) | Number of S-SSBs per period |
|---|---|---|
| FR1 | 15 kHz | 1 |
| | 30 kHz | 1, 2 |
| | 60 kHz | 1, 2, 4 |
| FR2 | 60 kHz | 1, 2, 4, 8, 16, 32 |
| | 120 kHz | 1, 2, 4, 8, 16, 32, 64 |

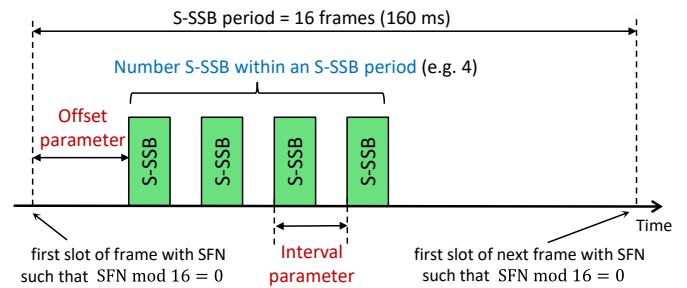

Fig. 8. Multiple S-SSB transmissions within an S-SSB period.

TABLE VI
OVERVIEW OF NR V2X PHYSICAL CHANNELS AND SIGNALS

| Physical Channels | Purpose | Associated Physical Signals | Additional Aspects |
|---|---|---|---|
| **PSCCH** (sent within first SL symbols of first sub-channel occupied by associated PSSCH) | PSCCH carries the 1st-stage SCI which indicates [53]: <br>• Resource allocation, modulation and coding scheme, and priority of associated PSSCH <br>• Resource reservation period <br>• Time pattern and number of ports for PSSCH DMRS <br>• Size and format of 2nd-stage SCI | • PSCCH DMRS [36]: based on Rel. 15 NR PDCCH DMRS | • Coding [53]: Polar coding used for Rel. 15 NR PDCCH <br><br>• Modulation: QPSK |
| **PSSCH** (sent in one or multiple sub-channels in a slot within a resource pool) | PSSCH carries a TB and the 2nd-stage SCI which indicates [53]: <br>• HARQ process ID, new data indicator and redundancy version <br>• Source ID and Destination ID <br>• HARQ enabled/disabled indicator <br>• Transmission type and CSI request (SCI format 2-A) <br>• TX UE's zone ID and required communication range (SCI format 2-B) | • PSSCH DMRS [36]: based on Rel. 15 NR PUSCH DMRS and PDSCH DMRS <br><br>• SL CSI-RS [36]: based on Rel. 15 NR CSI-RS <br><br>• SL PT-RS [36]: based on Rel. 15 NR PT-RS | 2nd-stage SCI: <br>• Coding [53]: Polar coding used for Rel. 15 NR PDCCH <br>• Modulation: QPSK <br><br>Transport block: <br>• Coding [53]: LDPC coding used for Rel. 15 NR PDSCH <br>• Supported modulation: QPSK, 16-QAM, 64-QAM, 256-QAM |
| **PSBCH** (sent periodically with the S-PSS and S-SSS in an S-SSB, not on slots of a resource pool) | PSBCH carries [44]: <br>• TDD configuration, <br>• Direct frame number <br>• Slot index <br>• in network/GNSS coverage indicator | • PSBCH DMRS [36]: based on Rel. 15 NR PBCH DMRS <br>• S-PSS [36]: $m$-sequence (based on Rel. 15 NR PSS) <br>• S-SSS [36]: Gold sequence (as Rel. 15 NR SSS) | • Coding for PSBCH [53]: LDPC coding used for Rel. 15 NR PBCH <br><br>• Modulation for PSBCH: QPSK <br><br>• Modulation for S-PSS and S-SSS: BPSK |
| **PSFCH** (sent periodically, in a symbol near end of a PSCCH/PSSCH slot) | As response to a PSSCH reception [66], PSFCH carries HARQ feedback: <br>• For unicast: ACK/NACK feedback <br>• For groupcast: NACK-only feedback (option 1) or ACK/NACK feedback (option 2) | | For a transmission of PSFCH, one PRB in a PSFCH symbol used carrying a <br>• Zadoff-Chu sequence [36]: based on Rel. 15 NR PUCCH format 0 |

As the SSB structure in NR Uu, the S-SSB structure supports the transmission of each S-SSB with a different beam, e.g. with beam sweeping [64]. Although the S-SSB structure is similar to the SSB structure, the SSBs transmission by a gNB are also part of the initial access procedure in NR Uu [66]. However, there is no such procedure in NR V2X sidelink. In NR Uu, after a UE has identified the best beamformed SSB (e.g., best beam from the gNB to reach the UE), the UE indicates this to the gNB via the physical random access channel (PRACH) in the uplink. This procedure is referred as random access and enables the gNB to be aware of the UE and of the best beam to reach the UE. As there is no similar procedure after the S-SSB transmissions in NR V2X SL, a SyncRef UE is not aware of the UEs that received its S-SSBs transmissions.

The main characteristics of the NR V2X SL physical channels and the S-SSB as well the associated signals and the two stage SCI design are summarized in Table VI.

### C.  Physical Layer Procedures for NR V2X sidelink

Rel. 16 NR V2X introduces unicast and groupcast communications for the sidelink. To support them, NR V2X introduces mechanisms for sidelink HARQ feedback and for determining the PSFCH to send the HARQ feedback. In addition, sidelink power control has also been introduced for NR V2X SL, as well as CSI acquisition for unicast transmissions. Similar to LTE V2X, NR V2X also supports procedures for SL transmission of synchronization information.

#### 1)  Sidelink HARQ feedback

HARQ increases the reliability of TB transmissions by employing forward error correction (FEC) and error detection codes in combination with a retransmission strategy [69]. With HARQ, a receiver requests a HARQ retransmission if it is not able to correct all transmission errors via the FEC code and it detects errors in the transmission via the error detection code. The HARQ retransmission can include data and parity bits to aid the successful reception of a TB. After a transmission, the transmitter waits for HARQ feedback from the receiver. The reply can consist of: (i) an acknowledgement (ACK) if the reception was successful; (ii) a negative acknowledgement (NACK) if the reception was unsuccessful; or (iii) no response if the control information associated with the transmission was not received successfully within a predefined time. In case of NACK or no response, the transmitter can perform a HARQ retransmission, with multiple retransmissions being possible.

Retransmissions in the SL can aid in meeting the reliability requirements for NR V2X use cases. For this purpose, blind retransmissions can also be considered [44], i.e., a certain number of retransmissions can be performed without any HARQ feedback. However, sidelink HARQ feedback can prevent unnecessary blind retransmissions that would otherwise waste SL resources and increase the channel load.

NR V2X supports ACK/NACK feedback for the SL HARQ feedback in unicast: an RX UE sends ACK if it has successfully decoded the TB carried in a PSSCH or sends NACK if it has not decoded the TB after decoding the $1^{st}$-stage SCI. For groupcast, two options (option 1 and option 2) are supported for the SL HARQ feedback in NR V2X. For option 1, an RX UE transmits NACK if it has not successfully decoded the TB (after decoding the $1^{st}$-stage SCI) and if its relative distance to the TX UE (referred as Tx-Rx distance) is less than or equal to the required communication range (indicated in the $2^{nd}$-stage SCI). Otherwise, the RX UE does not transmit any HARQ feedback. As the HARQ feedback for this option would only consist of NACK, option 1 is referred to as NACK-only feedback. Option 1 is illustrated in Fig. 9, which depicts scenarios where RX UEs do not transmit HARQ feedback because they have successfully decoded a TB or because they are outside the minimum required communication range (even if they did not successfully decode the TB). Option 2 for groupcast supports the ACK/NACK feedback from all RX UEs: an RX UE sends ACK if it has successfully decoded the TB or it sends NACK if it has not decoded the TB, after decoding the $1^{st}$-stage SCI. For unicast or any of the groupcast options, an RX UE does not send a reply if the RX UE does not decode the $1^{st}$-stage SCI.

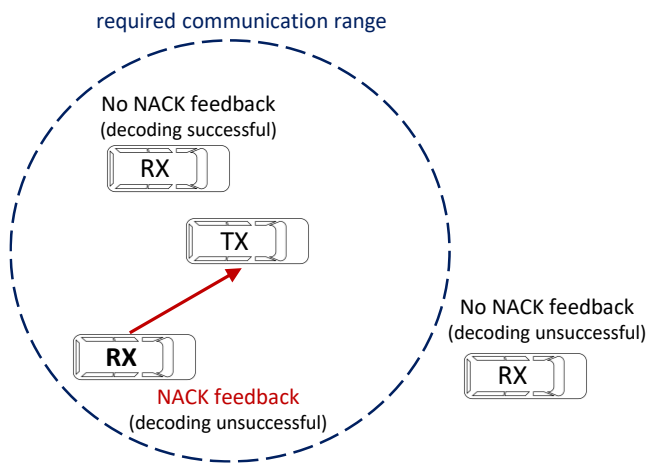

Fig. 9.  NACK-only feedback for groupcast NR V2X sidelink (option 1).

With option 1, the RX UEs of a transmission share a resource for sending their NACK-only feedback, while with option 2 each RX UE sends its ACK/NACK feedback on a separate resource. Thus, with option 1 a TX UE cannot identify which RX UEs sent NACK. If a TX UE receives at least one NACK with option 1, it is aware that at least one RX UE within the required communication range did not correctly decode the transmission. In addition, if a TX UE receives no reply with option 1, it cannot distinguish whether the RX UEs within the required communication range have successfully received the transmission or if some have not successfully decoded the corresponding $1^{st}$-stage SCI. On the other hand, a TX UE can distinguish the HARQ feedback of the RX UEs with option 2. This enables the TX UE to perform a retransmission tailored to specific RX UE(s). With option 2, if a TX UE receives no reply on the feedback resource corresponding to a given RX UE, it is aware that the RX UE has not successfully decoded the corresponding $1^{st}$-stage SCI. Consequently, option 2 enables higher reliability for the transmissions. The above advantages of option 2 over option 1, however, come at the expense of more resources required for the groupcast HARQ feedback. In NR V2X, the use of HARQ feedback and the choice whether to use option 1 or option 2 for groupcast HARQ feedback is up to UE implementation.

For groupcast communications, the TX UE indicates in the $2^{nd}$-stage SCI whether NACK-only feedback (option 1) or ACK/NACK feedback (option 2) should be used. NACK-only feedback may be considered for groupcast services where the provided information may not be so relevant for RX UEs outside the communication range, e.g., for the extended sensors use case (see Section III.A) [45]. For groupcast option 1, the Tx-Rx distance is obtained at the RX UE based on the location of the TX UE[15]. This location is indicated via the TX UE's zone ID carried in the $2^{nd}$-stage SCI. The Tx-Rx distance is derived at the RX UE based on its own location and the center of the nearest zone[16] with the indicated zone ID.

HARQ feedback for a TB sent on a PSSCH in a resource pool is carried on a PSFCH within the same resource pool. To disable the HARQ feedback for all SL transmissions in the resource pool, no resources for PSFCH are configured within the resource pool. Resources for PSFCH can be (pre-)configured periodically with a period of $N = 1$, 2 or 4 slot(s), i.e., there is a PSCCH/PSSCH slot with a PSFCH symbol every $N$ slots within the resource pool. Even if resources for PSFCH are (pre-)configured, the $2^{nd}$-stage SCI indicates whether HARQ feedback is enabled or not for a TB sent in a given PSSCH [56].

As discussed in Section V.B.4), one PRB is used for a PSFCH transmission. In one PRB, NR V2X supports code division multiplexing (CDM) between PSFCH transmissions of multiple RX UEs. With CDM, multiple transmissions share the same frequency resource simultaneously. To this aim, the transmissions should employ signals with low cross correlation (ideally orthogonal) [61]. This property allows a receiver to distinguish uncorrelated transmission signals that have been sent on the same time and frequency resources. Signals with low cross correlation can be generated from a periodic root or base sequence that has very good autocorrelation properties [70]. A set of uncorrelated signals for CDM can be generated from cyclically shifted versions of the base sequence [71]. The base sequence used for PSFCH is (pre-)configured per resource pool in NR V2X and it corresponds to a Zadoff-Chu sequence based on a format used for PUCCH in Rel. 15 NR Uu [36].

In a PRB used for PSFCH, the transmissions that can be multiplexed with CDM correspond to the HARQ feedback from several RX UEs. CDM is also used to distinguish the ACK or NACK feedback from an RX UE in the same PRB. To this aim, a pair of cyclic shifts (within a PRB) are used to distinguish the ACK or NACK from an RX UE [66]. The cyclic shift

---

UE's position with a zone ID. Out of all possible zones with the indicated zone ID, an RX UE assumes that the TX UE is in the zone closest to the RX UE.

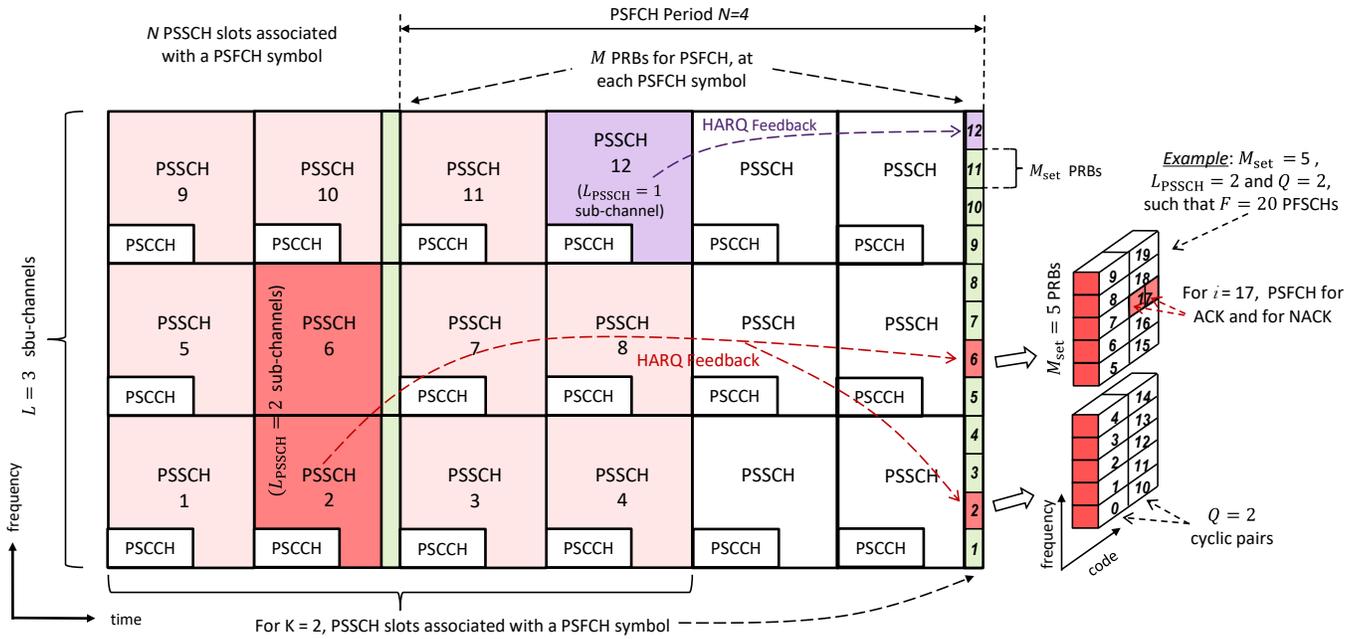

Fig. 10. PSFCHs for HARQ feedback associated with different transmissions.

corresponding to ACK is not defined in the case of NACK-only feedback for groupcast (option 1). Thus, each PSFCH in NR V2X is mapped to a time resource (PSFCH symbol), a frequency resource (one PRB), and a code resource (one cyclic shift among a cyclic shift pair). Next, we discuss how the PSFCH symbol, the PRB and cyclic shift are determined for the HARQ feedback for a given PSSCH transmission.

### 2) Resource selection for PSFCH

The PSFCH symbol that can be used for the HARQ feedback for a given PSSCH transmission corresponds to the PSFCH symbol in the first slot with PSFCH after a (pre-)configured number of $K$ slots after the PSSCH. $K$ represents the minimum number of slots within the resource pool between a slot with a PSSCH transmission and the slot containing PSFCH for the HARQ feedback of this transmission (see Fig. 11 for $K=3$). Let us consider that the last symbol of a PSSCH transmission is on slot $n$. The HARQ feedback for this transmission is expected in slot $n + a$, where $a$ is the smallest integer equal or higher than $K$ such that slot $n + a$ contains PSFCH. In the example shown in Fig. 11, the earliest possible slot for the HARQ feedback (slot $n+3$) does not contain PSFCH, so the HARQ feedback is sent at the next slot with PSFCH (slot $n+6$). The time gap of at least $K$ slots allows considering the RX UE's processing delay in decoding the PSCCH/PSSCH and generating the HARQ feedback. $K$ can be equal to 2 or 3, and a single value of $K$ can be (pre-)configured per resource pool[17]. This allows several RX UEs using the same resource pool to utilize the same mapping of PSFCH resource(s) for the HARQ feedback. With the parameter $K$, the $N$ PSSCH slots associated with a slot with PSFCH can be determined. In the example with $K=3$ shown in Fig. 11, the $N=4$ PSSCH slots associated with the PSFCHs in slot $n+6$ correspond to PSSCH slots $n$, $n+1$, $n+2$, and $n+3$.

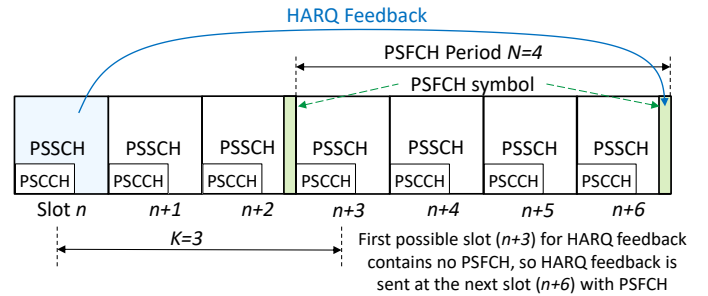

Fig. 11. PSSCH-to-HARQ feedback timing based on at least K=3 slots. For simplicity, the figure depicts only one sub-channel within the resource pool. We also omit the detailed structure of a PSCCH/PSSCH slot with or without PSFCH including PSSCH DMRS, AGC symbols and guard symbols.

With $L$ sub-channels in a resource pool and $N$ PSSCH slots associated with a slot containing PSFCH, there are then $N \cdot L$ sub-channels associated with a PSFCH symbol. With $M$ PRBs available for PSFCH in a PSFCH symbol, there are $M$ PRBs available for the HARQ feedback of transmissions over $N \cdot L$ sub-channels. With $M$ configured to be a multiple of $N \cdot L$ [49], then a distinct set of $M_{set} = M/(N \cdot L)$ PRBs can be associated with the HARQ feedback for each sub-channel within a PSFCH period. The first set of $M_{set}$ PRBs among the $M$ PRBs available for PSFCH are associated with the HARQ feedback of a transmission in the first sub-channel in the first slot. The second set of $M_{set}$ PRBs are associated with the HARQ feedback of a transmission in the first sub-channel in the second slot and so on. This is illustrated in Fig. 10 with $N = 4$, $L = 3$ and with all PRBs in a PSFCH symbol available for PSFCH. In this example, the HARQ feedback for a transmission at PSSCH $x$ is sent on the set $x$ of $M_{set}$ PRBs in the corresponding PSFCH

---

[17]In NR Uu, the PDSCH-to-HARQ timing (similar to the $K$ slots for the SL HARQ feedback) is signaled in the DCI for each DL transmission [53]. In NR

V2X, the PSSCH-to-HARQ feedback timing is not indicated in the SL control information, as it is (pre-)configured per resource pool with the parameter $K$.

symbol, with $x = 1, \ldots, 12$. For a transmission in a PSSCH with $L_{PSSCH} > 1$ sub-channels, $L_{PSSCH} \cdot M_{set}$ PRBs could be available for the HARQ feedback of this transmission. For example, let us consider the case of a transmission over the first and second sub-channel in the second slot (i.e., $L_{PSSCH} = 2$) in Fig. 10. There are then two sets of $M_{set}$ PRBs in the first and second sub-channel of the PSFCH symbol that could be available for the HARQ feedback of this transmission.

A set of $M_{set}$ PRBs associated with a sub-channel are shared among multiple RX UEs in case of ACK/NACK feedback for groupcast communications (option 2) or in the case of different PSSCH transmissions in the same sub-channel[18]. For each PRB available for PSFCH, there are $Q$ cyclic shift pairs available to support the ACK or NACK feedback of $Q$ RX UEs within the PRB. For a resource pool, the number of cyclic shift pairs $Q$ is (pre-)configured and can be equal to 1, 2, 3 or 6. We can compute the number $F$ of PSFCHs available for supporting the HARQ feedback of a given transmission. With each PSFCH used by one RX UE, $F$ available PSFCHs can be used for the ACK/NACK feedback of up to $F$ RX UEs[19]. The $F$ PSFCHs can be determined based on two options: either based on the $L_{PSSCH}$ sub-channels used by a PSSCH or based only on the starting sub-channel used by a PSSCH (i.e., based only on one sub-channel for the case when $L_{PSSCH} > 1$). Thus, $F$ can be computed based on: (i) $L_{PSSCH}$ sub-channels of a PSSCH; (ii) $M_{set}$ PRBs for PSFCH associated with each sub-channel; and (iii) $Q$ cyclic shift pairs available in each PRB. Depending on which of the two options above is (pre-)configured, there are either then $F = L_{PSSCH} \cdot M_{set} \cdot Q$ PSFCHs (associated with the $L_{PSSCH}$ sub-channels of a PSSCH) or $F = M_{set} \cdot Q$ PSFCHs (associated with the starting sub-channel of a PSSCH) available for multiplexing the HARQ feedback for the PSSCH.

Similarly to the PUCCH in Rel. 15 NR Uu [36], the available $F$ PSFCHs are indexed based on a PRB index (frequency domain) and a cyclic shift pair index (code domain). Depending on the (pre-)configured option, there are either $L_{PSSCH} \cdot M_{set}$ or $M_{set}$ PRBs available for PSFCH. The mapping of the PSFCH index $i$ ($i = 1, 2, \ldots, F$) to the $L_{PSSCH} \cdot M_{sub}$ or $M_{sub}$ PRBs and to the $Q$ cyclic shift pairs is such that the PSFCH index $i$ first increases with the PRB index until reaching the number of available PRBs for PSFCH (i.e., $L_{PSSCH} \cdot M_{set}$ or $M_{set}$). Then, it increases with the cyclic shift pair index, again with the PRB index and so on. An example of the PSFCH indexing is shown in Fig. 10 assuming $F = L_{PSSCH} \cdot M_{set} \cdot Q = 20$ PSFCHs are available, $M_{set} = 5$ PRBs, a PSSCH transmission with $L_{PSSCH} = 2$ sub-channels and $Q=2$ cyclic shift pairs available per PRB.

Among the $F$ PSFCHs available for the HARQ feedback of a given transmission, an RX UE selects for its HARQ feedback the PSFCH with index $i$ given by:

$$i = (T_{ID} + R_{ID}) \bmod (F) \tag{1}$$

where $T_{ID}$ is the Layer 1 ID of the TX UE (indicated in the 2nd-stage SCI). $R_{ID} = 0$ for unicast ACK/NACK feedback and groupcast NACK-only feedback (option 1). For groupcast ACK/NACK feedback (option 2), $R_{ID}$ is equal to the RX UE identifier within the group, which is indicated by higher layers. For a number $X$ of RX UEs within a group, the RX UE identifier is an integer between 0 and $X - 1$. An RX UE determines which PRB and cyclic shift pair should be used for sending its HARQ feedback based on the PSFCH index $i$. The RX UE uses the first or second cyclic shift from the cyclic shift pair associated with the selected PSFCH index $i$ in order to send NACK or ACK, respectively [66]. For instance, for the PSFCH index $i = 17$ shown in the example in Fig. 10, an RX UE would send its HARQ feedback on the eighth PRB using one of the cyclic shifts from the second cyclic shift pair. By RX UEs selecting PSFCHs with index $i$ (as given in (1)) a TX UE can distinguish the HARQ feedback of different RX UE(s) (via the RX UE identifier, e.g., for groupcast option 2) and the HARQ feedback intended for the TX UE (via the Layer 1 ID of the TX UE, e.g. for unicast). As $R_{ID} = 0$ for groupcast option 1, the RX UEs select the same PSFCH index $i$ for their NACK-only feedback based solely on the Layer 1 ID TX UE identifier $T_{ID}$.

### 3) Simultaneous Transmission and/or Reception of PSFCH

Fig. 12 illustrates examples of situations where a UE may have multiple PSFCHs to transmit simultaneously in a PSFCH symbol to the same or different TX UE(s) (e.g., for the HARQ feedback for multiple PSSCH transmissions). For the given example, the HARQ feedback for PSSCH transmissions on slot $n$ to slot $n+3$ are associated with the PSFCH symbol in slot $n+6$. As UE A receives a transmission from UE B and from UE C at slot $n$ and slot $n+1$, respectively, UE A would need to send PSFCH to UE B and to UE C in slot $n+6$. However, a UE may not be capable of sending multiple PSFCHs [72]. For simultaneous transmissions of PSFCH, a UE selects to transmit up to a configured number of PSFCH(s) based on the priority of the associated PSSCH(s) [66] (indicated in the 1st-stage SCI).

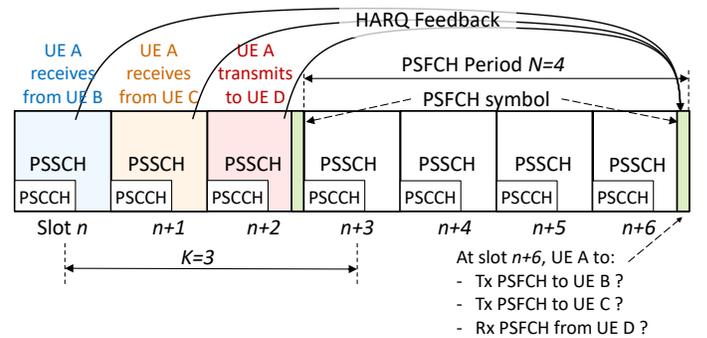

Fig. 12. Collisions between PSFCH transmissions or between a PSFCH transmission and a PSFCH reception.

Fig. 12 also illustrates the situation where a UE needs to receive a PSFCH (i.e., for the HARQ feedback of a transmission of the UE) and transmit a PSFCH (i.e., for the HARQ feedback of a transmission received by the UE) in the same PSFCH symbol. Since UE A transmits to UE D at slot $n+2$, it would then expect to receive the PSFCH from UE D at the same slot (slot $n+6$) where UE A is expected to send PSFCH to UE B and/or UE C. However, the half-duplex constraint prevents a UE from transmitting a PSFCH and receiving a PSFCH at the same time. In case a UE is expected to transmit and receive PSFCHs in the same PSFCH symbol, a UE selects to transmit or receive PSFCH(s) also based on the priority of the associated PSSCH(s) [66].

---

[18]This can happen, for example, when different TX UEs in mode 2 select the same sub-channel for a transmission (See Section VI.B).

[19]ACK/NACK feedback for groupcast (option 2) is used only when the group size is not larger than the number of PRBs available for PSFCH [49].

### 4) Sidelink CSI Reporting

NR V2X can use link adaptation to increase the efficiency and reliability of unicast transmissions. Link adaptation adjusts the MCS for a transmission according to the quality of the radio link[20] [73]. In NR V2X unicast[21], link adaptation can be based on the feedback of a channel quality indicator (CQI) from an RX UE to the TX UE. To determine the CQI, an RX UE measures the sidelink channel from the TX UE using the SL CSI-RSs transmitted by the TX UE within the PSSCH region of the slot. The CQI is selected from a CQI table that is derived based on the configured MCS table for a PSSCH [46]. The CQI provides an indication of the highest modulation and coding scheme that can be supported by the SL channel measured at the RX UE.

To improve the spectral efficiency, rank adaptation of unicast transmissions in NR V2X can also be applied by adjusting the number of streams that can be sent in a PSSCH with a multi-antenna transmission. Based on channel measurements of SL CSI-RSs sent from up to two antenna ports of the TX UE, the RX UE can determine the rank indicator (RI) that corresponds to the rank of the measured SL channel. The rank of the channel determines the number of streams that can be supported by the channel[22]. As up to two streams can be supported for a PSSCH transmission in NR V2X SL, the RI can be equal to 1 or 2. The RI is determined jointly with the CQI. The combination of CQI and RI represents the CSI that can be fed back from an RX UE to a TX UE for link and rank adaptation of a unicast PSSCH transmission. This is in contrast to the CSI that can be fed back by a UE to the gNB for adapting a downlink transmission in Rel. 15 NR Uu, where a UE can also feed back a precoding matrix indicator (PMI) besides the CQI and RI [41]. The PMI represents the index of the most suitable precoding matrix (out of a pre-defined codebook) that can be used to enable closed-loop multi-antenna transmission to the UE. As no PMI feedback is supported in NR V2X sidelink, open-loop multi-antenna SL transmissions can be performed by a TX UE based on the CQI and RI fed back by an RX UE.

An RX UE feeds back to the TX UE the CSI (CQI and RI) via CSI reporting. The CSI report is carried in the MAC control element (CE)) [74] over a PSSCH sent from the RX UE to the TX UE. The latency bound of the CSI report is signaled by the TX UE to the RX UE via PC5-RRC [54]. It is expected that the RX UE feeds back the CSI report within the latency bound [54].

The transmission of SL CSI-RS by a TX UE and the indication of CSI request in the $1^{st}$-stage SCI trigger the CSI reporting from an RX UE [42].

### 5) Selection of the Synchronization Reference

For NR V2X sidelink, there are two main sources for synchronization, namely GNSS and a gNB or eNB (referred to as gNB/eNB). In addition, a UE can use a SyncRef UE or its own internal clock as its synchronization reference [52]. Thus, there are four types of synchronization references for NR V2X sidelink (Fig. 13): gNB/eNB, GNSS, SyncRef UE, or a UE's own internal clock. In NR V2X, a UE selects its synchronization reference following the same procedure as in LTE V2X [55], i.e., based on the different priorities of gNB/eNB, GNSS and SyncRef UEs.

The priority of a SyncRef UE is given by its SLSS ID and the in coverage indicator that is carried on the PSBCH within the S-SSB transmissions of a SyncRef UE. We refer to the in coverage indicator as $I_{IC}$. $I_{IC} = 0$ indicates that a SyncRef UE is not in coverage of GNSS or a gNB/eNB and $I_{IC} = 1$ indicates that it is in coverage of GNSS or a gNB/eNB. The 672 SLSS IDs available in NR V2X are divided into two sets [44]:

1) In coverage SLSS IDs = $\{0, 1, ..., 335\}$,
2) Out of coverage SLSS IDs = $\{336, 337, ..., 671\}$.

SLSS IDs within the set of in coverage SLSS IDs are used by SyncRef UEs that are synchronized to a gNB/eNB, GNSS, or to another SyncRef UE which in turn is synchronized to a gNB/eNB or GNSS. As in LTE V2X, the SLSS ID = 0 is used in NR V2X by SyncRef UEs which are synchronized to GNSS or to another SyncRef UE that is synchronized to GNSS ([44]). SLSS IDs within the set of out of coverage SLSS IDs are used by SyncRef UEs that are synchronized to their own internal clock or to a SyncRef UE that is out of coverage of a gNB/eNB or GNSS. The SLSS IDs = 336, 337 are used by SyncRef UEs that are indirectly synchronized to GNSS ([44], [19]).

Based on the SLSS ID and $I_{IC}$, SyncRef UEs can be classified into five groups with different priorities:

- A first group with SyncRef UEs directly synchronized to GNSS (e.g., UE D in Fig. 13). SyncRef UEs in this group are identified by $I_{IC} = 1$ and SLSS ID = 0.
- A second group with SyncRef UEs that are out of coverage of GNSS or a gNB/eNB and are synchronized to a SyncRef UE from the first group (e.g., UE E in Fig. 13). SyncRef UEs in this group have $I_{IC} = 0$ and SLSS ID = 0.
- A third group with SyncRef UEs directly synchronized to a gNB/eNB (e.g., UE A and UE B in Fig. 13). SyncRef UEs in this group have $I_{IC} = 1$ and SLSS ID = $\{1, ..., 335\}$.
- A fourth group with SyncRef UEs that are out of coverage of GNSS or a gNB/eNB and are synchronized to a SyncRef UE from the third group (e.g., UE C in Fig. 13). SyncRef UEs in this group have $I_{IC} = 0$ and SLSS ID = $\{1, ..., 335\}$.

---

[20] For a poor quality link, a lower modulation scheme (e.g., QPSK) and lower code rate could be used for higher robustness. Higher modulation orders and code rates can be used when the SNR is high to increase the spectral efficiency.

[21] Adapting a groupcast transmission to RX UEs with different channel qualities requires CSI feedback from multiple RX UEs and could lead to adapting the transmission for the worst CSI, which may not be efficient. The reliability of a groupcast transmission in NR V2X can be improved by considering sidelink HARQ feedback and retransmissions.

[22] A channel with a high spatial correlation among the antennas has a lower rank compared to a channel with lower spatial correlation, which results in a larger rank. The rank of a channel is at most the minimum number of antennas ports at the TX UE or RX UE.

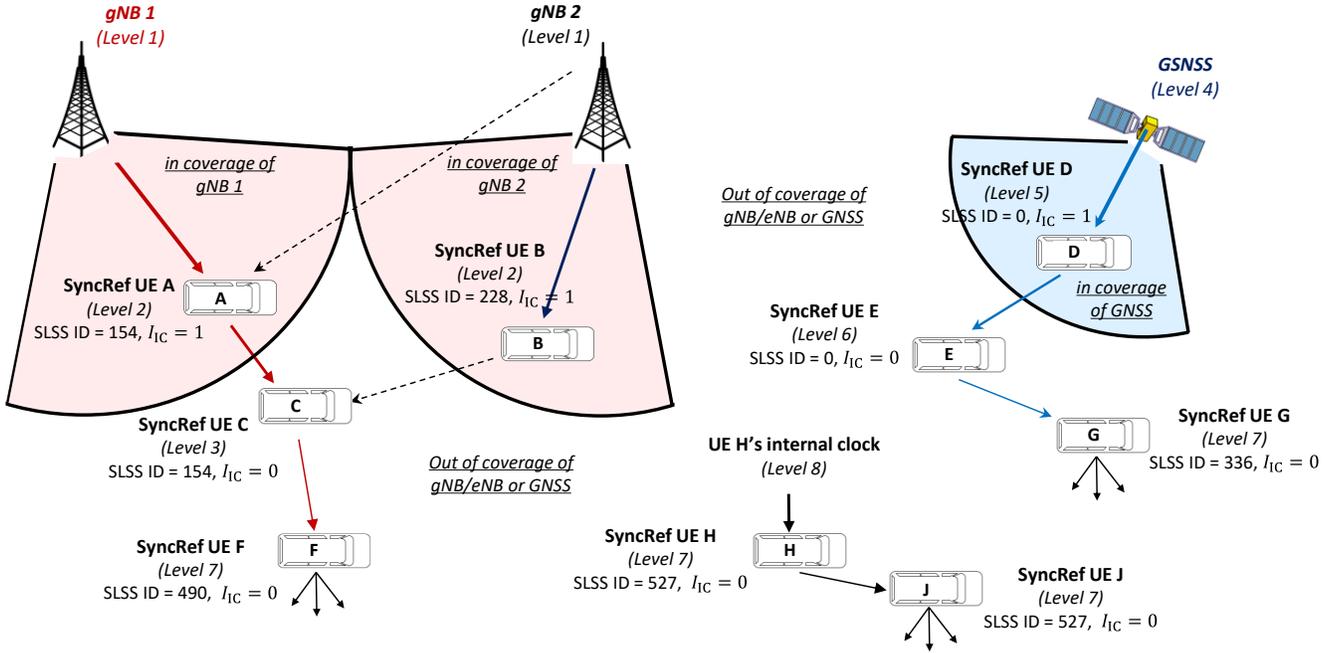

Fig. 13. Synchronization references with different priority levels.

- A fifth group with any other SyncRef UEs (e.g., UE F, UE G, UE H and UE J in Fig. 13). It includes SyncRef UEs that are out of coverage of GNSS or a gNB/eNB, and that are synchronized to their own internal clock (e.g., UE H in Fig. 13), to a SyncRef UE that uses its own internal clock or to a SyncRef UE from the second or fourth group. SyncRef UEs in this group have $I_{IC} = 0$ and SLSS ID = {336, ..., 671}.

The SLSS ID and $I_{IC}$ help distinguish whether a SyncRef UE is directly or indirectly synchronized to GNSS or a gNB/eNB. When a SyncRef UE is indirectly synchronized, the SLSS ID and $I_{IC}$ also indicate whether the SyncRef UE is either one hop or several hops away from a gNB/eNB or GNSS, where each hop corresponds to another SyncRef UE. The larger the number of hops to GNSS or a gNB/eNB, the lower is the quality of the synchronization provided by the SyncRef UE. Thus, SyncRef UEs that are directly synchronized to GNSS (i.e., from the first group) have a higher priority as synchronization reference than those that are one or more hops away from GNSS (i.e., from the second or fifth group). In the same way when considering a gNB/eNB, SyncRef UEs in the third group have a higher priority over SyncRef UEs in the fourth group and fifth group. A priority comparison between SyncRef UEs in the first or second group with SyncRef UEs in the third or fourth group depends on whether GNSS or a gNB/eNB has higher priority as synchronization source.

For the selection of a synchronization reference, NR V2X defines two sets of priority hierarchies (shown in Table VII) [45]: one with GNSS as the highest priority (GNSS-based synchronization) and another one with gNB/eNB as the highest priority (gNB/eNB-based synchronization). In both sets, SyncRef UEs are distinguished following the five groups previously described. Fig. 13 depicts examples of synchronization references according to the set of priority levels for gNB/eNB-based synchronization. Depending on which of the two set of priorities is (pre-)configured, a UE selects as its reference the available synchronization reference with the highest priority. The selection of a gNB/eNB is based on the cell selection procedure in NR Uu [66], while the selection of GNSS is based on the reliability of GNSS [52].

TABLE VII
DIFFERENT SETS OF PRIORITIES FOR A SYNCHRONIZATION REFERENCE

| Priority Level | GNSS-based synchronization | gNB/eNB-based synchronization |
|---|---|---|
| Level 1 | GNSS | gNB/eNB |
| Level 2 | SyncRef UE in network coverage and directly synchronized to GNSS, i.e., with $I_{IC} = 1$ and SLSS ID = {0} | SyncRef UE directly synchronized to gNB/eNB, i.e., with $I_{IC} = 1$ and with SLSS ID = {1, ...,335} |
| Level 3 | SyncRef UE out of GNSS/network coverage and one hop away from GNSS, i.e., with $I_{IC} = 0$ and SLSS ID = {0} | SyncRef UE out of GNSS/network coverage and one hop away from a gNB/eNB, i.e., with $I_{IC} = 0$ and with SLSS ID = {1, ...,335} |
| Level 4 | gNB/eNB | GNSS |
| Level 5 | SyncRef UE directly synchronized to a gNB/eNB, i.e., with $I_{IC} = 1$ and with SLSS ID = {1, ...,335} | SyncRef UE directly synchronized to GNSS, i.e., with $I_{IC} = 1$ and SLSS ID = {0} |
| Level 6 | SyncRef UE out of GNSS/network coverage and one hop away from a gNB/eNB, i.e., with $I_{IC} = 0$ and with SLSS ID = {1, ...,335} | SyncRef UE out of GNSS/network coverage and one hop away from GNSS, i.e., with $I_{IC} = 0$ and SLSS ID = {0} |
| Level 7 | SyncRef UE out of GNSS/network coverage and two or more hops away from a gNB/eNB or GNSS, i.e., with $I_{IC} = 0$ and with SLSS ID = {336,337, ...,671} | |
| Level 8 | UE's own internal clock | |

The selection of a SyncRef UE is based on the SLSS ID and the in coverage indicator $I_{IC}$. Depending on the priority level, a UE first searches for SyncRef UEs with certain SLSS IDs, i.e.,

SyncRef UEs transmitting a certain combination of S-PSS and S-SSS. For each SLSS ID, the UE measures an RSRP based on PSBCH DMRS sent by a SyncRef UE. Depending on UE implementation, the S-SSS can also be used for computing the RSRP [46]. For SLSS IDs with RSRP above a preconfigured threshold, the UE checks the value of the in coverage indicator $I_{IC}$ carried in the PSBCH. In this way, a UE can determine the SyncRef UEs which are its candidate synchronization reference for a given priority level. For multiple candidate SyncRef UEs in the same priority level, the SyncRef UE with the highest RSRP has higher priority. Independently of the set of priorities utilized, a UE that is unable to find any other synchronization reference (i.e., GNSS, a gNB/eNB or a SyncRef UE) uses its own internal clock as synchronization reference.

Examples for the selection of synchronization reference are given next. If a UE is (pre-)configured to use gNB/eNB-based synchronization and it is out of coverage of gNB/eNB, it checks if there are SyncRef UEs with SLSS ID = {1, ... ,335}. For the detected SLSS IDs with an RSRP above a preconfigured threshold, the UE checks whether $I_{IC} = 1$. If the UE has found multiple such SyncRef UEs (i.e., with priority level 2), the UE selects the SyncRef UE with the highest RSRP as its synchronization reference. In Fig. 13, this situation can be seen for UE C, which identifies that SyncRef UE A and B have priority level 2 but different RSRP at UE C. As SyncRef UE A is closer to UE C, the RSRP of SyncRef UE A is higher than that of SyncRef UE B. Thus, UE C selects SyncRef UE A as its synchronization reference. In another example, if a UE is out of coverage of GNSS or gNB/eNB, it selects as synchronization reference the SyncRef UE with the highest RSRP within the highest priority level (among priority levels 2, 3, 5, 6 and 7).

A gNB/eNB may not be synchronized to GNSS. The use of priority levels 4 to 6 can then be disabled for GNSS-based synchronization [45]. It can also happen that the situation of a UE changes regarding available synchronization references, e.g., a higher priority synchronization reference becomes available. Thus, a UE frequently searches for synchronization references with the objective to select always the highest priority synchronization reference that is available.

### 6) Triggering of S-SSB Transmissions

In NR V2X, there are two general procedures for triggering S-SSB transmissions at a UE (i.e., for the UE becoming a SyncRef UE) [63]: (i) a UE is configured by the network to become a SyncRef UE; or (ii) a UE decides on its own whether to become a SyncRef UE when in or out of network coverage. A SyncRef UE sends S-SSBs based on the timing reference provided by its synchronization reference.

A UE that is in coverage of a gNB/eNB can be configured by the network to transmit or not to transmit S-SSBs. In the former case, the UE is network configured to become a SyncRef UE, while in the latter case the UE is network configured not to act as a SyncRef UE. A network configured SyncRef UE sends S-SSBs irrespective of whether it has any data to transmit in the sidelink. The SyncRef UE knows which SLSS ID and resources to use for the S-SSB transmissions based on the sidelink system information provided by the network to UEs in the cell [63].

The network configuration of a UE to send or not to send S-SSBs is optional. Consequently, a UE in network coverage may not be configured to transmit or not to transmit S-SSBs. If a UE in network coverage has not received such configuration and the UE has data to transmit in the sidelink, the UE decides on its own whether to transmit S-SSBs or not [63]. This is in contrast to the network configured scenario where the decision regarding S-SSB transmissions is taken by the network, e.g., regardless whether the UE has data to transmit in the sidelink. A UE decides on its own to become a SyncRef UE, by comparing its RSRP of the serving gNB/eNB with a threshold provided by the network. The UE measures the RSRP based on reference signals (e.g., PBCH DMRS) associated with the synchronization signal sent by the serving gNB/eNB. If the RSRP is below the provided threshold, the UE can become a SyncRef UE and transmit S-SSBs. The SyncRef UE knows which SLSS ID and resources to use for the S-SSB transmissions based on the sidelink system information provided by the network in the cell [63]. If the RSRP of the serving gNB/eNB is above or equal to the threshold, the UE does not send S-SSBs. This RSRP-based triggering of S-SSB transmissions results in having UEs close to the cell edge becoming SyncRef UEs, if they have data to transmit in the sidelink; this is illustrated in Fig. 14 for UE A. Triggering S-SSB transmissions by a UE close to cell edge allows expanding the synchronization coverage of the serving gNB/eNB as shown in Fig. 14. This allows UEs with very poor network coverage or out of network coverage (e.g., UE C in Fig. 14) to use the timing reference of the serving gNB/eNB, i.e., for SL communication with UEs inside the cell. On the other hand, there may be little or no benefit in allowing UEs which are in good network coverage (i.e., close to the gNB/eNB and with the RSRP $\geq$ threshold) to transmit S-SSBs. The area that these UEs (e.g., UE B in Fig. 14) could cover with their S-SSB transmissions is already within the coverage of the serving gNB/eNB.

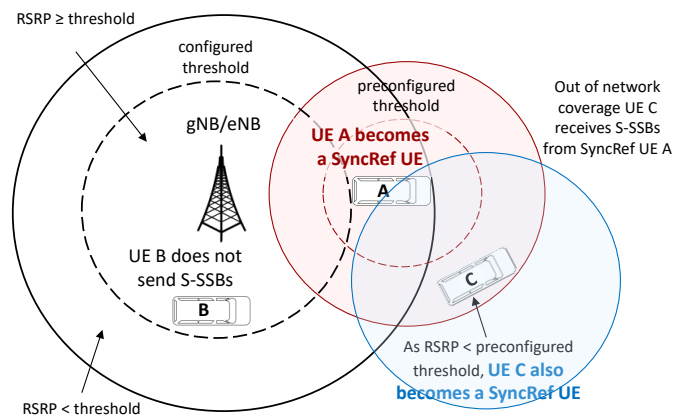

Fig. 14. Triggering of S-SSB transmissions when RSRP of synchronization reference is lower than a threshold, where the threshold is (pre-)configured.

If a UE is synchronized to a SyncRef UE and has data to transmit in the sidelink, it decides on its own whether to transmit S-SSBs or not [63]. A UE synchronized to a selected SyncRef UE can itself become a new SyncRef UE if the RSRP of its selected SyncRef UE is below a preconfigured threshold.

The RSRP is measured based on PSBCH DMRS sent in S-SSB transmissions of the selected SyncRef UE. The resources on which the new SyncRef UE sends S-SSBs are preconfigured [63], with the new SyncRef UE transmitting S-SSBs on different resources than the ones used by the selected SyncRef UE. If the selected SyncRef UE has an SLSS ID from the set of out of coverage SLSS IDs, the new SyncRef UE transmits S-SSBs with this same SLSS ID [63] (as shown for UE J in Fig. 13). If the selected SyncRef UE has an SLSS ID from the set of in coverage SLSS IDs, the new SyncRef UE transmits the S-SSBs with an SLSS ID equal to the SLSS ID of the selected SyncRef UE plus 336 [63] (as shown for UE F and UE G in Fig. 13). If the UE measures an RSRP of its selected SyncRef UE which is above or equal to the (pre-)configured threshold, the UE does not send S-SSBs. This RSRP-based triggering of S-SSB transmissions results in having UEs near the edge of the synchronization coverage becoming SyncRef UEs, if they have data to transmit in the sidelink. This enables expanding the synchronization coverage (e.g., as shown with UE C in Fig. 14) which aids nearby UEs in out of network coverage to share the same timing reference for SL communication.

If a UE has data to transmit in the sidelink and uses its internal clock as synchronization reference, the UE becomes a SyncRef UE. The SyncRef UE transmits S-SSBs with an SLSS ID randomly chosen from the set of out of coverage SLSS IDs excluding 336 and 337 [63]; this case is depicted in Fig. 13 with UE H. The resources used to send S-SSBs are preconfigured.

### 7) Sidelink Power Control

In NR V2X, power control is supported for PSCCH, PSSCH, PSFCH and S-SSB transmissions. As transmission power control (TPC) commands are not supported for NR V2X SL, the SL power control scheme is open-loop. For the SL power control, a maximum transmit power $P_{MAX}$ is (pre-)configured at the TX UE [44]. SL power control is supported for unicast and groupcast transmissions in NR V2X.

For a unicast transmission, the PSSCH power control can be configured to use the DL pathloss $PL_{DL}$ (between the gNB and TX UE) only, the SL pathloss $PL_{SL}$ (between TX UE and RX UE) only, or both DL pathloss $PL_{DL}$ and SL pathloss $PL_{SL}$ [66]. The PSSCH power control can be based on the DL pathloss $PL_{DL}$ when the TX UE is in network coverage. This allows mitigating the interference at the gNB (for uplink reception), similar to the SL power control in LTE V2X [75]. If the PSSCH power control is based on the DL pathloss only, TX UEs near the gNB transmit PSSCH at a lower power than TX UEs farther away from the gNB. The DL pathloss-based PSSCH power control can be enabled or disabled by the gNB [66]. The DL pathloss can be derived at the TX UE based on measurements of reference signals (e.g., CSI-RS or SSB) sent by the gNB [41].

For unicast, the PSSCH power control can also be based on the SL pathloss $PL_{SL}$ between the TX UE and the RX UE. This allows compensating the attenuation in the SL channel. For instance, a TX UE that is far away from the gNB may transmit

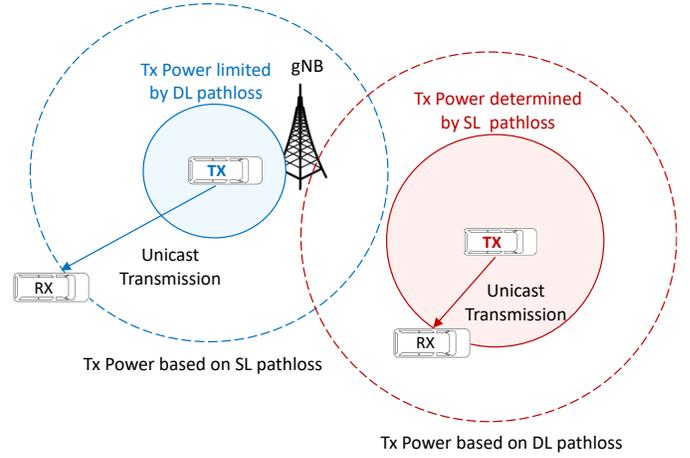

Fig. 15. Sidelink power control based on DL pathloss and SL pathloss.

PSSCH at a larger power than necessary when the PSSCH power control is configured to use DL pathloss only. However, if the PSSCH power control takes the SL pathloss also into account, this may avoid that a TX UE transmits at a large power as shown in Fig. 15. The SL pathloss-based PSSCH power control can be used when the TX UE is in or out of network coverage. The SL pathloss-based PSSCH power control can be enabled or disabled via (pre-)configuration. For this power control scheme, the TX UE requires an estimate of the SL pathloss that can be obtained from feedback of the RX UE [44]. Based on PSSCH DMRS transmitted by the TX UE, the RX UE can obtain an average RSRP over several RSRP measurements [76] in order to mitigate fluctuations on the received power. The RX UE cannot derive the SL pathloss based on the RSRP measurements since the transmit power of the PSSCH DMRS is not indicated to the RX UE [44]. Thus, the RX UE feeds back the average RSRP to the TX UE using higher layer signaling. The TX UE uses the fed back average RSRP along with the average transmit power of the PSSCH DMRS [49] to derive the sidelink pathloss $PL_{SL}$ (in dB) as follows:

$$PL_{SL} = \text{Average Tx power PSSCH DMRS (in dBm)} \\ - \text{Average RSRP (in dBm).} \tag{2}$$

For a groupcast transmission, the PSSCH power control can be configured to use the DL pathloss. Although the use of SL pathloss for groupcast PSSCH power control was discussed during the development of Rel. 16 NR V2X [77], it was finally agreed that SL pathloss-based PSSCH power control is not supported for groupcast in NR V2X. For groupcast PSSCH power control based on SL pathloss, RX UEs need to feed back their RSRP to the TX UE, which can lead to a large overhead[23].

When the PSSCH power control is configured to use both the DL pathloss and the SL pathloss, the transmit power for PSSCH is determined at the TX UE as follows (in dBm):

$$P_{PSSCH,1} = \min\big(P_{MAX}, \; P_{0,DL} + 10\log_{10}(2^{\mu} M_{PSSCH}) + \\ \alpha_{DL}, \; PL_{DL}, P_{0,SL} + 10\log_{10}(2^{\mu} M_{PSSCH}) + \; \alpha_{SL} \, PL_{SL}\big) \tag{3}$$

---

[23]To address this issue, some schemes were proposed to group RX UEs for their feedback. This enables several RX UEs to share a feedback resource [77]. The grouping of the RX UEs could be done based on the distance and/or RSRP.

TABLE VIII
POWER CONTROL FOR THE DIFFERENT CHANNELS AND TRANSMISSIONS IN NR V2X

| Physical Channel or Transmission | | Power control expression |
|---|---|---|
| **PSSCH** | PSSCH symbols without PSCCH | When in network coverage and configured to use DL pathloss (for unicast and groupcast):<br>$P_{\text{PSSCH},1} = \min\left(P_{\text{MAX}}, P_{0,\text{DL}} + 10 \log_{10}(2^{\mu} M_{\text{PSSCH}}) + \alpha_{\text{DL}} PL_{\text{DL}}\right)$ [dBm] |
| | | When in or out of network coverage and configured to use SL pathloss (for unicast):<br>$P_{\text{PSSCH},1} = \min\left(P_{\text{MAX}}, P_{0,\text{SL}} + 10 \log_{10}(2^{\mu} M_{\text{PSSCH}}) + \alpha_{\text{SL}} PL_{\text{SL}}\right)$ [dBm] |
| | | When in network coverage and configured to use DL pathloss and SL pathloss (for unicast):<br>$P_{\text{PSSCH},1} = \min\left(P_{\text{MAX}}, P_{0,\text{DL}} + 10 \log_{10}(2^{\mu} M_{\text{PSSCH}}) + \alpha_{\text{DL}} PL_{\text{DL}}, P_{0,\text{SL}} + 10 \log_{10}(2^{\mu} M_{\text{PSSCH}}) + \alpha_{\text{SL}} PL_{\text{SL}}\right)$ [dBm] |
| | PSSCH symbols with PSCCH | $P_{\text{PSSCH},2} = P_{\text{PSSCH},1} + 10 \log_{10}\left(\frac{M_{\text{PSSCH}} - M_{\text{PSCCH}}}{M_{\text{PSSCH}}}\right)$ [dBm] |
| **PSCCH** | | $P_{\text{PSCCH}} = P_{\text{PSSCH},1} + 10 \log_{10}\left(\frac{M_{\text{PSCCH}}}{M_{\text{PSSCH}}}\right)$ [dBm] |
| **PSFCH** | | When in network coverage: $\quad P_{\text{PSFCH}} = \min\left(P_{\text{MAX}}, P_{0,\text{DL}}^{\text{PSFCH}} + 10 \log_{10}(2^{\mu}) + \alpha_{\text{DL}}^{\text{PSFCH}} PL_{\text{DL}}\right)$ [dBm] |
| | | When out of network coverage: $\quad P_{\text{PSFCH}} = P_{\text{MAX}}$ [dBm] |
| **S-SSB** | | When in network coverage: $\quad P_{\text{S-SSB}} = \min\left(P_{\text{MAX}}, P_{0,\text{DL}}^{\text{S-SSB}} + 10 \log_{10}(2^{\mu} \cdot 11) + \alpha_{\text{DL}}^{\text{S-SSB}} PL_{\text{DL}}\right)$ [dBm] |
| | | When out of network coverage: $\quad P_{\text{S-SSB}} = P_{\text{MAX}}$ [dBm] |

where the second and third term in the argument of the min expression are associated with the DL pathloss and SL pathloss, respectively. The power control parameters (i.e., the nominal power $P_0$ and the parameter $\alpha$ used for fractional power control [78]), are (pre-)configured separately when considering the DL pathloss and SL pathloss: $P_{0,\text{DL}}$ and $\alpha_{\text{DL}}$ are associated with the DL pathloss and $P_{0,\text{SL}}$ and $\alpha_{\text{SL}}$ are associated with the SL pathloss. $\mu$ is the SCS configuration factor of the configured numerology. $M_{\text{PSSCH}}$ is the number of PRBs for the PSSCH (in symbols without PSCCH), where $M_{\text{PSSCH}} = L_{\text{subCH}} \cdot M_{\text{sub}}$ PRBs with $L_{\text{subCH}}$ sub-channels for PSSCH and $M_{\text{sub}}$ PRBs per sub-channel. If the sidelink power control is not based on the DL pathloss or the SL pathloss, then the second or the third term, respectively, is not included in the argument of the min expression. As mentioned before, SL pathloss-based PSSCH power control is only supported for unicast transmissions. In case both the DL pathloss and SL pathloss are disabled for the PSCCH power control, the transmit power for PSSCH is equal to the (pre-)configured maximum transmit power $P_{\text{MAX}}$.

For transmissions with two streams in PSSCH, the PSSCH transmit power is equally shared between the two streams [49]. The expression in (3) represents the PSSCH transmit power for PSSCH symbols without PSCCH. In NR V2X, a TX UE transmits PSSCH and PSCCH with the same power spectral density (i.e., with the same power over a PRB) in all symbols with PSCCH, PSSCH or PSCCH/PSSCH [49]. As a result, the transmit power $P_{\text{PSSCH},1}$ given by (3) is shared between PSSCH and PSCCH in the PSCCH/PSSCH symbols. With $M_{\text{PSCCH}}$ PRBs for PSCCH, there are $M_{\text{PSSCH}} - M_{\text{PSCCH}}$ PRBs available for PSSCH in a PSCCH/PSSCH symbol. Based on this, the PSSCH transmit power $P_{\text{PSSCH},2}$ and PSCCH transmit power $P_{\text{PSCCH}}$ in a PSCCH/PSSCH symbol is given by (in dBm) [66]:

$$P_{\text{PSSCH},2} = P_{\text{PSSCH},1} + 10 \log_{10}\left(\frac{M_{\text{PSSCH}} - M_{\text{PSCCH}}}{M_{\text{PSSCH}}}\right) \quad (4)$$

$$P_{\text{PSCCH}} = P_{\text{PSSCH},1} + 10 \log_{10}\left(\frac{M_{\text{PSCCH}}}{M_{\text{PSSCH}}}\right). \quad (5)$$

Reference signals sent in the PSSCH region and PSCCH region of the slot (i.e., SL CSI-RS, SL PT-RS, PSCCH DMRS, PSCCH DMRS) are also sent with the same power spectral density based on $P_{\text{PSSCH},1}$ [49]. The different variations for the PSSCH power control and the resulting PSCCH power control are summarized in Table VIII.

For a UE transmitting PSFCH, the PSFCH power control can be based on the pathloss between the gNB and the UE (i.e., the DL pathloss) if the UE is in network coverage [66]. While using the SL pathloss may avoid transmitting at a power larger than necessary, SL pathloss-based PSFCH power control is not supported in NR V2X [44]. For a SL pathloss-based power control for PSFCH, the RX UE(s) of a transmission need to know the SL pathloss to the TX UE. However, as a TX UE does not indicate its SL transmit power, an RX UE cannot derive the SL pathloss and cannot perform SL pathloss-based power control for PSFCH in NR V2X. The PSFCH power control parameters associated with the DL pathloss (i.e., $P_{0,\text{DL}}^{\text{PSFCH}}$ and $\alpha_{\text{DL}}^{\text{PSFCH}}$) are configured separately from the parameters used for the PSCCH/PSSCH power control [44]. The transmit power for PSFCH is determined at the UE as follows[24] (in dBm) [66]:

$$P_{\text{PSFCH}} = \min\left(P_{\text{MAX}}, P_{0,\text{DL}}^{\text{PSFCH}} + 10 \log_{10}(2^{\mu}) + \alpha_{\text{DL}}^{\text{PSFCH}} PL_{\text{DL}}\right). \quad (6)$$

If a UE sending a PSFCH is out of network coverage, it sends PSFCH with the (pre-)configured maximum transmit power $P_{\text{MAX}}$ as shown in Table VIII. If a UE is to send multiple PSFCH simultaneously, the power is equally shared among them in order of priority of the received PSSCH (see Section V.B.2)), up to the maximum number of PSFCHs the UE can transmit.

---

[24] A PSFCH consists of one PRB. The bandwidth factor is then $10 \log_{10}(2^{\mu})$.

For a SyncRef UE, the S-SSB power control can be based on the pathloss between the gNB and the SyncRef UE (i.e the DL pathloss) if the SyncRef UE is in coverage. The S-SSB power control parameters associated with the DL pathloss (i.e., $P_{0,DL}^{S-SSB}$ and $\alpha_{DL}^{S-SSB}$) are configured separately from the PSSCH or PSFCH power control parameters. The SyncRef UE sends S-SSBs with transmit power[25] (in dBm) [66]:

$$P_{S-SSB} = \min\left(P_{MAX}, P_{0,DL}^{S-SSB} + 10\log_{10}(2^{\mu} \cdot 11) + \alpha_{DL}^{S-SSB} PL_{DL}\right). \tag{7}$$

In case the SyncRef UE is out of network coverage, the SyncRef UE sends S-SSBs with the (pre-)configured maximum transmit power $P_{MAX}$ as shown in Table VIII.

In NR V2X sidelink, a UE transmits S-SSBs, PSSCH, PSCCH, or PSFCH with the same transmit power in all the corresponding symbols [66]. This is also the case when the UE transmits simultaneously on the uplink and sidelink. When a TX UE transmits simultaneously on the sidelink and uplink, it adjusts the transmit power based on the transmission that is prioritized [66]. If the sidelink is prioritized, the TX UE reduces the uplink transmit power before the start of the transmission if the UE's total transmit power would be larger than $P_{MAX}$ when the sidelink and uplink transmissions overlap. The same process is inversely applied if the uplink is prioritized. In either case, it is up to UE implementation how to reduce the uplink or sidelink transmit power. Sidelink transmissions can also be dropped in some overlapping symbols when the uplink transmission is prioritized and the UE cannot achieve the same sidelink transmit power in the symbols. It is up to UE implementation the symbols that are dropped within the overlapping symbols. If a UE is not capable to simultaneously transmit on the sidelink and uplink, it transmits only on the link with highest priority.

## VI. RESOURCE ALLOCATION FOR 5G NR V2X SIDELINK

Rel. 16 defines two new modes (modes 1 and 2) for the selection of sub-channels in NR V2X SL communications using the NR V2X PC5 interface. These two modes are the counterparts to modes 3 and 4 in LTE V2X (Section II). However, LTE V2X only supports broadcast SL communications while NR V2X supports broadcast, groupcast, and unicast SL communications.

### A. Mode 1

Similar to mode 3 in LTE V2X, the gNB or eNB assigns and manages the SL radio resources for V2V communications under mode 1 using the NR (or LTE) Uu interface. UEs must therefore be in network coverage to operate using mode 1. SL radio resources can be allocated from licensed carriers dedicated to SL communications or from licensed carriers that share resources between SL and UL communications. The SL radio resources can be configured so that mode 1 and mode 2 use separate resource pools. The alternative is that mode 1 and mode 2 share the resource pool. Pool sharing can result in a more efficient use of the resources, but it is prone to potential collisions between mode 1 and mode 2 transmissions. To solve this, mode 1 UEs notify mode 2 UEs of the resources allocated for their future transmissions as it is described below.

Mode 1 uses dynamic grant (DG) scheduling like LTE V2X mode 3, but replaces the semi-persistent scheduling in LTE V2X mode 3 with a configured grant scheduling [45]. With DG, mode 1 UEs must request resources to the base station for the transmission of every single TB (and possible blind or HARQ retransmissions[26]) [44]. To this aim, the UEs must send a Scheduling Request (SR) to the gNB using the PUCCH, and the gNB responds with the DCI over the PDCCH. The DCI indicates the SL resources (i.e., the slot(s) and sub-channel(s)) allocated for the transmission of a TB and up to 2 possible retransmissions of this TB. The UE informs other UEs about the resources it will use to transmit a TB and up to 2 possible retransmissions using the $1^{st}$-stage SCI. Nearby UEs operating under mode 2 can then know which resources UEs in mode 1 will utilize.

Requesting resources for each TB increases the delay. Mode 1 includes the configured grant scheduling option to reduce the delay by pre-allocating SL radio resources. With this scheme, the gNB can assign a set of SL resources to a UE for transmitting several TBs. This set is referred to as a configured grant (CG). The UE sends first a message with UE assistance information to the gNB indicating information about the expected SL traffic including: periodicity of TBs, TB maximum size and QoS information. The QoS information includes KPIs such as the latency and reliability required by the TBs and their priority. This information is used by the gNB to create, configure and allocate a CG to the UE that satisfies the requirements of the SL traffic. The CG is configured using a set of parameters that includes the CG index, the time-frequency allocation and the periodicity of the allocated SL resources. Fig. 16 shows two examples of configured CGs. The two CGs have different periodicities and different time-frequency allocations. The time-frequency allocation indicates the slot(s) and sub-channel(s) that are assigned periodically to the UE in a CG. A UE can be assigned a maximum of 3 SL resources during each period of the CG. The UE informs other vehicles of the resources allocated by the gNB for a particular CG period using again the $1^{st}$-stage SCI[27]. The UE can decide how to use the SL resources of an assigned CG. However, it can only transmit one new TB in each CG period. The SL resources of a CG period can also be used to retransmit the new TB transmitted in this CG, or to retransmit other TBs initially transmitted in previous CG periods. The maximum number of retransmissions per TB in a CG is associated with the priority of the TB. HARQ retransmissions in a CG are only possible if the resource pool of the CG has a PSFCH configured by the gNB (see Section

---

[25] An S-SSB consists of 11 PRBs. The bandwidth factor is then $10\log_{10}(2^{\mu} \cdot 11)$.

[26] This is in contrast to mode 3 that only allows one blind retransmission.

[27] This SCI only informs about the resources allocated in the current CG period. Extensions to inform about the resources allocated in the next CG period are being studied in [119].

V.B.4). It should be noted that a gNB can assign multiple SL CGs to a UE. The configuration of each CG can be adapted to the characteristics or demands of different V2X applications. However, the transmission and retransmissions of a TB must always utilize resources of a single CG [44].

Mode 1 defines two types of CG schemes for SL: CG type 1 and CG type 2. Both are configured using Radio Resource Control (RRC) signaling (Section IX.B). CG type 1 can be utilized by the UE immediately until it is released by the base station (also using RRC signalling). SL CG type 2 can be used only after it is activated by the gNB and until it is deactivated. To this aim, the gNB notifies of the activation and deactivation using DCI signalling. The DCI also includes the CG index and time-frequency allocation of CG type 2. CG type 2 can configure multiple CGs for a UE and only activate a subset of the CGs based on the UE needs. Resources in non-active CGs can be allocated to other UEs. CG type 1 can also configure multiples CGs. However, it forces UEs to activate CGs at the time of their configuration. CG type 1 reduces the signalling and the time needed to initiate a transmission compared to CG type 2. However, if any of the CGs type 1 are not used by the UE, the resources cannot be allocated to other UEs.

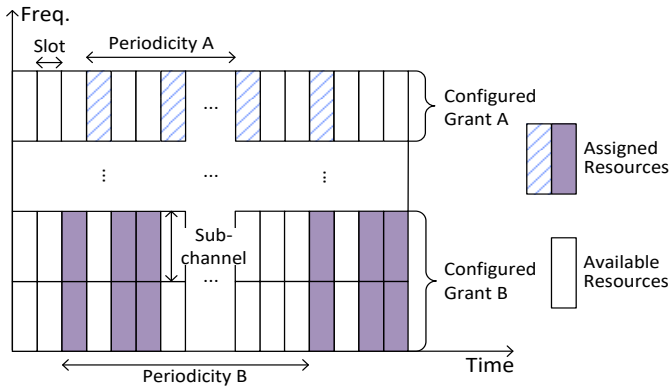

Fig. 16. Examples of CGs under NR V2X SL mode 1.

Fig. 17 shows an example that compares how transmissions are organized and scheduled when utilizing the DG and CG schemes to transmit two TBs (TB$_1$ generated at $t_0$ and TB$_2$ generated at $t_5$). We discuss first the case of the DG scheme. The UE sends an SR at $t_1$ to request resources for transmitting TB$_1$. The gNB responds with a DCI at $t_3$ that indicates the resources the user can utilize at $t_4$. The same process takes place starting at $t_5$ when TB$_2$ is generated. TB$_2$ can be transmitted using the next set of resources allocated at $t_{10}$. For CG, the UE does not request resources but waits until the gNB grants it a CG. We suppose the UE receives the granted CG at $t_4$. This CG includes a set of resources assigned periodically to the UE at $t_4$, $t_8$ and $t_{11}$. The time period is adjusted to the time between TBs indicated by the UE in the provided UE assistance information. The UE uses the resources at $t_4$ to transmit TB$_1$ and the resources at $t_8$ to transmit TB$_2$. The CG scheme reduces the time needed to transmit the two TBs compared to DG. However, the DG scheme can utilize resources more efficiently when handling non periodic traffic since resources are only allocated when needed to transmit TBs.

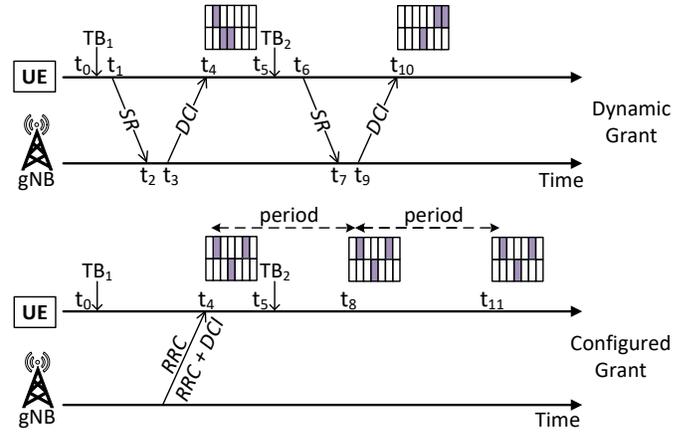

Fig. 17. Mode 1 dynamic and configured grant schemes.

Mode 1 defines three different MCS tables that are used to determine the MCS of a TB (Table 5.1.3.1-1, Table 5.1.3.1-2 and Table 5.1.3.1-3) in Section 5.1.3.1 of [42]. There are different ways for determining the MCS of a TB. The gNB can configure a UE to use one or more MCSs. In the latter case, the UE autonomously selects the MCS. The gNB can also configure a UE to use one, two or three of the MCS tables. In this case, the UE again autonomously selects the MCS from the configured tables. The UE indicates the selected MCS and MCS table (when applicable) for a TB in the associated 1st-stage SCI.

In mode 1, the gNB can enable HARQ retransmissions for groupcast and unicast communications. In this case, a RX UE can use a PSFCH associated with a received PSSCH to send HARQ feedback as described in Section V.C.2). The TX UE can inform the gNB about the feedback received from RX UEs with a feedback report. The gNB configures a PUCCH for the TX UE to send the feedback report (Section 16.5 in [66]). This feedback report in the PUCCH assists the gNB in the allocation of SL resources. The feedback report informs whether the SL transmissions of the last allocated SL resources were successful or not. Therefore, it is transmitted in the PUCCH configured after the PSFCH associated to the last allocated resource in the case of DG. For the case of the CG, the feedback report is transmitted in the PUCCH configured after the PSFCH associated to the last resource allocated for each CG period. It should be noted that the feedback report uses a single bit of the PUCCH to inform the gNB about the success of the potentially last three SL transmissions (both for DG, or a CG period). When the feedback report includes information of the transmission and retransmissions of the same TB, the TX UE reports an ACK in the PUCCH only if all RX UEs receive correctly at least one of the (re-)transmissions and they positively acknowledge it to the TX UE. Otherwise the TX UE reports a NACK in the PUCCH. When the feedback report includes information of the transmission and retransmissions of different TBs, the TX UE only reports an ACK if all RX UEs receive correctly the different TBs and they positively acknowledge them to the TX UE. After receiving the feedback report, the gNB evaluates if it has to allocate new SL resources to the TX UE for additional HARQ retransmissions.

## B. Mode 2

Like with mode 4 in LTE V2X, UEs can autonomously select their SL resources (one or several sub-channels) from a resource pool when using mode 2 in NR V2X. In this case, UEs can operate without network coverage. The resource pool can be (pre-)configured by the gNB or eNB when the UE is in network coverage. Mode 2 and mode 4 differ on the scheduling scheme. Mode 4 operates following the sensing-based SPS scheme explained in Section II. On the other hand, mode 2 can operate using a dynamic or a semi-persistent scheduling scheme that differs from the one designed for mode 4. The dynamic scheme selects new resources for each TB and can only reserve resources for the retransmissions of that TB. We should note that we distinguish in this section between a selected resource (using dynamic or semi-persistent schemes) and a reserved resource. A reserved resource is a selected resource that a UE reserves for a future transmission by notifying neighboring UEs using the 1st-stage SCI. A UE can select and reserve resources for the transmission of several TBs (and their retransmissions) when utilizing the semi-persistent scheme [45]. It is important to note that the semi-persistent scheme can be enabled or disabled in a resource pool by (pre-)configuration.

Mode 2 uses almost the same procedure to select resources for the dynamic and semi-persistent schemes [45]. The differences result from the fact that the dynamic scheme only selects resources for a TB while the semi-persistent scheme selects resources for a number of consecutive *Reselection Counter* TBs. The time period between the resources selected for the transmission of consecutive TBs in the semi-persistent scheme is defined by the Resource Reservation Interval (*RRI*). The possible values of the *RRI* are {0, [1:99], 100, 200, 300, 400 500, 600, 700, 800, 900, 1000} ms. It should be emphasized that NR V2X mode 2 provides higher flexibility to fit the requirements and characteristics of different eV2X services by allowing any integer *RRI* between 1 and 99 ms. This is in contrast to LTE V2X mode 4 that only considered the following RRI values: {0, 20, 50, 100, 200, 300, 400, 500, 600, 700, 800, 900, 1000} ms. Like in mode 4, a list of permitted *RRIs* is (pre-)configured in the resource pool for mode 2 [63]. The list can include a maximum of 16 different *RRIs*. A UE selects an *RRI* from the list when it selects new SL resources[28]. The selection of the *RRI* is left to UE implementation. The UE can select the *RRI* that best suits the characteristics of the traffic to be transmitted. The selected *RRI* also determines the *Reselection Counter* that is randomly set within an interval that depends on the selected *RRI*. If *RRI*≥100 ms, this counter is randomly set within the interval [5,15]. If *RRI*<100 ms, the counter is randomly set within the interval [5*C,15*C] where C=100/max(20,*RRI*) (Section 5.22.1 in [74]).

A UE can select new SL resources when it generates a new TB[29]. A new selection can also be triggered (only for the semi-persistent scheme) because a new TB does not fit in the previously reserved resources. To select new SL resources (for both dynamic and semi-persistent schemes), a UE first defines the selection window where it looks for candidate resources to transmit a TB. The selection window includes all resources within the range of slots [$n+T_1$, $n+T_2$] (Fig. 18) (Section 8.1.4 in [42]), where $n$ is the resource (re-)selection trigger or slot at which new resources must be selected. $T_1$ is the processing time (in slots) required by a UE to identify candidate resources and select new SL resources for transmission. $T_1$ is equal to or smaller than $T_{proc,1}$. $T_{proc,1}$ is equal to 3, 5, 9 or 17 slots for a SCS of 15, 30, 60 or 120 kHz, respectively[30]. The value of $T_2$ is left to UE implementation but must be included within the range $T_{2min} \le T_2 \le$ PDB, where PDB is the Packet Delay Budget (in slots). PDB is the latency deadline by which the TB must be transmitted. The deadline is established by the V2X application generating the packet to be transmitted in the TB. The value of $T_{2min}$ depends on the priority of the TB and the SCS. Possible values of $T_{2min}$ include {1,5,10,20}*$2^\mu$ slots, where μ is the SCS configuration factor as defined in Section V.A.1). This results in values of $T_{2min}$ equivalent to {1,5,10,20} ms. NR V2X mode 2 can then guarantee minimum latency levels of 1 ms compared to 10 ms in LTE V2X mode 4.

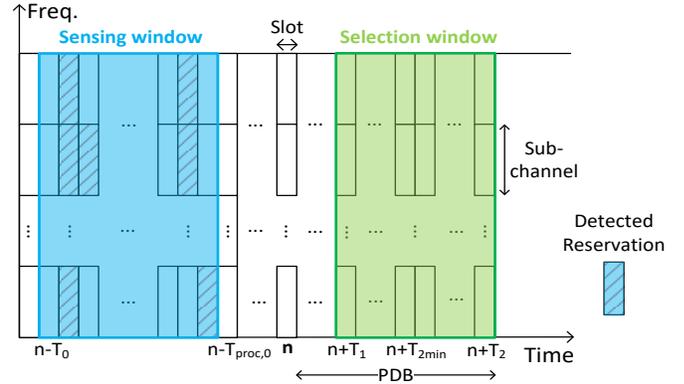

Fig. 18. Sensing and selection windows of NR V2X SL mode 2 when $T_2$=PDB.

Once the selection window is defined, the UE must identify the candidate resources within the selection window. A candidate resource is defined by a slot in time and $L_{PSSCH}$ contiguous sub-channels in frequency. $L_{PSSCH}$ is an integer in the range 1≤$L_{PSSCH}$≤max($L_{PSSCH}$), where max($L_{PSSCH}$) is the total number of sub-channels per slot in the selection window. However, the value of max($L_{PSSCH}$) can be modified by the congestion control process explained in Section VI.C. $L_{PSSCH}$ must be selected such that the TB and its associated SCI (including the 1st-stage and 2nd-stage SCI)[31] fit in the candidate resource. $L_{PSSCH}$ therefore depends on the size of the TB and the SCI as well as on the utilized MCS. For mode 2, a UE is (pre-)configured to use one, two or three of the MCS tables that are supported for TB and mode 1 (see Section VI.A). The UE selects the MCS from the (pre-)configured MCS tables. The UE indicates the MCS used for the TB in the associated 1st-stage SCI. The 1st-stage SCI also includes the selected MCS table if the UE is (pre-)configured with more than one table.

---

[28] This does not apply for the selection of new SL resources with the re-evaluation or pre-emption mechanisms that are explained below.

[29]It may also select new SL resources in mode 2 with the re-evaluation or pre-emption mechanisms that are explained below.

[30] $T_{proc,1}$ is equivalent to 3, 2.5, 2.25 or 2.125 ms for a SCS of 15, 30, 60 or 120 kHz, respectively.

[31]The TB and SCI are transmitted in the same slot as discussed in Section V.

When a UE is not transmitting, it senses the SL resources during the sensing window (Fig. 18) in order to identify available candidate resources. The sensing window is the time interval defined by the range of slots $[n\text{-}T_0, n\text{-}T_{proc,0})$, where $n$ is the resource (re-)selection trigger or slot at which new resources must be selected. $T_0$ is an integer defined in number of slots that depends on the SCS configuration but that must be set to a value (in number of slots) equivalent to 1100 ms or 100 ms. The selected value is determined by the (pre-)configuration of the resource pool. $T_{proc,0}$ is the time required to complete the sensing procedure and is equal to one slot for a SCS of 15 or 30 kHz, and equal to 2 or 4 slots for a SCS of 60 or 120 kHz, respectively[32]. During the sensing process, the UE decodes the 1st-stage SCI received from other UEs in the sensed SL resources. The 1st-stage SCIs indicate the SL resources that other UEs have reserved for their TB and SCI transmissions in the PSSCH and PSCCH. In particular, the 1st-stage SCIs can indicate the SL resources reserved for retransmissions of the TB associated to the 1st-stage SCI, and resources reserved for the initial transmission and retransmissions of the next TB. The UE also measures the RSRP [45] of the transmissions associated to the 1st-stage SCIs received from other UEs. The UE stores the sensed information (the decoded 1st-stage SCI and the RSRP measurements) and uses it to determine which candidate resources from the selection window should be excluded when a new selection is triggered.

To select new resources, mode 2 defines a 2-step algorithm that is used by both the dynamic and semi-persistent scheduling schemes [44] with small differences that are highlighted when applicable. Step 1 excludes candidate resources in the selection window. The first exclusions relate to the half-duplex operation since a UE cannot sense the reservations from other UEs announced in the slots of the sensing window where the UE was transmitting. Let us suppose the UE was transmitting in a slot $s_i$ in the sensing window. The UE could not sense the reservations of other UEs at slot $s_i$, and it excludes all candidate resources at any slot $s_i + q*RRI_i$ in the selection window in case other UEs have reserved them. $RRI_i$ represents all the possible values of the $RRI$ (represented here in slots) based on the list of permitted $RRIs$ in the resource pool. The parameter $q$ is an integer that takes values in the range $1 \leq q \leq Q$. $Q$ is equal to $\left\lceil \frac{T_2}{RRI_i} \right\rceil$ when the following two conditions are fulfilled: 1) $RRI_i < T_2$ ($RRI_i$ and $T_2$ in ms) and 2) $n\text{-}s_i \leq RRI_i$ ($RRI_i$ in slots), where $n$ is the resource (re-)selection trigger or slot at which new resources must be selected and the value of $T_2$ (in slots) is left to UE implementation but must be included within the range $T_{2min} \leq T_2 \leq PDB$. Otherwise $Q$ is set equal to 1. $Q$ indicates the estimated number of periodic transmissions from other UEs for each potential $RRI$. If the UE operates with the semi-persistent scheme, it should exclude additionally all candidate resources in any slot $s_j$ in the selection window such that any slot $s_j + j*RRI_{TX}$ overlaps with any slot $s_i + q*RRI_i$ where: $RRI_{TX}$ is the $RRI$ (in slots) selected by the UE that is selecting new resources, $s_i$ is the slot during which the UE was transmitting in the sensing window, $j$ is an integer that can take values in the range $1 \leq j \leq 10*ReselectionCounter\text{-}1$[33], and $RRI_i$ represents all the possible values of the $RRI$ (in slots) based on the list of permitted $RRIs$ in the resource pool (Section 8.1.4 in [42]).

The UE also excludes the candidate resources based on the reservations received from other UEs in the 1st-stage SCIs detected during the sensing window. In this case, the candidate resources are only excluded if the UE has measured an RSRP associated with the reservation that is higher than an RSRP threshold. Note that a list of RSRP thresholds is (pre-)configured in the resource pool. The utilized RSRP threshold depends on the priority of the TB for which the UE is selecting new resources and on the priority (included in the 1st-stage SCI) of the other UE that reserved the resource [44]. The procedure to exclude resources based on the resources reserved by other UEs for transmitting their next TB can be formally expressed as follows. Let us consider that a 1st-stage SCI was received (with a measured RSRP higher than the corresponding RSRP threshold) by the UE in a slot $s_k$ in the sensing window, and that the 1st-stage SCI is reserving a resource $R_x$ for the slot $s_k + RRI_{RX}$. $RRI_{RX}$ is the $RRI$ included in the 1st-stage SCI and it is represented in number of slots[34]. Then, the UE excludes every candidate resource that overlaps in frequency with $R_x$ in any slot $s_k + q*RRI_{RX}$, where $q$ is the variable previously defined but substituting $s_i$ by $s_k$ and $RRI_i$ by $RRI_{RX}$ in the related expressions. If the UE that is selecting the SL resources is operating with the semi-persistent scheme, it excludes additionally every candidate resource that overlaps in frequency with $R_x$ and is located in any slot $s_p$ in the selection window such that any slot $s_p + j*RRI_{TX}$ overlaps with any slot $s_k + q*RRI_{RX}$ (Section 8.1.4 in [42]). Remember that $RRI_{TX}$ is the $RRI$ (in slots) selected by the UE that is selecting new resources, and $j$ is an integer that can take values in the range $1 \leq j \leq 10*ReselectionCounter\text{-}1$. This exclusion is applied so that the UE avoids that any of its following $10*ReselectionCounter\text{-}1$ transmissions collide with a potential transmission from the UE that sent the 1st-stage SCI.

After executing all exclusions in step 1, the UE (with dynamic or semi-persistent scheme) checks whether the percentage of remaining available candidate resources (i.e., those that have not been excluded in step 1) in the selection window is equal or higher than $X$%. If not, the RSRP thresholds are increased by 3 dB, and the process is repeated iteratively until the percentage of available candidate resources in the selection window is at least equal to $X$%. Possible values of $X$ are 20, 35 or 50. The selected value depends on the priority of the TB for which the UE is selecting new SL resources.

In Step 2, the UE randomly selects the SL resource from the list of available candidate resources. Mode 2 eliminates the the UE transmits *Reselection Counter* TBs and then maintains the same selected resources based on the probability parameter $P$ that is defined later.

---

[32] $T_{proc,0}$ is equivalent to 1 ms for a SCS of 15 kHz and 0.50 ms for the rest of SCS configurations.

[33] Note that even though the UE would perform a number of consecutive *Reselection Counter* transmissions, $j$ can take up to 10 times this value. This is the case to avoid potential collisions with transmissions from other UEs in case

[34] Note that in contrast to the previous procedure that considered the list of $RRIs$ (i.e. $RRI_i$) to exclude candidate resources, in this case only the $RRI$ received in the 1st-stage SCI is taken into account.

third step included in mode 4 in LTE V2X for selecting resources. This third step estimated the average RSSI of the candidate resources to select one from a list of candidate resources with the lowest RSSI.

A UE selects $N$ candidate resources ($N \leq N_{MAX}$ with $1 \leq N_{MAX} \leq 32$) within the same selection window for the initial transmission of a TB and its $N$-$1$ blind or potential HARQ retransmissions[35]. The process to select $N$ candidate resources applies to both the dynamic and semi-persistent schemes and follows the 2-step algorithm previously explained. A UE can only perform HARQ retransmissions for groupcast or unicast transmissions if a PSFCH is (pre-)configured in the resource pool (see Section V.C.1). If not, only blind retransmissions are possible. We should note that if a TB is positively acknowledged when using HARQ retransmissions, the TX UE drops all next scheduled retransmissions of the TB in the selected resources (including the reserved ones) (Section 5.22.1 in [74]). The value of $N$ is left to UE implementation. The value of $N_{MAX}$ is (pre-)configured and can vary depending on the channel utilization or load following the congestion control mechanism (see Section VI.C). The UE cannot select for $N$ a value higher than the number of available candidate resources after step 1. Once the UE has selected the value of $N$, it considers the limitations of the 1st-stage SCI for the selection and reservation of the $N$ candidate resources. In particular, a 1st-stage SCI can only notify about resource reservations located within a window W of 32 slots, which conditions the candidate resources that can be selected. The first slot of the window W is where the 1st-stage SCI is transmitted. In addition, a 1st-stage SCI can only notify about a maximum number of $N_{SCI}$ resources. The maximum value of $N_{SCI}$ (i.e. max($N_{SCI}$)) is (pre-)configured in the resource pool and can be equal to 2 o 3 resources. A 1st-stage SCI therefore notifies about the transmission (or retransmission) of the TB happening in the same slot as the transmission of this 1st-stage SCI, and $N_{SCI}$-1 reserved resources for the following $N_{SCI}$-1 retransmissions of the TB (Figure 19) (Section 8.1.5 in [42]).

Once a UE has identified the list of available candidate resources following the step 1 previously described, it executes step 2 to select the $N$ candidate resources. To do so, the UE first selects randomly one of the $N$ candidate resources[36]. Let us consider that the first candidate resource is selected at the slot $m_1$ (see Fig. 19). Then, the UE selects also randomly the second candidate resource, but with the restriction that the gap between this candidate resource and the first selected candidate resource must be smaller than a window W of 32 slots. This means that the second candidate resource will be located (at the slot $m_2$) within the range of slots [$m_1$-31, $m_1$+31]. This guarantees that the 1st-stage SCI of one of the two selected candidate resources is able to reserve the other candidate resource. If $N$>2, the UE selects also randomly the third candidate resource but with the restriction that it is located (at the slot $m_3$) within the range [$m_1$-31, $m_1$+31] or the range [$m_2$-31, $m_2$+31]. This chain procedure is repeated with the aim that all candidate resources for the TB retransmissions can be reserved by a previous SCI. However, if the UE can only select a subset of the $N$ candidate resources following this procedure, the remaining ones are selected randomly within the selection window even if they do not meet the previous 1st-stage SCI limitations. It is also important to note that 3GPP standards specify that there should be a minimum time gap $t_{GAP}$ (in number of slots) between any pair of consecutive selected resources for HARQ retransmissions so that the TX UE can receive and process the acknowledgements from the RX UEs, and prepare the next HARQ retransmission (Section 5.22.1 in [74]). This minimum gap does not apply to blind retransmissions if a PSFCH is not (pre-)configured in the resource pool. If it is (pre-)configured, the gap applies although blind retransmissions do not use HARQ feedback.

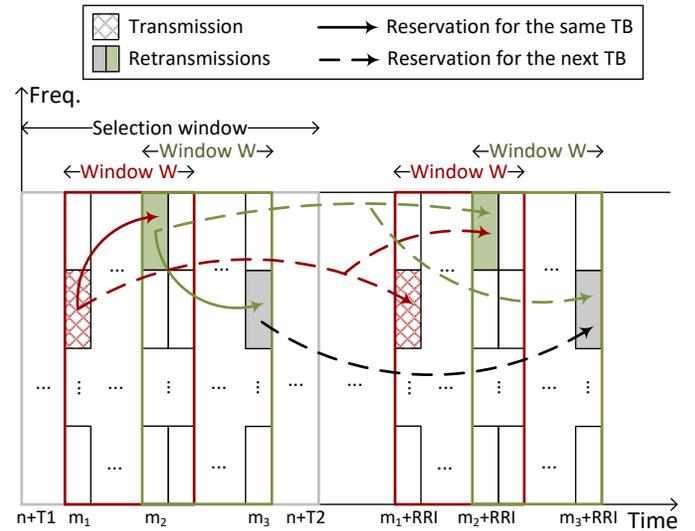

Fig. 19. Resource reservation for the retransmissions of the same TB and for the transmission and retransmissions of the next TB in NR V2X mode 2 (illustrative example when $N$=3, max($N_{SCI}$)=3 and RRI (in number of slots)). The transmission of the TB at $m_1$ does not announce the reservation of the retransmission at $m_3$ because it is outside of the window W. This reservation for the second retransmission is announced by the first retransmission at $m_2$.

When a UE operates with the dynamic scheme, it needs to select new SL resources for the transmission of each TB and its retransmissions. On the other hand, a UE selects resources for a number of consecutive *Reselection Counter* TBs in case of the semi-persistent scheme. Remember that the time period between the selected resources for consecutive TBs is determined by the *RRI*. The *Reselection Counter* is randomly chosen every time new resources must be selected[37]. The *Reselection Counter* is decremented by one after finishing the transmission of a TB, i.e. after transmitting the TB and all its possible retransmissions. When *Reselection Counter* is equal to zero, the UE must select new resources with probability (1-*P*). Otherwise, the UE continues using the same SL resources with

---

the same *RRI* for a number of consecutive *Reselection Counter* TBs[38]. Each UE can set up *P* between 0 and 0.8. Before transmitting the last TB that sets the *Reselection Counter* to zero, the TX UE evaluates (i.e. with probability 1-*P*) whether it must select new resources for the next TB. If this is the case, the UE sets the *RRI* equal to 0 ms in the 1st-stage SCI to indicate other UEs that it is not reserving the same resources for the next TB. If not, the UE keeps the same resources for the next TB and the same *RRI* included in the 1st-stage SCI.

Even if the *Reselection Counter* is higher than zero, new resources must be selected if the new TB to be transmitted does not fit in the resources previously selected, or if those resources cannot fulfil the latency requirement for transmitting the new TB. A TX UE implementing the semi-persistent scheme utilizes the *RRI* included in the "resource reservation period" field of each transmitted 1st-stage SCI (Section 8.1.4 in [42]) to inform nearby UEs about the resources it is reserving for the initial transmission or a retransmission of the next TB. The 1st-stage SCI can indicate up to $N_{SCI}$ resources, so the TX UE can indicate with each transmitted 1st-stage SCI the resources reserved for the following $N_{SCI}$-1 retransmissions of the next TB (Fig. 19).

### 1) Re-evaluation mechanism

A UE that has selected one new SL resource (e.g. at the slot *m*) will continue sensing the transmissions from other UEs during the selection window. A UE can decide whether to execute again step 1 to check if the selected resource is still available. The decision to execute again step 1 (and how often it is executed) is left to UE implementation. The UE can execute again step 1 at slot *m*-$T_3$ or before. $T_3$ is the maximum time allowed for a UE (in slots) to complete the resource selection process and is equal to $T_{proc,1}$. A UE can execute again step 1 after slot *m*-$T_3$ only if its processing capabilities are sufficient to complete step 1 and step 2 before *m*. The process to execute again step 1 works as follows. We denote *n'* as the slot at which a UE initiates a new execution of step 1. The UE defines a new selection window SW' that starts at slot *n'*+$T_1$ and ends at slot *n'*+$T_2$'. $T_2$' must be within the range $T_{2min} \leq T_2' \leq$ PDB-(*n'-n*)[39]. The UE executes then step 1 over the candidate resources in SW' in order to evaluate the currently available and excluded resources. If the selected resource at slot *m* is now excluded, then the UE has detected what is called in 3GPP standards a re-evaluation (Section 8.1.4 in [42]). Re-evaluation applies to both dynamic and semi-persistent schemes. This re-evaluation triggers the execution of step 2 to select a new SL resource among the currently available resources in SW' (Section 5.22.1 in [74]) (Fig. 20). The UE does not execute step 2 if the initially selected resource remains available. We should note that the UE could have selected *N* candidate resources in the initial selection window SW (for the transmission of the TB and its retransmissions as described previously). If this is the case, the UE can decide again whether to execute again step 1 to check if the selected resources are still available. If the UE detects that a subset of *M* ($M \leq N$) of the selected resources are not available anymore (i.e. they have been excluded after executing again step 1), then the UE executes step 2 to select *M* new candidate resources within the selection window SW'. It is important to highlight that the re-evaluation is an important novelty introduced in NR V2X SL mode 2 compared to LTE V2X mode 4. It provides higher flexibility in the management of resources and better capacity to cope with the variability in transmitted messages and previously undetected interference problems between vehicles.

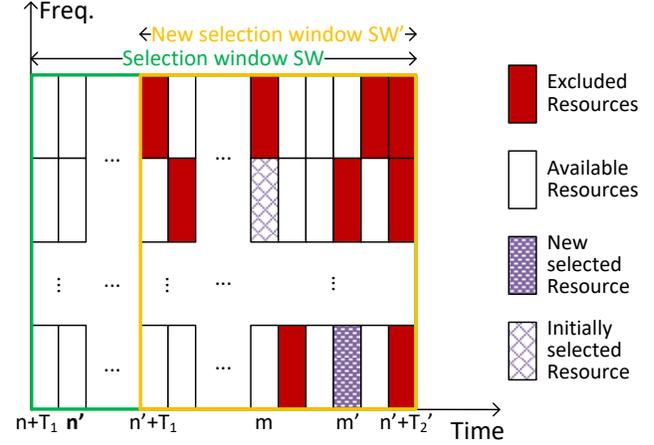

Fig. 20. New step 2 execution after a new step 1 execution and re-evaluation detection in NR V2X SL mode 2.

### 2) Pre-emption mechanism

Mode 2 introduces a pre-emption mechanism to prioritize traffic[40] [45] that is an important novelty compared to mode 4. It offers a more flexible management of resources under the presence of traffic with different priorities. With pre-emption, a UE with low priority traffic must free its reserved resource if it estimates that another UE with higher priority will use the reserved resource[41]. When a priority threshold is (pre-)configured in the resource pool, then the UE only frees its reserved resource if the priority of the other UE is higher than the priority threshold. The UE frees its reserved resource (e.g. at the slot *r*) to avoid a collision with the estimated transmission of the other UE that has higher priority. Note that pre-emption applies to both the dynamic and semi-persistent schemes. Note also that the UE can only identify that other UEs will use its reserved resource if it executes again the step 1. The decision to execute the step 1 (and how often it is executed) is left to UE implementation. With pre-emption, the step 1 is only executed after a TB is generated and it is waiting for the reserved resource to be transmitted. This is the case because the UE can only know the priority of its TB once this is generated. The UE can execute the step 1 at slot *r*-$T_3$ or before. It could execute the step 1 after slot *r*-$T_3$ only if its processing capabilities are sufficient to complete step 1 and step 2 before *r*. Let us consider that the UE initiates the step 1 at the slot *n''*. The UE defines a new selection window SW'' that starts at slot *n''*+$T_1$ and ends at slot *n''*+$T_2$''. $T_2$'' must be within the range $T_{2min} \leq T_2'' \leq$ PDB-(*n''-$n_G$*). $n_G$ is the slot at which the current TB has been generated and PDB is the deadline for the transmission of the TB. If the reserved resource at slot *r* is now excluded with step 1, then the

---

UE checks the pre-emption condition (Section 8.1.4 in [42]), i.e. if the other UE that wants to utilize the reserved resource has higher priority or not (and higher than the priority threshold if it is (pre-)configured). If it has higher priority, the UE must free the reserved resource at slot $r$ and execute again step 2 to select a new SL resource among the currently available resources in SW'' (Section 5.22.1 in [74]). We should note that the UE could have initially reserved $N_R$ resources in SW''. If this is the case, the UE checks whether the $N_R$ resources are available or not after executing step 1 again. If the UE detects that a subset of $M_R$ ($M_R{\leq}N_R$) resources have been excluded after step 1 and they fulfil the pre-emption condition, then the UE frees the $M_R$ resources and selects new $M_R$ resources within the selection window SW'' following step 2.

*C. Congestion Control*

Rel. 16 supports congestion control for NR V2X SL communications in mode 2[42]. Like in Rel. 14 [23], Rel. 16 does not specify a particular congestion control algorithm but defines related metrics and possible countermeasures to reduce the channel congestion. These metrics are the Channel Busy Ratio (CBR) and the Channel occupancy Ratio (CR) that are referred to in Rel. 16 as SL CBR and SL CR.

In LTE V2X, the CBR is defined as the ratio of occupied sub-channels within the previous 100 subframes (Section 5.1.30 in [24]). A sub-channel is considered occupied if the measured RSSI exceeds a (pre-)configured threshold[43]. In NR V2X, the size of the time window for estimating the SL CBR is equal to 100 slots (for any $\mu$) or $100{\cdot}2^{\mu}$ slots by (pre-)configuration per resource pool [76], where $\mu$ is the SCS configuration factor. Rel. 16 also adapts the calculation of the RSSI to the granularity of NR V2X in the time and frequency domains (Section 5.1.25 in [76]).

The CR estimates the channel occupancy generated by a TX UE. In Rel. 14 LTE V2X, the CR is computed as the ratio of sub-channels that the TX UE utilizes or selects within a period or window of 1000 subframes or 1 second (Section 5.1.31 in [24]). The CR at subframe $n$ is estimated over subframes [$n$-$a$,$n$+$b$]. It is up to the UE to decide the values of $a$ and $b$ subject to $b{\geq}0$, $a$+$b$+1=1000, and $a{\geq}500$. Rel. 16 adapts the computation of SL CR so that the [$n$-$a$,$n$+$b$] range is defined in slots rather than subframes. Rel. 16 also adapts the size of the window to compute the SL CR (i.e., $a$+$b$+1) that is set equal to 1000 slots (for any $\mu$) or $1000{\cdot}2^{\mu}$ slots by (pre-)configuration per resource pool [76]. $a$ and $b$ are determined by UE implementation considering that $a$ must be positive, $b <$ ($a$+$b$+1)/2 and $n$+$b$ shall not exceed the last selected resource (Section 5.1.26 in [76]). When the UE evaluates the SL CR in slots [$n$+1, $n$+$b$], it shall assume that the transmission parameters used at slot $n$ are also utilized in the scheduled (re-)transmissions within this range. The UE shall also assume that packet dropping will not occur in the range [$n$+1, $n$+$b$].

A TX UE uses the measured SL CBR and SL CR to identify whether it has to modify its transmission parameters to reduce the channel load. This is done using a (pre-)configured lookup table that includes up to 16 CBR ranges. Each range is linked with a maximum SL CR ($CR_{limit}$) that the TX UE cannot surpass[44]. $CR_{limit}$ increases as the CBR range decreases. 3GPP establishes that the value of the $CR_{limit}$ for each CBR range should be a function of the priority of the TB and the absolute speed of the TX UE [63]. The standard specifies that the TX UE evaluates whether it is exceeding the $CR_{limit}$ and has to modify its transmission parameters at each (re-)transmission. To this aim, the TX UE computes the SL CR and SL CBR at slot $n$-$N_{proc}$ for a scheduled (re-)transmission at slot $n$[45]. $N_{proc}$ depends on the UE processing capability. The standard differentiates UEs with fast processing capability or processing capability 1, and UEs with slow processing capability or processing capability 2 (Section 8.1.6 in [42]). When $\mu$=0, $N_{proc}$ is equal to 2 slots for any UE processing capability. For any other $\mu$, $N_{proc}$ is equal to $2^{\mu}$ slots for a UE with processing capability 1 and equal to $2{*}2^{\mu}$ slots for a UE with processing capability 2. A TX UE has to modify its transmission parameters if the SL CR exceeds the $CR_{limit}$. The objective is to reduce the SL CR and control the channel load generated by a TX UE. A TX UE can modify the following transmission parameters per resource pool:

1) MCS: the TX UE can reduce the channel load using a higher order MCS that reduces the number of sub-channels necessary to transmit a TB.
2) Number of sub-channels: the UE can reduce its CR by limiting max($L_{PSSCH}$) or the number of sub-channels it can utilize. If a UE needs to fit a TB to a reduced number of sub-channels, it can utilize, e.g., a higher order MCS.
3) Number of (re-)transmissions: the UE can reduce its CR by limiting $N_{MAX}$ or the number of (re-)transmissions.
4) Transmission power: the UE can decrease the CBR by reducing its transmission power. If the CBR decreases to values within lower CBR ranges, the UE can utilize a higher $CR_{limit}$.

## VII. QoS Framework for 5G NR V2X

V2X communications over the LTE-Uu and LTE-PC5 interfaces have QoS support. For LTE-Uu based V2X communications, the LTE QoS model that is based on QoS class identifiers (QCIs) is utilized. The QCI is a scalar that references specific QoS characteristics (e.g., packet loss rate) to be provided to a service data flow. This LTE QoS model was extended to introduce standardized QCI values specific for unicast and multicast/broadcast V2X messages (Section 4.4.5.2 in [22]). V2X SL communications over the LTE-PC5 interface leverage the QoS handling specified by ProSe. In particular, LTE V2X manages the QoS of SL communications on a per-packet basis. V2X packets generated at the Application Layer are associated to priority (PPPP, Proximity Service Per-Packet Priority) and, optionally, reliability (PPPR, Proximity Service Per-Packet Reliability) values [81], [34]. It is up to UE-implementation how to map at the Access Stratum (AS) Layer (i.e., PHY and Layer 2) the PPPP and PPPR values to the Sidelink Radio Bearer (SLRB) configuration as well as how to configure the corresponding SL logical channels. This means there are no unified rules for managing QoS among UEs in LTE V2X SL, and different UEs may handle the prioritization of logical channels differently.

---

[42]The gNB controls the channel load under mode 1. To this aim, it can request each UE to report (periodically or on demand) its measured CBR [62].

[43]A range of values can be found in Rel. 14 [79]. Rel. 16 also specifies a range of values for this threshold that are defined as (-112 + $n$*2) dBm, where $n$ is an integer in the range $0 \leq n \leq 45$.

[44]For LTE V2X, this table is defined in Europe by ETSI in Table 1 of Section 4.4 in [80]. This table has not yet been defined for NR V2X. However, a new Work Item (NR-V2X access layer - REN/ITS-00446) is defined in ETSI to extend the LTE V2X access layer specifications to include NR V2X.

[45]This is in contrast to LTE V2X that uses the CBR and CR values computed at subframe $n$-4 for each scheduled (re-)transmission at subframe $n$.

NR V2X is designed to support diverse and stringent QoS requirements of eV2X services. These requirements are defined in terms of priority, transmission rate, latency, reliability, data rate, and communication range [51]. Rel. 16 identified that using per-packet QoS management for NR V2X SL based on PPPP and PPPR only might not be sufficient to account for all QoS requirements of advanced eV2X services. Consequently, 3GPP has adopted for NR V2X SL communications the more advanced 5G QoS model that is defined for NR Uu (Section 5.7 in [35], Section 5.4 in [51], [82]). The 5G QoS model is described in Section VII.A. QoS management in NR V2X SL is configured by the network and is not dependent on UE implementation. It is based on *QoS Flows*. QoS Flows are associated with the QoS requirements of the eV2X applications following QoS Profiles (Section VII.B). The QoS Profiles are specified using QoS parameters and QoS characteristics [35]. Section VII.C describes the PC5-RRC sublayer that is introduced in NR V2X for QoS profiling and management of unicast NR V2X SL communications. For NR V2X communications over the Uu interface, Rel. 16 introduces advanced mechanisms to support service continuity in case of QoS changes (Section VII.D) and report analytics to monitor expected changes of the application QoS (Section VII.E).

### A. QoS management for NR V2X sidelink

Rel. 16 NR V2X QoS model is based on the 5G QoS model. The model is based on 5G QoS flows that are defined in [35] as "*the finest granularity for QoS forwarding treatment in the 5G System. All traffic mapped to the same 5G QoS Flow receive the same forwarding treatment (e.g., scheduling policy, queue management policy, rate shaping policy, RLC configuration, etc.). Providing different QoS forwarding treatment requires separate 5G QoS Flow.*". The 5G QoS model for NR V2X SL is illustrated in Fig. 21. First, the V2X application packets are mapped to QoS flows at the V2X layer taking into account the applications' QoS requirements (i.e., QoS parameters and QoS characteristics that are defined in Section VII.B). The mapping is done using the PC5 QoS rules following the mapping configuration provided by the 5G Core Network (Section VII.B). Each QoS flow is identified with a PC5 QoS Flow ID (PFI). The QoS flows are then mapped to sidelink radio bearers[46] using rules provided to the UE by the network via NR Uu RRC signaling. The mapping from QoS flows to radio bearers is performed at the SDAP (Service Data Adaptation

Protocol) layer. SDAP was first introduced in Rel. 15 and it is used in Rel.16 to support the QoS management of NR V2X SL communications. SDAP is also responsible for marking the transmitted packets with corresponding PFI. Finally, the SLRB is established with the peer node (e.g., another UE for SL unicast communications) and configured. Configuring an SLRB implies configuring all parameters of the AS Layer (i.e., Packet Data Convergence Protocol (PDCP), Radio Link Control (RLC), MAC, and PHY layer). It should be noted that the 5G QoS model applies when UEs are in network coverage and in RRC connected state. Rel. 16 NR V2X extends the 5G QoS model to also support out-of-coverage operation and QoS provisioning when the UE is in RRC the idle state ([51], Section 7 in [62]). In these cases, the UE could use (pre-)configured mapping rules or SLRB configurations provided by the network. Rel. 16 adapts the application of the 5G QoS model to NR V2X SL based on the cast type (unicast, broadcast and groupcast). Information on the specific SLRB configuration process (based on the 5G QoS model) for NR V2X SL unicast, broadcast, and groupcast communications is provided below.

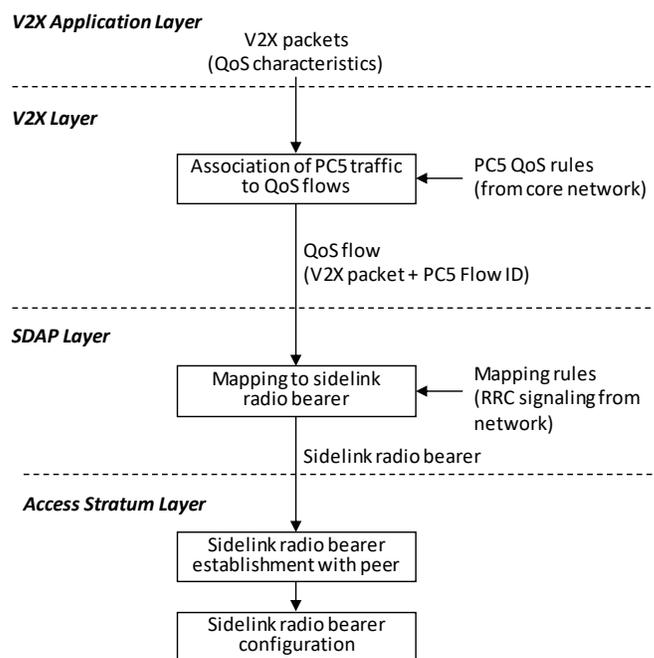

Fig. 21. Per-flow 5G QoS Model.

#### 1) SL unicast

For SL unicast, two procedures are defined for configuring the SLRB (Section 6.3 in [51]). The first one is based on the 5G QoS model described in Fig. 21, and is referred to as a UE-specific SLRB configuration. The process starts with the provisioning to the UE of the PC5 QoS rules as part of the service authorization process. The PC5 QoS rules are used at the UE to map the application layer eV2X packets to PFI(s). The UE uses the derived PFI(s) to request from the network a configuration for the SLRB(s). The network uses the provided PFI(s) to derive the PC5 QoS profile(s) (i.e., a set of specific

---

[46]A radio bearer is defined in [83] as "*the service provided by the Layer 2 for transfer of user data between UE and radio access network.*" The radio

bearer specifies the configuration of Layer 2 and PHY layer to meet the QoS requirements.

PC5 QoS parameters) and notifies the configuration of the SLRB to the UE. This UE-to-network request/response process is executed via NR Uu RRC signaling. With the configuration received from the network, the UE establishes and configures the SLRB with the peer UE via specific PC5-RRC signaling (see Section VII.C), and maps the PC5 QoS flow(s) to the configured SLRB. The mapping from PC5 QoS flow to SLRB ocurrs in the SDAP layer of the UE. The SL unicast communication between the two UEs is then performed as per the SLRB configuration. In the second procedure, the UE leverages NR SL (pre-)configured mapping rules (either signaled by the network or pre-configured in the UE) to configure the SLRB. This second procedure is used when the UE is out-of-coverage. The second procedure is referred to as pre-configured based SLRB configuration. This procedure maps (at the V2X layer) the eV2X application packets to PFI(s) based on (pre-)configured PC5 QoS rules. The NR SL (pre-)configuration is also used at the UE's AS layer to configure the SLRB based on the derived PFI(s).

*2) SL groupcast and broadcast*

Rel. 16 has defined three different procedures for the configuration of the SLRB for groupcast and broadcast communications (Section 6.3 in [51]). SL groupcast and broadcast transmissions are managed on a per-packet basis but with the advanced QoS profiles presented in Section VII.B. The V2X layer identifies the adequate PC5 QoS profile for each V2X packet to be transmitted. In the first procedure (referred to as UE-specific), the UE requests the network via RRC signaling for a specific SLRB configuration for the V2X packets, which are characterized by a PC5 QoS profile (e.g., range). The received SLRB configuration is used to establish the SLRB(s) and map the packets to the established SLRB(s). The second method (referred to as cell-specific) is based on a cell-specific configuration of the SLRB. In this case, the network uses V2X specific System Information Block (SIB) messages to broadcast the SLRB configuration associated with each possible PC5 QoS profile. The packets to be transmitted are then mapped to the SLRB configuration included in the SIB. Finally, NR SL can also use (pre-)configured based SLRB for SL groupcast and broadcast communications. In this procedure, the SLRB configuration associated with all PC5 QoS profiles is (pre-)configured either via signaling messages from the network or pre-configured in the UE. The UE uses this (pre-)configuration to map the V2X packets to be transmitted to the most adequate SLRB.

*B. PC5 QoS Profiles for V2X sidelink*

As described in Section VII.A, PC5 QoS Rules are used to classify and mark the eV2X application packets of SL user plane traffic, i.e., associate them to PC5 QoS Flows. A PC5 QoS Rule consists of the PFI of the associated PC5 QoS Flow, a precedence value, and a PC5 Packet Filter Set. The precedence value determines the order in which the PC5 QoS Rules are evaluated when classifying the traffic. The PC5 Packet Filter Set contains the information on the V2X service type (e.g., the ITS-AID[47] or PSID), Source/Destination Layer-2 (L2) ID[48], and Application Layer ID (e.g., Station ID).

TABLE IX
STANDARDIZED PQI TO QoS CHARACTERISTICS MAPPING (BASED ON TABLE 5.4.4-1 IN [51])

| PQI Value | Resource Type | Default priority Level | Packet Delay Budget | Packet Error Rate | Example V2X Services |
|---|---|---|---|---|---|
| 21 | GBR | 3 | 20 ms | $10^{-4}$ | Platooning between UEs – Higher level of automation (LoA); Platooning between UE and RSU – Higher LoA |
| 22 | | 4 | 50 ms | $10^{-2}$ | Sensor sharing – Higher LoA |
| 23 | | 3 | 100 ms | $10^{-4}$ | Information sharing for automated driving – between UEs or UE and RSU - Higher LoA |
| 55 | Non-GBR | 3 | 10 ms | $10^{-4}$ | Cooperative lane change – Higher LoA |
| 56 | | 6 | 20 ms | $10^{-1}$ | Platooning information exchange – low LoA; Platooning – information sharing with RSU |
| 57 | | 5 | 25 ms | $10^{-1}$ | Cooperative lane change – lower LoA |
| 58 | | 4 | 100 ms | $10^{-2}$ | Sensor information sharing – lower LoA |
| 59 | | 6 | 500 ms | $10^{-1}$ | Platooning – reporting to an RSU |
| 90 | Delay Critical GBR | 3 | 10 ms | $10^{-4}$ | Cooperative collision avoidance; Sensor sharing – Higher LoA; Video sharing – Higher LoA |
| 91 | | 2 | 3 ms | $10^{-5}$ | Emergency trajectory alignment; Sensor sharing – Higher LoA |

A UE checks for any service request or packet coming from the V2X application layer if there is any existing PC5 QoS Flow matching the request, based on the PC5 QoS Rules for the existing PC5 QoS Flows. If not, the UE derives the PC5 QoS parameters taking into account the V2X application requirements and service type according to the mapping configuration. The mapping configuration could be pre-configured at the UE or provided/updated by the V2X application server or the core network. The UE then creates a new PC5 QoS Flow and assigns it a PFI [51].

QoS Flows are characterized by QoS Parameters and QoS Characteristics that, overall, are referred to as QoS Profiles. For each cast type (unicast, broadcast, and groupcast), the UE maintains the mappings of PFIs to the PC5 QoS parameters and PC5 QoS Rules per destination (identified by Destination L2 ID). The assigned PFI and the associated PC5 QoS parameters are provided from the V2X layer to the AS layer. For unicast communications, the peer UEs negotiate the PFI and PC5 QoS parameters as described in Section VII.C.

Rel. 16 defines the following PC5 QoS Parameters associated to the PC5 QoS Flows (Section 5.4.2 in [51]):

1) *PC5 5G NR Standardized QoS Identifier (PQI).* PQI is an identifier (a scalar value) that refers to specific PC5 QoS characteristics associated with V2X services. It is introduced in Rel. 16 as a special 5G NR Standardized QoS Identifier (5QI).

---

[47]Intelligent Transportation Systems Application Identifier (ITS-AID) is a globally unique number used to identify an ITS application (e.g., 36 corresponds to CAM), as specified by ISO [84]. ITS-AID is named as PSID (Provider Service Identifier) in the IEEE WAVE specifications.

[48]Source and Destination Layer-2 IDs identify the sender and the target of the NR SL communication, respectively. One of three types of transmission modes (unicast, groupcast, and broadcast) is supported for a pair of a Source L2 ID and a Destination L2 ID.

2) *PC5 Flow Bit Rates*. This parameter is specified for QoS Flows with guaranteed bit rate (GBR) only. It consist of Guaranteed Flow Bit Rate (GFBR) and Maximum Flow Bit Rate (MFBR) parameters that are used to control the bitrate on the PC5 link. GFBR denotes the lowest bitrate that the V2X service can tolerate. MFBR sets the maximum bitrate expected by the QoS Flow. The excess traffic may get discarded or delayed by a rate shaping or policing function at the UE, RAN or CN. The measurements of GFBR and MFBR are done over an Averaging Time Window that is specified as a part of the associated QoS characteristics.

3) *PC5 Link Aggregated Bit Rates*. This parameter is defined for non-GBR links, and includes per link Aggregate Maximum Bit Rate (PC5 LINK-AMBR). PC5 LINK-AMBR limits the aggregate bit rate (measured over an AMBR averaging window) for all non-GBR QoS flows over a single PC5 unicast link with a peer UE.

4) *Range*. It is defined for groupcast communications over PC5. It indicates a minimum range (in meters) between the TX UE and the RX UEs for which the rest of PC5 QoS parameters must be guaranteed. For RX UEs beyond the indicated range, the communication is best effort[49].

The PQI identifies the following PC5 QoS characteristics that define the QoS profile and control how packets are managed from the QoS perspective:

1) *Resource Type* GBR, Delay-critical GBR or Non-GBR GBR flows require the dedicated allocation of network resources. Delay-critical GBR is introduced in 5G, and has specific definitions of Packet Delay Budget and Packet Error Rate that differ from the ones defined for GBR flows. Delay-critical GBR is also characterized by Maximum Data Burst Volume (MDBV).

2) *Priority Level*. It is similar to the priority value of LTE PC5 (PPPP, [34]), but improves backward compatibility for NR PC5. It is utilized to prioritize PC5 services (with the lowest value corresponding to the highest priority) if QoS requirements cannot be fulfilled for all of them.

3) *Packet Delay Budget (PDB)*. It indicates the upper bound delay. For Delay-Critical GBR flows, a packet delayed more than PDB is counted as lost (if the data burst does not exceed the MDBV within PDB, and the flow is not exceeding the GFBR). For GBR QoS flows, 98% of the packets shall not experience a delay over the PDB (if the flow is not exceeding GFBR). Services using Non-GBR QoS flows can experience packet drops and delays during congestions. Yet, in uncongested scenarios, 98% of the packets should not experience a delay over PDB.

4) *Packet Error Rate (PER)*. It is the upper bound for the rate of packets processed by the RLC layer of the TX UE but not successfully delivered by the RX UE to its PDCP layer. PER is not related to congestion losses but to radio errors. For Delay-critical GBR QoS Flows, packets delayed more than PDB are counted as lost and included in the PER estimation (unless the GFBR or MDBV is exceeded).

5) *Averaging window* (for GBR and Delay-critical GBR resource type only). It specifies the duration over which the GFBR and MFBR are calculated.

6) *Maximum Data Burst Volume* (for Delay-critical GBR resource type only). MDBV denotes the largest amount of data that the PC5 link is required to serve for QoS Flows with Delay-critical GBR within a period of PDB.

Table IX provides the mapping between standardized PQI values and PC5 QoS characteristics following Table 5.4.4-1 in [51]. The mapping is done for PC5 QoS characteristics that correspond to QoS requirements of V2X services provided as examples. PQI values of 21-23 refer to GBR flows mapped to services with higher degree of vehicular automation. Non-GBR flows are represented by PQI values of 55-59, and provide a wider range of options for the PDB and PER constraints. Delay Critical GBR flows are indicated by PQIs 90 and 91, and correspond to the most stringent PDB and PER requirements. Note that GBR and Delay Critical GBR PQIs can only be used for unicast PC5 communications. MDBV is specified for all Delay Critical GBR flows by default as 2000 bytes. The averaging window for GBR and Delay Critical GBR is set to 2000 ms. The default values of MDBV and averaging window could be overridden if indicated by the application. Similarly, a PQI could be used with a priority level (indicated by an application) that is different than its default value.

## C. PC5-RRC

PC5-RRC is introduced in NR V2X to provide functionalities to support SL unicast communications [62]. Note that in LTE V2X, PC5-RRC has a single functionality, which is to support exchanging synchronization-related information between the UEs on the Sidelink Broadcast Control Channel (SBCCH) [19]. PC5-RRC comprises RRC protocol and signaling that runs over the AS PC5 control plane protocol stack composed of the RRC, PDCP, RLC, MAC and PHY sublayers [45]. The PC5-RRC functionalities in NR V2X mainly consist of exchanging AS-level information that needs to be aligned between TX UE and RX UE to support SL unicast communications. These include:

- SL UE capability information to indicate additional, i.e., non-mandatory, features and parameters that the UE supports for each AS radio protocol (only utilized for the unicast NR SL communication), such as the frequency band, SCS, and MCS.

- SLRB configuration information to establish, modify, and release SLRBs (cf. Section VII.B).

- SL measurement configuration provided by the TX UE to the RX UE. The RX UE should be able to report RSRP measurements to the TX UE (without necessarily notifying the gNB of the result). Based on the RSRP measurements, the TX UE could then adjust its SL transmission power. The reporting could be event-triggered or periodical (for more information, see

---

[49] For example, transmission reliability is improved in HARQ groupcast feedback option 1 with HARQ retransmissions based on NACK sent only by RX UEs within the required communication range.

Section V.C.7).

For the signaling of the above information between the peer UEs, different PC5-RRC procedures and messages are specified [87]. UEs can request each other's capability information via Capability Enquiry message along with sending their own, whereas, SL RRC Reconfiguration procedure is used for the exchange of SLRB and SL measurement configurations. Signaling is done over the Sidelink Signaling Radio Bearer (SRB) on the logical channel SCCH. In case a configuration fails, UEs can utilize explicit failure message and timer based indications to notify the failure to the peer UE.

A PC5-RRC connection is defined as a logical connection between a pair of Source and Destination L2 IDs for unicast NR SL communication [87]. A UE may have multiple PC5-RRC connections, i.e., unicast connections, with one or more UEs for different pairs of Source and Destination L2 IDs. Signaling for establishing a PC5-RRC connection is initiated after a corresponding unicast link is established. The PC5-RRC connection and the corresponding SLRBs are released when the unicast link is released. Establishment and release of the unicast link is specified in [51].

PC5-RRC also supports the detection of SL Radio Link Failures (RLF) over the unicast NR SL communications. This is important to determine whether or when to release an SL unicast connection, for example, due to the degradation of the link as UEs move away from each other. The RLC triggers the SL RLF declaration when the maximum number of retransmissions to a specific destination has been reached. Upon the declaration, the UE releases the PC5-RRC connection immediately and discards any associated SL UE context.

### D. Alternative QoS Profiles and Service Requirements for V2X communication over Uu

Rel. 16 also introduces advanced mechanisms to support service continuity in NR V2X communications over Uu. Service continuity is an important requirement of critical V2X services. Rel. 16 facilitates service continuity by supporting V2X applications with a range of different configurations and QoS characteristics (e.g., different bitrates or delay requirements). This is useful since the applications can continue to be operational even if the initial QoS profile is not available and only an alternative QoS profile (i.e., with a lower QoS) can be used. An alternative QoS profile is a combination of QoS parameters and QoS characteristics to which the application traffic is able to adapt and that has the same format as the QoS profile for that QoS Flow (as described in Section VII.B). Note that Section VII.B presents PC5 QoS Profiles, i.e., PQI. Equivalent profiles for 5G communications over the Uu are reported in Section 5.7.4 of [35] and are referred to as 5QI.

To support V2X applications that can operate with different configurations, the V2X AS provides to the 5GS the requested level of service requirements and the alternative service requirements. The alternative QoS profile contains one or more QoS reference parameters in a prioritized order for service operation. This enables the 5GS to use the alternative service profiles in case of the QoS changes. The NG-RAN notifies the 5GC that an alternative QoS profile can be supported, and then the 5GC can provide this notification to the V2X AS.

The V2X AS includes the alternative service requirements as specified in Section 6.1.3.22 of [88] when providing service information to the PCF (i.e., through the NEF in case the V2X AS is outside of the 3GPP network domain, see Section IV). The V2X AS subscribes to receive notifications from the PCF when the QoS targets (the initial or the alternative ones) can no longer/again be fulfilled. The PCF enables QoS notification control and includes the derived alternative QoS parameter sets (in the same prioritized order indicated by the V2X AS) in the information sent to the SMF. If the NG-RAN receives from 5GC (through the SMF) a list of alternative QoS profile(s) for a QoS flow, it checks if a QoS profile of the list can be supported and sends a notification to the PCF. When the PCF notifies the V2X AS that the QoS targets can no longer/again be fulfilled, it includes the alternative QoS parameter set.

### E. QoS Sustainability Analytics for V2X communication over Uu

The experienced QoS over the Uu interface may be affected by various factors (e.g., UE density, interference, mobility, handover, and roaming transitions). Rel. 16 introduces in the 5GS the mechanisms to monitor, collect and report information of the experienced QoS. In particular, the 5GS can notify the V2X application (upon request and through a V2X AS) of an expected or estimated change of QoS before it actually occurs [51]. This procedure is referred to as QoS Sustainability Analytics in 3GPP standards and helps the V2X application to decide in a proactive and safe manner if there is need for an application change (e.g., safely stop a service, adapt an application, etc.) when the QoS degrades. A V2X AS may request notifications on QoS sustainability analytics for an indicated geographic area and time period. The V2X AS then provides the network with location information in the form of a path of interest or a geographical area where to receive notifications of potential changes in QoS. The V2X AS also sends to the 5GS the QoS parameters that should be monitored as well as the corresponding QoS thresholds (e.g., minimum acceptable data rate, maximum acceptable packet error rate) for efficient and safe operation of an application. The 5GS compares the QoS thresholds with predicted values of QoS parameters to decide if it should notify tthe V2X AS of an expected change of QoS.

The V2X AS can either subscribe to notifications (i.e., a Subscribe-Notify model) or request a single notification (i.e., a Request-Response model) by the 5GS. Fig. 22 depitcs the procedure to provide notifications on QoS Sustainability Analytics. This procedure is described in detail in Section 6.4.1 of [51] and Section 6.9 of [89] and consists of the next steps:

1) The V2X AS collects application layer information (e.g., V2X service, path, path start time and QoS requirements and thresholds).
2) The V2X AS requests or subscribes to analytics information on QoS Sustainability provided by the NWDAF through the NEF (see Section IV). NWDAF is responsible for on demand provision of analytics.
3) The NWDAF collects statistics provided by the

Operations, Administration and Maintenance (OAM) entity that is responsible for management plane functions including network performance monitoring.

4) The NWDAF computes the requested analytics or prediction about expected change of QoS for the requested area and time period. The NWDAF can detect the need to notify a potential QoS change by comparing the requested analytics of the target QoS profile against the threshold(s) provided by the V2X AS.

5) The NWDAF replies to the V2X AS (via the NEF) with the following information:
   - Applicable Area. A list of Tracking Area Identifiers (TAIs) or Cell IDs where the provided analytics applies.
   - Applicable Time Period. The time period during which the analytics applies.
   - Crossed Threshold(s). The QoS parameters with thresholds that are met or exceeded compared with the analytics or prediction computed by the NWDAF.
   - Confidence. Confidence of computed analytics or prediction.

6) The V2X application can take the necessary decisions based on the notification received from the network (these decisions are outside of 3GPP scope).

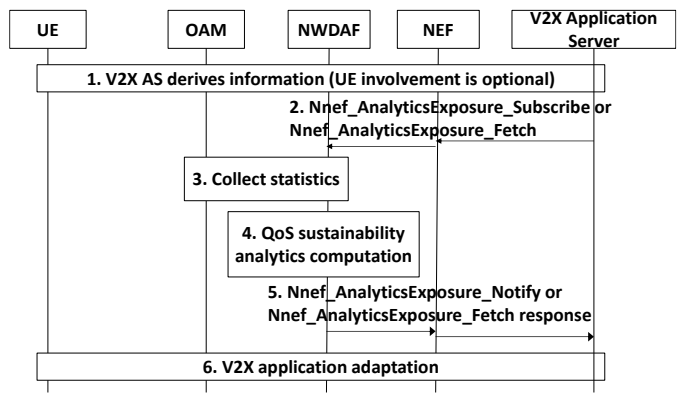

Fig. 22. Notification on QoS Sustainability Analytics to the V2X Application Server (based on 3GPP TS 23.287 and 3GPP TS 23.288).

## VIII. 5G NR V2N ENHANCEMENTS

### A. Uu enhancements

Rel. 16 NR V2X supports V2N communication over the Uu interface. Rel. 16 introduces enhancements to the Rel. 15 NR Uu and LTE Uu interfaces to meet the requirements of advanced eV2X services. It should be noted though, that neither Rel. 15 nor Rel. 16 NR Uu support broadcast and multicast V2N communication. If any eV2X service requires broadcast or multicast transmissions over the Uu interface, this can be supported using the MBSFN (Multimedia Broadcast Single Frequency Network) and SC-PTM (Single-Cell Point-To-Multipoint) technologies introduced in Release 13 (Rel. 13) [90]. This section presents the main enhancements to the NR Uu interface introduced for NR V2X in Rel. 16.

### 1) Multiple active UL configured grants

Uplink configured grants (UL CGs) are a set of periodic resources utilized to schedule V2N transmissions semi-persistently. Similarly to CG in sidelink (Section VI.A), UL CGs avoid the signaling necessary to dynamically schedule each transmission. Uplink semi-persistent scheduling (UL SPS) was first introduced in Release 8 (Rel. 8) [91]. Rel. 15 LTE V2X supports an enhanced version of Rel. 8 UL SPS with up to eight UL SPS configurations. These configurations are established by the eNB using the UEs' reports on UL traffic characteristics (Section 5.2.2 in [15]). A Rel. 15 UL SPS configuration is defined, among others, by the semi-persistent scheduling periodicity (indicated in number of subframes or short transmission time intervals (TTIs)) and the number of permitted blind retransmissions (see [19] for a complete list of configuration parameters). Out of the eight UL SPS configurations, a UE can only have one configuration active at a time (Section 6.1.2 in [62]). The eNB dynamically selects the most suitable configuration based on the information sent on the UE's report (including the periodicity of TBs and their maximum size) [19]. The eNB activates and releases an UL SPS configuration using the DCI (Section 5.3.3.1.1 in [93]). The DCI also indicates the resources assigned to a UE and the MCS used for all its UL transmissions.

Rel. 15 LTE V2X UL SPS may not efficiently support eV2X services with varying traffic patterns and strict QoS requirements. Supporting eV2X services with UL SPS from Rel. 15 LTE V2X may result in an overprovisioning of resources to satisfy the most demanding QoS requirements. It can also result in frequent re-configurations of the UL CGs to adjust the resource allocation and MCS to the packets to be transmitted. In this context, Rel. 16 includes the support for multiple active UL CGs in order to efficiently and simultaneously support eV2X services with distinct requirements (e.g., latency, packet size, and reliability). These grants are also activated, configured, and released using the DCI. Fig. 23 shows an example in which a UE has two active UL CGs. Each grant is characterized by a time-frequency domain allocation, periodicity, and number of blind retransmissions of a TB (the number of blind retransmissions can be 2, 4, or 8). A complete characterization of each grant for Rel. 16 UL CG is available in [63], and it includes (among others) the MCS, TB size, DMRS configuration, and an indication of whether power control should be utilized or not (Section 7.1.1 in [66]). A UE with multiple active CGs can support more efficiently periodic and aperiodic transmissions with different requirements. For every UL TB transmission, the UE would select the active UL CG that most closely satisfies the service requirements. It can occur that if a UE has multiple active UL CGs, there are resources from different grants in the same slot (i.e., they coincide in time). In this case, the UE selects for this TB transmission the UL SPS configuration (from the grants that overlap on the same slot) that best fits the service requirements (Section 6.1.2 in [62]). It should be noted that a UE can only perform one transmission using any of the UL CGs at a time (i.e., at the same slot).

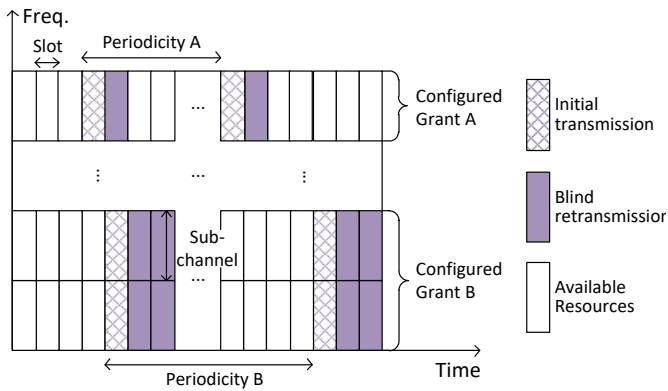

Fig. 23. Multiple active UL configured grants for NR V2N communications.

### 2) UE assistance information to the gNB

Rel. 16 allows UEs to report relevant assistance information to the gNB. This includes information about the Uu and SL traffic characteristics such as periodicity, packet latency requirements, and maximum TB size among others (see Section 5.7.4 in [63] for a complete list). The gNB can exploit this information to identify the UL CGs that best match the characteristics and requirements of the Uu traffic. The assistance information can also be utilized to improve V2N communications in scenarios where SL and Uu transmissions share the same radio resources. In this case, the gNB can exploit the information about the SL traffic characteristics for scheduling UL transmissions and to identify the adequate UL CGs that can also minimize the interference to the SL transmissions. The UE can also report (over the Uu interface) assistance information to the gNB that can be utilized by mode 1 and mode 2 SL scheduling [94]. These reports include information about the SL traffic (e.g., SL CBR of the SL resource pool) and UE-related geographic information (e.g., position, speed). The gNB could, for example, use UE's location to avoid assigning under mode 1 the same resources to UEs that are close to each other. On the other hand, the gNB could assign the same resources to UEs that are far enough apart in order to improve the efficiency of resource use. For mode 2 SL scheduling, [94] shows that geo-location information could be used, for example, to allocate the same resource pool to vehicles driving in the same direction on a highway.

### B. Mobility Enhancements

Rel. 16 introduces mobility enhancements to manage handovers (HOs) of advanced eV2X applications supported over the Uu interface (e.g., remote driving). The HO mechanism in NR Rel. 15 is based on LTE (legacy HO). In a legacy HO, the source cell configures the moment at which the UE must execute a HO by sending a HO command. These commands are generally sent when the source cell detects that the UE receives a signal level from a potential target cell higher than from the source cell. The transmission of the HO command may fail (e.g., due to low signal level). If this happens, the UE will not perform the HO to the target cell and may suffer a RLF with the source cell and consequently a service interruption. Even if the HO command is correctly received, the legacy HO includes a mobility interruption time during the time the UE disconnects from the source cell and connects to the target cell [16]. During this time, a UE cannot exchange user plane packets with any base station. LTE includes solutions such as Make Before Break and RACH-less HO that reduce the mobility interruption time to ~5ms [19]. However, this is still not sufficient for certain eV2X applications with strict QoS requirements such as remote driving. To overcome the limitations of legacy HOs, Rel. 16 introduces for NR the Dual Active Protocol Stack (DAPS) and conditional HOs (CHO). Rel. 16 supports intra-frequency and inter-frequency HOs from NR {FR1, FR2} to NR {FR1, FR2} [52]. For all possible options, [52] defines the handover delay (or time it takes to the UE to respond to the HO command) and maximum interruption time as a function of the utilized SCS.

With DAPS, UEs can have two active links during a HO, one with the source cell and the other one with the target cell [95]. The UE can simultaneously receive data from (or transmit data to) the two cells as depicted in Fig. 24. When the UE simultaneously transmits to both cells, it must share the available transmission power between both connections. In this case, the UE receives power control commands from the source and target cells [95]. The total transmission power configured via the power control commands by the two cells cannot be higher than the UE maximum transmission power. If it is, then the UE prioritizes the power control command from the target cell and may reduce the transmission power to the source cell. The capacity to transmit (and/or receive) simultaneously to two cells is UE implementation dependent. The UE must inform the network whether it can execute DAPS for its DL and/or UL transmissions. In case a UE supports DAPS HO for DL communication only, the UE must switch the UL connection from the source to the target cell after the Random Access (RA) procedure is completed with the target cell (Section 8 in [66]).

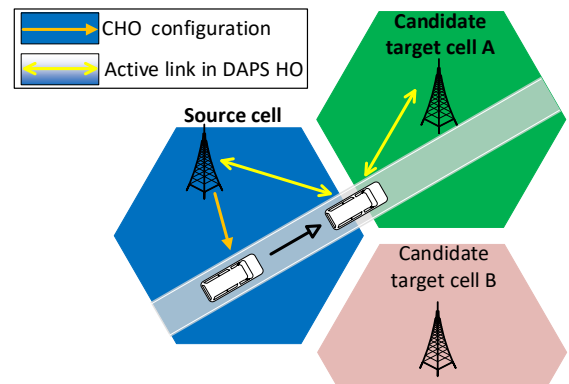

Fig. 24. Rel. 16 mobility enhancements: DAPS and CHO mechanisms.

A UE can receive ACKs or NACKs from the source and target cells during the DAPS HO. These acknowledgments may result in HARQ UL retransmissions by the UE. With DAPS, a UE can continue sending HARQ UL retransmissions to the source cell after starting the UL connection with the target cell [95]. Managing these retransmissions is not a challenge if the UE can maintain active links with both cells. However, conflicts can appear if the UE cannot simultaneously transmit to both cells. For example, a conflict occurs if both cells schedule a UL transmission from the UE in resources that overlap in time. If such conflicts arise, the UE prioritizes the transmission to the target cell and discards the scheduled UL transmission to the source cell. The UE can perform UL HARQ retransmissions to the source cell until it receives an explicit indication from the target cell to stop the exchange of information with the source cell.

Rel. 16 also introduces the concept of CHO to improve the reliability of the handover process [95]. CHO avoids the risk of not receiving correctly the HO command. With CHO, the UE is notified of the conditions for executing a CHO before the link quality. This process is referred to as CHO configuration (Fig. 24) and is implemented through the exchange of RRC messages between the source cell and the UE. The UE constantly compares the HO measurements (i.e. RSRP, Reference Signal Received Quality (RSRQ) and Reference Signal-Signal to Noise and Interference Ratio (RS-SINR)) with the conditions on the CHO configuration to decide when to execute the HO [95]. A condition included in the CHO configuration is associated to a reference signal type (SSB or CSI-RS sent by a gNB), a trigger quantity[50] (RSRP, RSRQ or RS-SINR), and an event[51]. An example of an event (referred to as A3 event in the standard, Section 5.5.4 in [63]) is when the UE detects that the signal level of a target cell is higher than the source cell by a given offset. Each condition included in the CHO configuration is associated with a candidate target cell for the execution of the CHO. Multiple conditions can be associated with a candidate target cell as long as: 1) all the conditions associated with a candidate target cell must refer to a single reference signal type; and 2) all conditions use up to two triggering quantities (out of the three available ones). The UE continuously compares the conditions on the CHO with its HO measurements to decide when to execute the HO. When several candidate target cells simultaneously fulfill the conditions to execute a CHO, the selection of the candidate target cell with which to execute the CHO is left to UE implementation. Once a UE starts executing a CHO with a candidate target cell, it stops checking the conditions of other candidate target cells and ignores any new CHO configuration from the source cell [96]. It should be noted though that CHO is optional for both UEs and cells [96]. Rel. 16 still considers the use of legacy HO that is compatible with CHO. It is then possible that a source cell configures and sends a CHO configuration but then sends a HO command (i.e., legacy HO) to the UE. In this case, the legacy HO is prioritized

The execution of a CHO or a legacy HO may fail. If this happens, the UE has to select a cell following the process described within Section 5.2.3 in [97]. This process includes two procedures depending on the information (e.g., frequencies and cell parameters) stored in the UE about the neighboring cells. This information could be obtained, for example, as part of the CHO configuration for the candidate target cells. If the UE can leverage stored information, it selects a cell that fulfils the received signal level and quality selection criterion defined in Section 5.2.3.2 of [97]. If the selected cell belongs to the set of CHO candidate target cells, the UE tries (again) to execute the CHO. If the CHO fails again, the UE does not try anymore to execute the CHO. Instead, it performs an RRC connection re-establishment procedure with the selected cell (Section 5.3.7 in [63]). The RRC connection re-establishment is also performed if the selected cell is not a CHO candidate target cell. This is also the case when the UE selects the cell without leveraging stored information about the surrounding cells. However, when the selected cell is not a CHO candidate target cell, the selected cell might not be aware of the current UE context (i.e., information needed to maintain the service). Retrieving the UE context increases the delay to resume the Uu data transmission (Section 5.3.7 in [63]).

In addition to DAPS and CHO, Rel. 16 also introduces a mechanism to improve the reliability of legacy HOs in NR V2X [95]. The mechanism is referred to as fast recovery handover failure. It can be applied when the UE is moving away from the source cell and does not receive correctly the HO command. The fast recovery handover failure includes three different timers: TTT, T310 and T312. First, the TTT timer is activated when the UE is approaching a target cell and the conditions specified in the A3 event are fulfilled. The T310 timer activates (after the activation of the TTT timer) when the UE detects a number of N310 out-of-sync signals[52]. The T312 timer activates when the TTT timer expires. The T312 timer is introduced in Rel. 16 for fast recovery handover failure, following the early RLF recovery mechanism of LTE described in [98]. The introduction of this timer forces the execution of the RRC connection re-establishment procedure when the T312 expires. Otherwise, as the T310 timer is usually longer than the T312 timer, the RRC connection re-establishment procedure would be executed when the T310 timer expires. In case the T310 timer expires earlier than the T312 timer, the T312 timer is cancelled. It should be also noted that when the TTT timer expires, the UE sends to the source cell a measurement report, which may trigger a HO command from the source cell. If this HO command is received while the T312 timer is active, the UE performs a legacy HO to the target cell. In this case, the T310 and T312 timers are cancelled, since there is no need to execute the RRC connection re-establishment procedure.

## IX. COEXISTENCE BETWEEN LTE V2X AND NR V2X SIDELINK

From the start of the work on NR V2X [9], 3GPP decided that NR V2X will complement LTE V2X and not replace it. LTE V2X is envisioned to support basic active safety applications, whereas NR V2X will support more advanced applications including connected and automated driving (Section III). Vehicles will select the adequate Radio Access Technology (RAT) based on the active V2X application. Fig.

---

[50] Trigger quantity is the term used in the CHO mechanism to refer to the metric used to trigger the HO execution.

[51] An event defines the condition that should be fulfilled between the HO measurements obtained from the source and the target cell to trigger the HO execution in the CHO mechanism.

[52] The UE detects an out-of-sync signal when the signal level of the source cell's PDCCH channel is below the threshold Qout [99]. The N310 variable can be equal to {1, 2, 3, 4, 6, 8, 10, 20} and it is configured by the source cell (Section 6.3.2 in [63]).

25 illustrates the coexistence of NR V2X and LTE V2X. Vehicles exchange data and/or control information with other vehicles through the LTE or NR PC5 interfaces and with the infrastructure using the LTE or NR Uu interface. Referring to Fig. 25, vehicles A, B, and C can establish multi-RAT and multi-link V2V communications and still be able to communicate with vehicle D. In this context, Rel. 16 has defined mechanisms to facilitate the coexistence of NR V2X and LTE V2X at the vehicle/device level (referred to as in-device coexistence) in order to manage SL resources [45]. Rel. 16 also considers the coexistence and cooperation between the two RATs at the network level [45]. This coexistence is referred to as cross-RAT control.

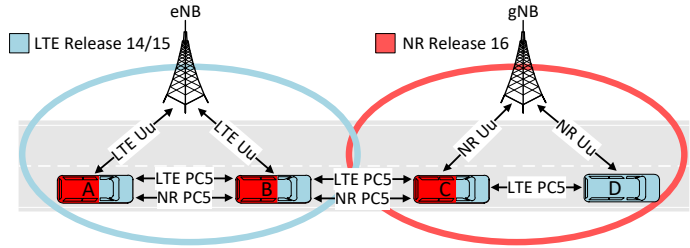

Fig. 25. Coexistence between NR V2X and LTE V2X.

### A. In-Device Coexistence

The integration of two RATs within a vehicle creates a number of in-device coexistence challenges, especially if the two RATs share a part of the radio chain needed to perform transmission/reception or if the two RATs need to share the same channel/carrier/resources. Co-channel coexistence is defined in 3GPP as the concurrent usage of the same time/frequency resources by the two RATs (NR V2X and LTE V2X). Co-channel coexistence is not supported by Rel. 16 V2X since each RAT will utilize a different resource pool [100]. This decision was made at the start of the work item [10] due to the complexity that co-channel coexistence entails. However, the configuration and selection of resources in one RAT affects the other RAT for two main reasons [44]. The first reason relates to the transmission power. A vehicle has a maximum transmission power and the available power can be utilized by one RAT only during a certain amount of time (coordination is necessary between the two RATs to alternate) or it can be shared if both RATs transmit simultaneously. The first challenge is then related to the simultaneous transmission (Tx/Tx) over both RATs. The second reason relates to the potential interference at a given vehicle caused by the two RATs. This interference can appear if the LTE and NR V2X resource pools are not sufficiently separated in frequency and the two RATs are simultaneously utilized. If such separation cannot be guaranteed, a vehicle will might not be able to correctly receive data through RAT$_1$ if it is transmitting at the same time through RAT$_2$. The second challenge therefore occurs when the transmission of one RAT temporarily overlaps with the reception of the other RAT (Tx/Rx). Simultaneous reception (Rx/Rx) might be possible depending on the implemented solution to address the in-device coexistence challenges [68].

To address these challenges, Rel. 16 proposes TDM (Time Division Multiplexing) and FDM (Frequency Division Multiplexing) solutions [62]. The solutions are applicable independently of whether NR V2X and LTE V2X operate using mode 1 and mode 3, or mode 2 and mode 4, respectively [44].

### 1) TDM Solutions

TDM solutions prevent vehicles from simultaneously transmitting over both RATs, thus requiring the synchronization of the two RATs [62]. Rel. 16 defines long-term and short-term TDM solutions for the coordination of the two RATs [62]. Both options are illustrated in Fig. 26.

The long-term TDM solution assigns statically the time during which each RAT can utilize its own resources (Fig. 26(a)). This allocation is pre-configured or determined by the gNB or eNB. Fig. 26(a) shows an example where LTE V2X is allowed to operate (Tx or Rx) during subframes $SF_1$ and $SF_3$ while NR V2X is allowed to operate (Tx or Rx) during subframes $SF_2$ and $SF_4$. A vehicle can only transmit or receive in the same SF since the radios operate in half-duplex. The other RAT cannot transmit or receive during this time. This solution is simple to design, implement, and operate, but can negatively impact the QoS experienced over each RAT (e.g., the latency), since a RAT can only utilize the resources within its pool during the assigned subframes.

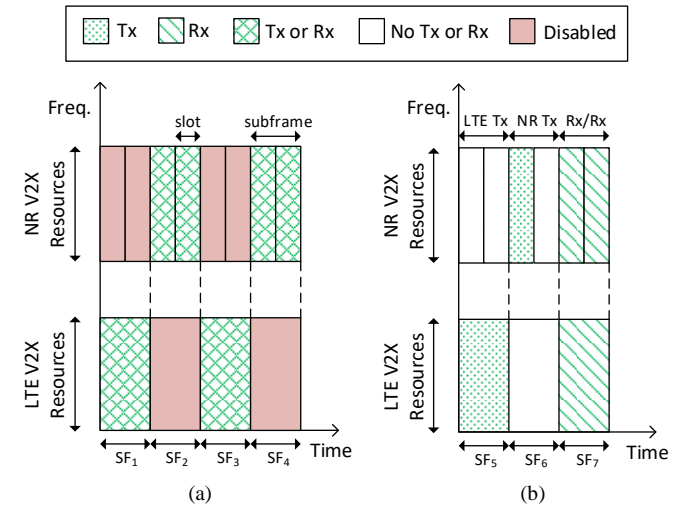

Fig. 26. TDM solutions for the coordination between RATs. (a) Long-term. (b) Short-term.

With the short-term TDM solution, each RAT can transmit or receive data in any slot or subframe of the assigned resource pool. LTE V2X and NR V2X need to dynamically coordinate the usage of their radio resources to avoid inter-RAT interference. This dynamic coordination enables a more flexible and granular usage of the resources that can better suit the requirements of applications served by each RAT. For the coordination, RATs exchange information about the resources they intend to use and the priority associated to this usage. In particular, RATs notify each other about [62]: 1) all subframes (SFs)[53] required for their planned transmissions (i.e., their reserved SFs); and 2) all SFs in which they expect to receive a transmission (i.e., all SFs with detected reservations by other vehicles); and 3) the priority of these transmissions and expected receptions (if available). With this information, the RATs can detect if they were both planning to be active (Tx or Rx) at the same time. If the RAT with higher priority is planning to use its resources for a transmission, then the other RAT cannot transmit or receive data at the same time. If the RAT with higher priority is planning to use its resources for receiving data, then the other RAT can also receive data at the same time but cannot transmit [45]. The possible scenarios for the short-term TDM solution are illustrated in Fig. 26(b). In subframe $SF_5$, a transmission of LTE V2X is prioritized (LTE Tx) and NR V2X cannot transmit or receive in any of the slots within $SF_5$. In subframe $SF_6$, a transmission of NR V2X is prioritized during the first slot of the subframe (NR Tx), but LTE V2X cannot transmit or receive during the complete $SF_6$. Subframe $SF_7$ represents the case in which a reception (for any of the two RATs) is prioritized. In this case, the other RAT can also receive (Rx/Rx) but not transmit.

---

[53]The co-existence with LTE V2X requires the time granularity to be at the subframe level and not the slot level.

The short-term TDM solution leaves certain decisions for UE implementation. For example, let's consider the scenario where a UE has a potential conflict between the two RATs because they both want to transmit at time $t_n$. The conflict can be avoided using the short-term TDM coordination mechanism if the two RATs exchange their coordination information before $t_n$-$T$. The value of $T$ is up to UE implementation subject to an upper bound of 4 ms[54]. If the coordination information is not exchanged before the deadline, the approach to resolve the conflict between the two RATs is left to UE implementation [56]. For example, the vehicle could always prioritize LTE V2X transmissions as LTE V2X supports basic safety applications [44]. It is also left up to UE implementation the resolution of the conflict when both RATs want to transmit at the same time and have the same priority [44]. Conflicts between RATs can be frequent if the network load is high. These conflicts can be particularly delicate if a RAT is generally prioritized (e.g., due to the criticality of the data it transmits) and prevents the other RAT from being able to transmit, thus implying a QoS degradation. To prevent this from happening, 3GPP considers that the short-term TDM solution should only be utilized when the load of both RATs is below an acceptable level to avoid high performance degradation to any of the two RATs [62].

A vehicle is not required to support the short-term TDM solution. In this case, the vehicle will implement the long-term solution if it operates under TDM. The gNB or eNB configures the time during which each RAT is active (and can utilize its resource pool) if the vehicle is under cellular coverage. To this aim, the vehicle must report to the gNB or eNB whether it can execute or not the short-term TDM solution [44].

It should be highlighted that the TDM solutions help combat the power and interference challenges of in-device coexistence. Indeed, the TDM solutions (long-term and short-term) allow only one RAT to transmit at a given time. Consequently, the transmitting RAT can operate at maximum transmission power. For the same reason, the TDM solutions prevent the interference between RATs. The two RATs can be active simultaneously only when they are in reception mode.

### 2) FDM Solutions

A vehicle can simultaneously transmit over NR V2X and LTE V2X if it implements FDM solutions for in-device coexistence [62]. In this case, the maximum transmission power of a vehicle or UE must be shared between the RATs. Similar to TDM, the sharing can be static or dynamic. In the latter case, the power is dynamically shared between RATs based on the priority of the data to be transmitted by each RAT [62]. An advantage of static power sharing is that it does not require any synchronization between RATs. Synchronization and coordination are necessary with dynamic power sharing to ensure each RAT adequately utilizes the corresponding transmission power at each point in time [62]. FDM solutions can also be inter-band or intra-band, depending on whether NR V2X and LTE V2X operate on the same or different frequency bands [62]. The frequency separation between the resource pools of LTE V2X and NR V2X (whether inter- or intra-band) can have a significant impact on the interference between RATs and hence on the operation and performance of FDM solutions. Sufficient spectral separation is needed not only to be able to transmit using both RATs, but also to be able to simultaneously transmit over one RAT and receive over the other RAT. Fig. 27 illustrates inter-band and intra-band FDM solutions with power sharing and inter/intra-band deployments. Static and dynamic power sharing as well as intra-band and inter-band operations were discussed during the standardization process. However, Rel. 16 specifications only include static power allocation in inter-band deployments [100]. This is the case because the inter-band FDM solution with sufficient spectral separation between the RATs is the only viable solution for avoiding interference and allowing simultaneously transmitting over one RAT and receiving over the other RAT (Tx/Rx) ([101]).

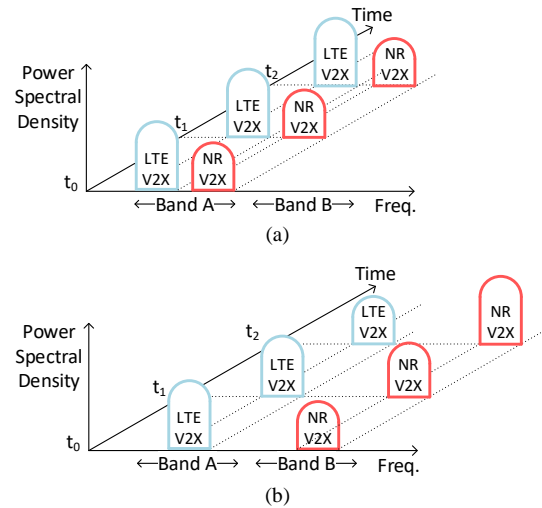

Fig. 27. FDM solutions. (a) Intra-band operation and static power sharing. (b) Inter-band operation and dynamic power sharing.

### B. Support for Cross-RAT control

Rel. 16 has standardized cross-RAT control so that eNBs and gNBs can cooperate to manage the radio resources of NR V2X and LTE V2X SL communications when vehicles are under cellular coverage. In particular, the eNB can manage the resources associated with the LTE and NR PC5 interface through the LTE Uu interface. Furthermore, the gNB can manage the resources associated with both the LTE and NR PC5 interfaces through the NR Uu interface [45]. This can reduce the need to deploy gNBs to support NR V2X SL communications. With cross-RAT control, the gNB or eNB can also configure the CBR measurement and UE assistance information reporting on LTE V2X or NR V2X sidelink, respectively (see Section VIII) [63]. However, the cross-RAT control has some restrictions that depend on the type of signaling utilized and the resource allocation modes [102].

Base stations (eNBs and gNBs) send RRC signaling messages to the vehicles for managing the cross-RAT resources. Base stations can use two types of RRC signaling for cross-RAT control: dedicated RRC signaling and V2X-specific SIB [102]. With dedicated RRC signaling, a base station of $RAT_1$ can configure the resource pool of $RAT_2$ (cross-resource pool configuration) and also allocate its radio resources (cross-resource allocation). Dedicated RRC signaling can only be utilized if the vehicle has established an RRC connection with the serving base station, and it is in RRC connected state. Establishing a RRC connection is not necessary when utilizing V2X-specific SIB signalling. However, the base station can only execute the cross resource pool configuration (and not cross-resource allocation) when operating with V2X-specific SIB signalling.

Rel. 16 introduces changes to the RRC signalling so that vehicles can correctly decode the information transmitted by eNBs and gNBs [102]. In particular, Rel. 16 defines new RRC containers within the RRC messages of dedicated RRC signalling to be able to execute cross resource allocation and cross resource pool configuration. The standard introduces a container in NR RRC [63] that is compatible with LTE RRC [19] so that a gNB can control and manage the LTE V2X resources. The standard also updates the LTE RRC and includes a new container compatible with NR RRC so that eNBs can control and manage the NR V2X resources [102]. Similarly, Rel. 16 introduces two new V2X-specific SIBs for cross-RAT control: a V2X-Specific SIB compatible with LTE RRC is defined for gNBs and a V2X-Specific SIB compatible with NR RRC is defined for eNBs [102].

Cross-RAT control can be used to manage the radio resources under NR V2X mode 1 and LTE V2X mode 3, and also under NR V2X mode 2 and LTE V2X mode 4. For mode 2 and mode 4, a base station can (but does not need to) configure the resource pools of both RATs (cross-resource pool configuration). This is the case because vehicles autonomously select their radio resources and hence do not require cross-resource allocation from the base stations. For mode 1 and mode 3, a base station of RAT1 can configure the resource pool of RAT2 and also allocate its radio resources. However, it should be noted that an eNB can only allocate resources of NR V2X mode 1 when operating with CG type 1 scheduling (and not with dynamic scheduling or CG type 2) [102].

## X. EVALUATION METHODOLOGY

Rel. 16 NR V2X introduces new communication modes (unicast, groupcast) compared to Rel. 14/15 LTE V2X, as well as more stringent use case requirements (increased reliability and reduced latency) and new data traffic patterns requiring higher message and data rates (Section III). These differences required 3GPP to define an evaluation methodology for LTE V2X and NR V2X [103] that has been utilized during the standardization work and that should guide the community when evaluating the performance of NR V2X. The evaluation methodology includes: i) new channel models, in particular for V2V SL; and ii) assumptions for system and link level simulations.

### A. Channel models

The SI that defined the evaluation methodology placed particular attention to the V2V SL channel models. For scenarios of interest (urban and highway), the goal of the study was to define propagation conditions, path loss and shadow fading, large scale parameters, and small scale parameters for V2V SL channels. Specifically, the approach to generate the V2V SL channels in Rel. 16 is based on the 12-step procedure defined in the Rel. 15 NR channel modeling study [104] and depicted in Fig. 28. However, Rel. 16 adapts several important components to the peculiarities of V2V channels. For example, the Rel. 16 V2X SL channel models introduce the NLOSv state in addition to the LOS and NLOS states used for SL and UL/DL. The NLOSv state describes V2V SL channels where the LOS path is blocked by another vehicle. There were several reasons leading to the introduction of NLOSv: i) measurements reported in [105] and [106] showed that blockage of SL LOS by vehicles exhibits considerably different properties compared to NLOS channels since NLOS assumes blockage by objects considerably larger than vehicles (e.g., buildings); ii) in highways, there are no other objects that block the LOS other than vehicles; and iii) dynamics of NLOSv blockage (e.g., in terms of temporal and spatial variation of blockage) is different to NLOS blockage.

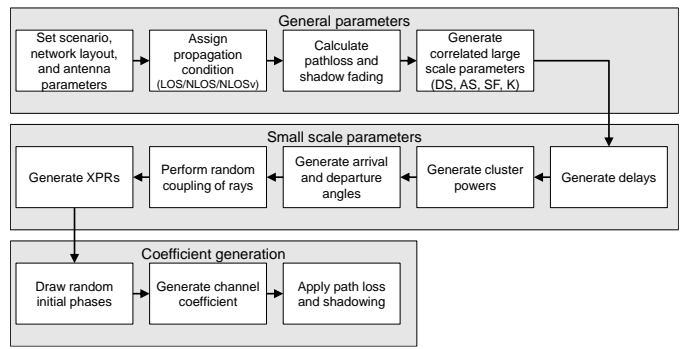

Fig. 28. Channel generation procedure for V2V SL (adapted from [104]).

Other important novelties introduced by the NR V2V SL channel models defined in Rel. 16 include:

1) New equations defining probability of LOS, NLOS, and

TABLE XI
PATHLOSS FOR V2V SL LINKS [103]

| LOS/NLOS/NLOSv | Pathloss [dB] | | Shadow fading std. deviation $\sigma_{SF}$ [dB] |
|---|---|---|---|
| LOS, NLOSv | **Highway** $PL = 32.4 + 20 \log_{10}(d) + 20 \log_{10}(f_c)$ | **Urban grid** $PL = 38.77 + 16.7 \log_{10}(d) + 18.2 \log_{10}(f_c)$ | $\sigma_{SF} = 3$ |
| NLOS | $PL = 36.85 + 30 \log_{10}(d) + 18.9 \log_{10}(f_c)$ | | $\sigma_{SF} = 4$ |

NLOSv state for urban grid and highway.

2) New shadow fading and fast fading parameters for NLOSv.

3) New dual mobility (i.e., both transmitter and receiver are mobile) Doppler calculations that also include variable scatterer velocities.

### 1) LOS probability, pathloss model, and shadow fading

Pathloss and LOS probability equations have been defined separately for all three channel states (LOS, NLOS, NLOSv). Formally, the three channel states are defined as follows [103]:

1) LOS: V2V link is in LOS state if the two vehicles are in the same street and the LOS path is not blocked by vehicles.

2) NLOS (LOS path blocked by buildings): V2V link is in NLOS state if the two vehicles are in different streets.

3) NLOSv (LOS path blocked by vehicles): V2V link is in NLOSv state if the two vehicles are in the same street and the LOS path is blocked by vehicles.

LOS probabilities were not previously defined for SL communications. In fact, the LOS probabilities included in [104] were only defined for UL/DL. This, along with introduction of a new SL channel state (NLOSv), required new LOS probability equations that have been defined for the evaluation methodology of NR V2X. These equations have been defined based on an extensive study reported in [108] that analyzed LOS probabilities and state transitions using maps of real cities and highways. 3GPP adopted a probabilistic approach for calculation of LOS probabilities for V2V SL communications when vehicles are in the same street. For urban grid scenarios, 3GPP assumes that the LOS path is blocked by buildings when vehicles are not in the same street and want to establish a V2V SL link (i.e., they operate under NLOS state). Consequently, LOS probability equations are only required for LOS and NLOSv states and are defined following Table X for V2V SL communications. In the table, $d$ denotes the relative distance between TX UE and RX UE in meters.

Furthermore, a new set of pathloss models for V2V SL

TABLE X
PROBABILITY OF LOS AND NLOSv STATES.

| Highway | |
|---|---|
| LOS | If $d \leq 475$ m, $P(LOS) = \min\{1, ad^2 + bd + c\}$ where $a = 2.1013 \times 10^{-6}$, $b = -0.002$ and $c = 1.01093$. If $d > 475$m, $P(LOS) = \max\{0, 0.54 - 0.001(d-475)\}$ |
| NLOSv | $P(NLOSv) = 1 - P(LOS)$ |
| **Urban** | |
| LOS | $P(LOS) = \min\{1, 1.05 \ast \exp(-0.0114d)\}$ |
| NLOSv | $P(NLOSv) = 1 - P(LOS)$ |

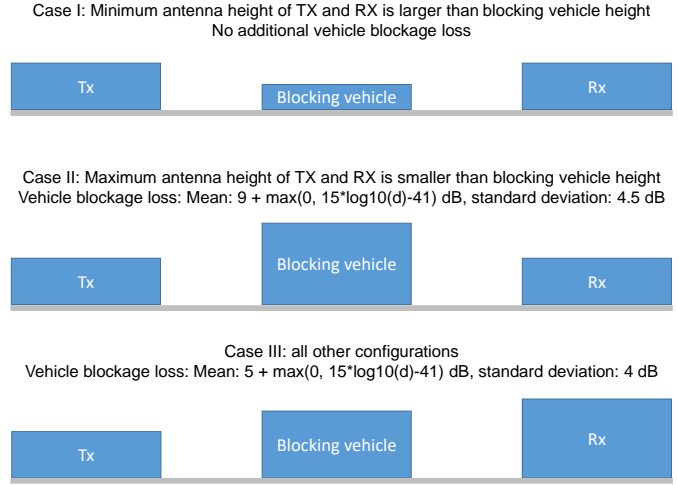

Case I: Minimum antenna height of TX and RX is larger than blocking vehicle height
No additional vehicle blockage loss

Case II: Maximum antenna height of TX and RX is smaller than blocking vehicle height
Vehicle blockage loss: Mean: 9 + max(0, 15*log10(d)-41) dB, standard deviation: 4.5 dB

Case III: all other configurations
Vehicle blockage loss: Mean: 5 + max(0, 15*log10(d)-41) dB, standard deviation: 4 dB

Fig. 29. Additional vehicle blockage loss under NLOSv: scenarios and loss calculation. Vehicle blockage loss is modeled by the normal distribution for all cases.

communications in highway and urban grid have been introduced in Rel. 16, along with new values for the shadow fading standard deviation. The shadow fading is modeled with a random variable according to a lognormal distribution with zero mean. The shadow fading models for all types of SL (V2V, V2P, P2P[55], V2R, and R2R[56] links) are taken over from LTE V2X, as described in [15]. Effectively, for each V2V link, shadow fading is an independent and identically distributed lognormal random variable. The LOS shadowing model from [15] applies to NLOSv as well. The model for spatial correlation of shadow fading defined for LTE V2X in [15] also applies to NR V2X.

Table XI contains the pathloss expressions for highway and urban grid scenarios where $f_c$ denotes the carrier frequency in GHz and $d$ denotes the Euclidean distance between a TX UE and a RX UE in 3D space in meters, i.e., considering also the heights of the Tx antenna and Rx antenna. Note that LOS and NLOSv states use the same pathloss equation. However, additional *vehicle blockage loss* is introduced for NLOSv following Fig. 29. The figure describes how much additional loss (if any) is added in case of NLOSv for different relationships of the height of Tx antenna, Rx antenna and the blocking vehicle. The resulting pathloss plots are shown in Fig. 30 for all link types and in urban grid and highway scenarios.

### 2) Fast fading model

A completely new set of fast fading parameters were introduced for SL under Rel. 16. Each of the states (LOS, NLOS, NLOSv) have a specific set of parameters associated with them, most of which are dependent on the assumed center

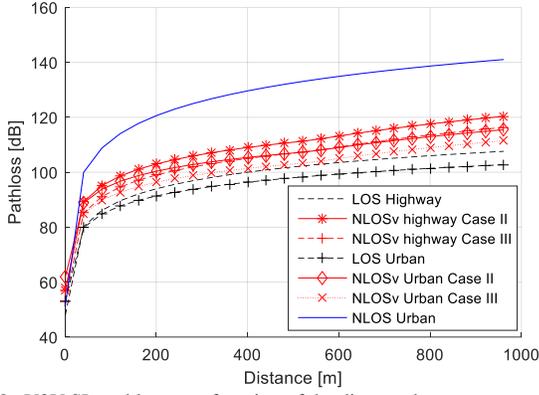

Fig. 30.  V2V SL pathloss as a function of the distance between transmitter and receiver. The pathloss for NLOSv is plotted as an average value for each case.

frequency $f_c$. Table 6.2.3-1 in [103] lists all of the fast fading parameters.

### 3) Modeling dual mobility for V2V SL

In V2V SL, transmitter and receiver are highly mobile. To account for this condition, Rel. 16 NR V2X introduces for the first time a dual mobility model addressing different Doppler components. The model is an extension of the single mobility model from [104]. It takes into account the relative speed difference between a TX UE and RX UEs as well as the relative speed of delayed paths coming from scatterers. Specifically, the Doppler for the LOS path is expressed as follows [103]:

$$v_{n,m} = \frac{\hat{r}^T_{rx,n,m} \cdot \bar{v}_{rx} + \hat{r}^T_{tx,n,m} \cdot \bar{v}_{tx}}{\lambda_0} \tag{8}$$

$$\bar{v}_{rx} = v_{rx}[\sin\theta_{v,rx}\cos\phi_{v,rx},\ \sin\theta_{v,rx}\sin\phi_{v,rx},\ \cos\theta_{v,rx}]^T \tag{9}$$

$$\bar{v}_{tx} = v_{tx}[\sin\theta_{v,tx}\cos\phi_{v,tx},\ \sin\theta_{v,tx}\sin\phi_{v,tx},\ \cos\theta_{v,tx}]^T \tag{10}$$

whereas the Doppler for the delayed (scattered) paths is:

$$v_{n,m} = \frac{\hat{r}^T_{rx,n,m} \cdot \bar{v}_{rx} + \hat{r}^T_{tx,n,m} \cdot \bar{v}_{tx} + 2\alpha_{n,m}D_{n,m}}{\lambda_0} \tag{11}$$

where $\lambda_0$ denotes the wavelength of the carrier, $\theta$ is the elevation angle, $\Phi$ is the azimuth angle, $\hat{r}_{rx,n,m}$ is the spherical unit vector with azimuth arrival angle $\phi_{n,m,AOA}$ and elevation arrival angle $\theta_{n,m,ZOA}$, given by:

$$\hat{r}_{rx,n,m} = \begin{bmatrix} \sin\theta_{n,m,ZOA}\cos\phi_{n,m,AOA} \\ \sin\theta_{n,m,ZOA}\sin\phi_{n,m,AOA} \\ \cos\theta_{n,m,ZOA} \end{bmatrix}. \tag{12}$$

$n$ denotes a cluster and $m$ denotes a ray within cluster $n$. $\hat{r}_{tx,n,m}$ is the spherical unit vector with azimuth departure angle $\phi_{n,m,AOD}$ and elevation departure angle $\theta_{n,m,ZOD}$, given by:

$$\hat{r}_{tx,n,m} = \begin{bmatrix} \sin\theta_{n,m,ZOD}\cos\phi_{n,m,AOD} \\ \sin\theta_{n,m,ZOD}\sin\phi_{n,m,AOD} \\ \cos\theta_{n,m,ZOD} \end{bmatrix}. \tag{13}$$

Furthermore, $D_{n,m}$ is a random variable with uniform distribution from $-v_{scatt}$ to $v_{scatt}$ m/s. $v_{scatt}$ is the maximum speed (in m/s) of the vehicle in the simulation, and $\alpha_{n,m}$ ($0 \le \alpha_{n,m} \le 1$) is a random variable with uniform distribution. $D_{n,m}$ ensures that scatterers can have a range of speeds varying from minimum to maximum speed values, whereas $\alpha_{n,m}$ determines the proportion of scatterers that are mobile relative to the TX UE and RX UEs.

### B. System level simulations

#### 1) Evaluation scenarios, vehicle types, and dropping options

Similar to evaluation settings in Rel. 14 and Rel. 15 LTE V2X, Rel. 16 considers two evaluation environments for NR V2X: urban grid and highway, which are shown in Fig. 31 and Fig. 32, respectively. The requirement for urban grid is that the simulation is performed over at least 3x3 road grids in order to reduce the lower interference values at the edge where there are fewer communicating vehicles. The same problem in the highway scenario is tackled with the so-called wrap-around that reintroduces the simulated vehicles at one end of the highway at the other end of the simulated area. For each environment, the evaluation scenarios are divided into below and above 6

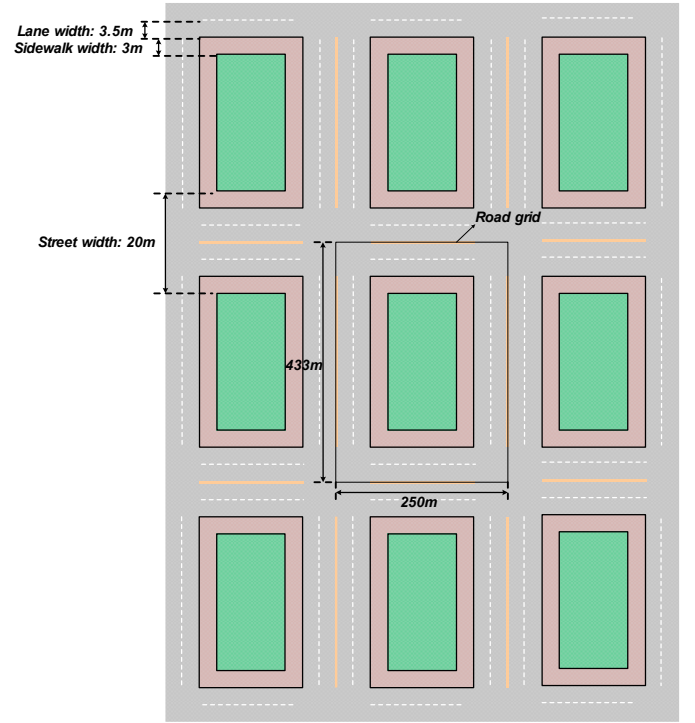

Fig. 31.  Road configuration for urban grid (adapted from [103]).

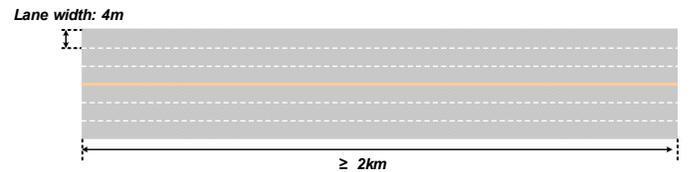

Fig. 32.  Road configuration for highway scenario (adapted from [103]). Three lanes per direction. Wrap-around applied to simulation area.

TABLE XII
EVALUATION SCENARIOS AND PARAMETERS FOR NR V2X (ADAPTED FROM [103])

| Parameters | Below 6 GHz | | Above 6 GHz | |
|---|---|---|---|---|
| | Urban grid for eV2X | Highway for eV2X | Urban grid for eV2X | Urban grid for eV2X |
| **Carrier frequency** | Macro BS to/from vehicle/pedestrian UE: 4 GHz<br>Between vehicle/pedestrian UE: 6 GHz<br>Micro BS to/from vehicle/pedestrian UE: 4 GHz<br>UE-type-RSU to/from vehicle/pedestrian UE: 6 GHz<br>Note: Agreed value does not mean non-ITS band is precluded for real deployment for sidelink | Macro BS to/from vehicle/pedestrian UE: 2 GHz or 4GHz<br>Between vehicle/pedestrian UE: 6 GHz<br>Micro BS to/from vehicle/pedestrian UE: 4 GHz<br>UE-type-RSU to/from vehicle/pedestrian UE: 6 GHz<br>Note: Agreed value does not mean non-ITS band is precluded for real deployment for sidelink | Macro BS to/from vehicle/pedestrian UE: 30 GHz<br>Between vehicle/pedestrian UE: 30 or 63 GHz<br>Micro BS to/from vehicle/pedestrian UE: 30 GHz<br>UE-type-RSU to/from vehicle/pedestrian UE: 30 or 63 GHz<br>Note: Agreed value does not mean non-ITS band is precluded for real deployment for sidelink | Macro BS to/from vehicle/pedestrian UE: 30 GHz<br>Between vehicle/pedestrian UE: 30 or 63 GHz<br>Micro BS to/from vehicle/pedestrian UE: 30 GHz<br>UE-type-RSU to/from vehicle/pedestrian UE: 30 or 63 GHz<br>Note: Agreed value does not mean non-ITS band is precluded for real deployment for sidelink |
| **Aggregated system bandwidth** | Up to 200 MHz (DL+UL)<br>Up to 100 MHz (SL) | Up to 200 MHz (DL+UL)<br>Up to 100 MHz (SL) | Up to 1 GHz (DL+UL)<br>Up to 1 GHz (SL) | Up to 1 GHz (DL+UL)<br>Up to 1 GHz (SL) |
| **Simulation bandwidth** | 20 or 40 MHz (DL+UL)<br>10 and 20 MHz (baseline for SL)<br>100 MHz (optional for SL) | 10 and 20 MHz (DL+UL)<br>10 and 20 MHz (baseline for SL)<br>100 MHz (optional for SL) | 200 MHz (DL+UL)<br>200 MHz (SL) | 200 MHz (DL+UL)<br>200 MHz (SL) |
| **BS Tx power** | Macro BS: 49 dBm PA scaled down proportionally with simulation BW when system BW is higher than simulation BW. Otherwise, 49 dBm<br>Micro BS: 24 dBm PA scaled down with simulation BW when system BW is higher than simulation BW. Otherwise, 24 dBm<br><br>Note: 33 dBm for RSU is not precluded | Macro BS: 49 dBm PA scaled down proportionally with simulation BW when system BW is higher than simulation BW. Otherwise, 49 dBm<br>Micro BS: 24 dBm PA scaled down with simulation BW when system BW is higher than simulation BW. Otherwise, 24 dBm<br><br>Note: 33 dBm for RSU is not precluded | Macro BS: 43 dBm PA scaled down proportionally with simulation BW when system BW is higher than simulation BW. Otherwise, 43 dBm. EIRP should not exceed 78 dBm and is also subject to appropriate scaling | Macro BS: 43 dBm PA scaled down proportionally with simulation BW when system BW is higher than simulation BW. Otherwise, 43 dBm. EIRP should not exceed 78 dBm and is also subject to appropriate scaling |
| **UE Tx power** | Vehicle/pedestrian UE or UE-type-RSU: 23 dBm<br><br>Note: 33 dBm is not precluded | Vehicle/pedestrian UE or UE type RSU: 23 dBm<br><br>Note: 33 dBm is not precluded | Vehicle/pedestrian UE or UE type RSU: 23 dBm for 30 GHz, 21 dB baseline for 63 GHz, 27 dBm optional for 63 GHz. For both 30 and 63 GHz, EIRP should not exceed 43 dBm. | Vehicle/pedestrian UE or UE type RSU: 23 dBm for 30 GHz, 21 dB baseline for 63 GHz, 27 dBm optional for 63 GHz. For both 30 and 63 GHz, EIRP should not exceed 43 dBm. |
| **BS receiver noise figure** | 5 dB | 5 dB | 7 dB | 7 dB |
| **UE receiver noise figure** | 9 dB | | 13 dB (baseline), 10 dB (optional) | |

GHz with the related parameters shown in Table XII. Beyond the different carrier frequencies, a notable difference is the assumed total system bandwidth. This bandwidth is up to 200 MHz for UL+DL and 100 MHz for SL below 6 GHz. It is up to 1 GHz for UL+DL and 1 GHz for SL above 6 GHz. This indicates that the expectation of 3GPP is to have significantly larger bandwidths available in above 6 GHz (especially mmWave) frequencies.

3GPP guidelines for system level simulations distinguishes three types of vehicles:

1) Type 1 (passenger vehicle with lower antenna position, e.g., at the bumpers): length of 5 meters, width of 2.0 meters, height of 1.6 meters, antenna heightof 0.75 meters.

2) Type 2 (passenger vehicle with higher antenna position, e.g., at the roof): length of 5 meters, width of 2.0 meters, height of 1.6 meters, antenna height of 1.6 meters.

3) Type 3 (truck/bus): length of 13 meters, width of 2.6 meters, height of 3 meters, antenna height of 3 meters (e.g., antenna at the roof).

The vehicles are dropped independently in each lane. They are dropped following an exponential distribution with mean equal to the average of the speed per lane multiplied by 2 seconds. Additionally, vehicles dropped in the same lane must always respect a minimum distance of 2 meters between the bumpers. Once dropped, vehicles maintain a fixed speed according to the speed assigned to the lane. Furthermore, to support the platooning use case, a clustered vehicle dropping is supported for Type 3 vehicles in the highway scenario. The cluster consists of vehicles in the same lane and with the same speed and a predefined distance between them. Further details on vehicle dropping options are shown in Table XIII.

*2) Traffic Models*

Traffic models have been defined in order to support at best as possible the diversity of the use case requirements defined in [12], while keeping the simulation complexity manageable. Three periodic and two aperiodic traffic models have been defined, as indicated in Table XIV.

TABLE XIII
VEHICLE DROPPING OPTIONS (ADAPTED FROM [103][103])

| Vehicle Dropping | Urban grid | Highway |
|---|---|---|
| Option A | • Vehicle type distribution: 100% vehicle type 2.<br>• Clustered dropping is not used.<br>• Vehicle speed is 60 km/h in all the lanes.<br>• In the intersection, a UE goes straight, turns left, turns right with the probability of 0.5, 0.25, 0.25, respectively. | • Vehicle type distribution: 100% vehicle type 2.<br>• Clustered dropping is not used.<br>• Vehicle speed is 140 km/h in all the lanes as baseline and 70 km/h in all the lanes optionally. |
| Option B | • Vehicle type distribution: 20% vehicle type 1, 60% vehicle type 2, 20% vehicles type 3.<br>• Clustered dropping is not used.<br>• Vehicle speed in each lane is as follows:<br>  ○ In the East-West/West-East direction:<br>    ▪ Speed in Lane 1: 60km/h (east-west left lane)<br>    ▪ Speed in Lane 2: 50km/h (east-west right lane)<br>    ▪ Speed in Lane 3: 25km/h (west-east left lane)<br>    ▪ Speed in Lane 4: 15km/h (west-east right lane)<br>  ○ In the North-South/South-North direction:<br>    ▪ 0 km/h in all the lanes.<br>• No vehicles are dropped at the intersections in the North-South direction, so as not to block the traffic in the intersection. Vehicles do not change their direction at the intersection. | • Vehicle type distribution: 20% vehicle type 1, 60% vehicle type 2, 20% vehicle type 3.<br>• Clustered dropping is not used.<br>• Vehicle speed in each lane is as follows:<br>  ○ Speed in Lane 1: 80km/h (east-west right lane)<br>  ○ Speed in Lane 2: 100km/h (east-west center lane)<br>  ○ Speed in Lane 3: 140km/h (east-west left lane)<br>  ○ Speed in Lane 4: 40km/h (west-east left lane)<br>  ○ Speed in Lane 5: 30km/h (west-east center lane)<br>  ○ Speed in Lane 6: 20km/h (west-east right lane) |
| Option C | N/A | • Vehicle type distribution: 0% vehicle type 1, 67% vehicle type 2, 33% vehicle type 3.<br>• Clustered dropping is used. Each cluster consists of 6 Type 3 vehicles with a gap of 2 meters.<br>• Vehicle speed is 140 km/h in all the lanes. |

TABLE XIV
DATA TRAFFIC MODELS [103]

| Model | Periodic | | | Aperiodic | |
|---|---|---|---|---|---|
| | Model 1 (low traffic intensity) | Model 2 (medium traffic intensity) | Model 3 (high traffic intensity) | Model 1 (medium traffic intensity) | Model 2 (high traffic intensity) |
| Inter-packet arrival time (ms) | 100 ms | 10 ms | 30 ms | 50 ms + an exponential random variable with the mean of 50 ms | 10 ms + an exponential random variable with the mean of 10 ms |
| Packet size (bytes) | Pattern of {300 bytes, 190 bytes, 190 bytes, 190 bytes, 190 bytes} with random starting point for each UE | 1200 bytes with probability of 0.2 and 800 bytes with probability of 0.8 | Uniformly random in the range between 30000 bytes and 60000 bytes with the quantization step of 10000 bytes | Uniformly random in the range between 200 bytes and 2000 bytes with the quantization step of 200 bytes | Uniformly random in the range between 10000 bytes and 30000 bytes with the quantization step of 4000 bytes |
| Latency requirement (ms) | 100 ms | 10 ms | 30 ms | 50 ms | 10 ms |

*3) Antenna settings*

Antenna patterns for BSs have been largely carried over from [104], with heights set to 25 meters for macro BS and 5 meters for micro BS in urban grid. For highway, the height is set to 35 meters for macro BS with inter site distance (ISD) of 1732 meters, 25 meters height for ISD of 500 meters, and 5 meter height for micro BS. Antenna patterns for pedestrians are also largely carried over from [104] and the height is set to 1.5 meters. Antenna heights for vehicles are defined according to the vehicle type. Antenna patterns for vehicles are defined separately for roof and bumper locations for both rear and front, as well as for different frequencies. For full details on the patterns, we refer the reader to Section 6.1.4 of [103].

*4) Performance metrics*

Three performance metrics have been defined and utilized in 3GPP as shown in Table XV.

*C. Link level simulations*

Whenever applicable, the assumptions for system level simulation are used for link level simulation as well. The following list contains the most relevant link level simulation parameters [103]:
1) Carrier frequency.
2) Channel model (e.g., fast fading model).
3) PHY packet size.
4) Channel codes (for control and data channels).
5) Modulation and code rates (for control and data channels).
6) Signal waveform (for control and data channels).
7) SCS.
8) CP length.
9) Frequency synchronization error.
10) Time synchronization error.
11) Channel estimation (e.g., DMRS pattern).
12) Number of retransmission and combining (if applied).
13) Number of antennas (at UE and BS).

TABLE XV
PERFORMANCE METRICS [103]

| Packet reception ratio (PRR) | Packet Inter-Reception (PIR) | Throughput (defined in [109]) |
|---|---|---|
| **PRR type 1**: For one Tx packet, the PRR is calculated by $X/Y$, where $Y$ is the number of UE/vehicles that are located in the range $(a, b)$ from the TX, and X is the number of UE/vehicles with successful reception among $Y$. CDF of PRR and the following average PRR are used in evaluation.<br>• CDF of PRR with $a = 0$, $b$ = baseline of 320 meters for highway and 150 meters for urban. Optionally, $b = 50$ meters for urban with 15 km/h vehicle speed.<br>• Average PRR, calculated as $(X1+X2+X3….+Xn)/(Y1+Y2+Y3….+Yn)$ where n denotes the number of generated messages in simulation. with $a = i*20$ meters, $b = (i+1)*20$ meters for $i$=0, 1, …, 25 | **PIR type 1**: For a given distance $d$, PIR is the time $Ti$ elapsed between two successive successful receptions of two different packets transmitted from node A to node B for the same application, if the distance between node A and node B at the two packets' receiving time is within the range of $(0,D]$.<br>• Average PIR within given distance $d$, calculated as $(T1+T2+T3+…+Tn)/n$ where n denotes the number of collected PIR in simulation.<br>• CDF of PIR with given distance $D$. | **User throughput** = amount of data (file size) / time needed to download data<br>• Time needed to download data starts when the packet is received in the transmit buffer, and ends when the last bit of the packet is correctly delivered to the receiver |
| **PRR type 2**: For one Tx packet, the PRR is calculated by S/Z, where Z is the number of UEs in the intended set of receivers, and S is the number of UE with successful reception among Z.<br>Unicast is the special case where Z includes a single UE, where the PRR is the average of packets of the unicast link. | **PIR type 2**: PIR is the time $Ti$ elapsed between two successive successful receptions of two different packets transmitted from node A to node B for the same application, if the node B is one of the intended set of receivers of the node A.<br>• Average PIR with intended set of receivers, calculated as $(T1+T2+T3+…+Tn)/n$ where $n$ denotes the number of collected PIR in simulation.<br>• CDF of PIR with intended set of receives. | |

14) Transmission diversity scheme (if applied).
15) UE receiver algorithm.
16) AGC settling time and guard period.

For all above parameters, [103] contains the suggested values or range of values. Furthermore, a set of Cluster Delay Line (CDL) models[57] has also been developed for the purpose of link level simulations. CDL models can be used for computationally efficient simulation of wireless channels, with the angular component allowing the incorporation of aspects particularly relevant for MIMO systems. The CDL models generated for Urban LOS, Urban NLOS, Urban NLOSv, Highway LOS, and Highway NLOSv channels are available in [103].

## XI. OUTLOOK FOR FUTURE RELEASES

After completing Rel. 16, 3GPP has already identified new study and work items for Rel. 17. This section presents the most relevant study and work items for NR V2X SL communication as well as other possible enhancements that can benefit NR V2X SL communication, even if they are not included in the officially approved items.

### A. Beamforming in Sidelink

In Rel. 16, NR V2X SL communication has been designed with a predominant focus on FR1. FR2 can be considered by using the design for FR1, but no specific optimization for FR2 (except for SL PT-RS) or beam management is supported under Rel. 16 [10]. In this context, enhancements to support high data rates (e.g., based on FR2) were initially considered in the preparation of the new work item on NR SL enhancement for Rel. 17. However, the agreed work item description for NR SL enhancement does not finally consider optimizations for FR2 [110]. Nonetheless, certain eV2X use cases such as vehicles platooning, advanced driving and extended sensors require high date rates (50-700 Mbps) for long distances (200 meters or beyond) [12] (see Table II). These use cases could be supported with FR2, in particular with beamforming, to compensate the pathloss at higher frequencies. Beamforming is not only applicable to higher frequencies, but can also be supported for FR1. As beamforming enables directional transmission, this allows for a spatial reuse of available resources due to reduced interference. This can be leveraged for supporting a high connection density as required in certain V2X scenarios [12].

Although no specific optimization for FR2 is supported in Rel. 16 NR V2X, the S-SSB structure enables the transmission of synchronization information by a SyncRef UE in various directions using different beams (as discussed in Section V.B.5)) [111]. For a configured number of S-SSBs, the synchronization information can be sent by a SyncRef UE with the same beam multiple times or with different beams at each S-SSB, e.g., through beam sweeping. However, as the goal of transmitting S-SSBs is to expand the synchronization coverage, sending S-SSBs in all directions may not be efficient. For instance, S-SSB transmissions by a SyncRef UE in the direction of a gNB is not necessary as this area is already under the gNB's synchronization coverage as shown in Fig. 33(a). In fact, S-SSB transmissions towards the gNB may only lead to interference as a UE (which receives synchronization information from the gNB and the SyncRefUE) anyway selects the gNB as its synchronization reference, as it has higher priority as explained in Section V. To increase the synchronization coverage, a SyncRef UE should transmit S-SSBs away from the gNB as depicted in Fig. 33(b). This spatially selective transmission of synchronization information can be enabled with beamforming, i.e., by the SyncRef UE transmitting S-SSBs on certain beams.

---

[57]Clustered Delay Line is a type of channel model where the received signal is composed of a number of separate delayed clusters. Each cluster contains a number of multipath components with the same delay but different Angle of Departure and Angle of Arrival.

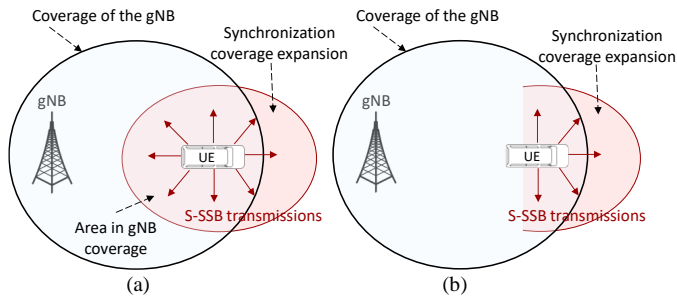

Fig. 33. Synchronization coverage expansion. (a) SyncRef UE transmits S-SSBs omnidirectionally. (b) SyncRef UE transmits S-SSBs away from the gNB.

This mechanism can also be beneficial for synchronization coverage expansion in out of coverage scenarios.

Further optimization includes extending the PSCCH/PSSCH power control to consider beam-based transmissions, similar to how the PUSCH and PUCCH power control support beamforming in NR Rel. 15 [66]. In NR Rel. 15, a UE determines the initial transmit power for sending PUSCH or PUCCH with a given beam based on the DL pathloss measured on that beam[58]. With beam-based SL power control, the transmit power for a SL transmission on a given beam would depend on the SL pathloss and/or DL pathloss measured on that beam. This enables optimizing the PSCCH/PSSCH transmit power of a TX UE in the spatial domain. Consider for example a TX UE that is in network coverage with the PSCCH/PSSCH power control based on the DL pathloss (see Section V.C.7). Assuming in this case that the strongest signal (i.e., smallest DL pathloss) from the gNB is received by the TX UE on beam $X$ (as seen in Fig. 34(b)), the transmit power of a PSCCH/PSSCH sent on another beam (i.e., beam $Y$) would not depend on the smallest DL pathloss which was measured on beam $X$. This is in contrast to having no beam-based power control as depicted in Fig. 34(a). , where the transmit power of PSCCH/PSSCH (in any direction) is limited by the DL pathloss (see Fig. 15).

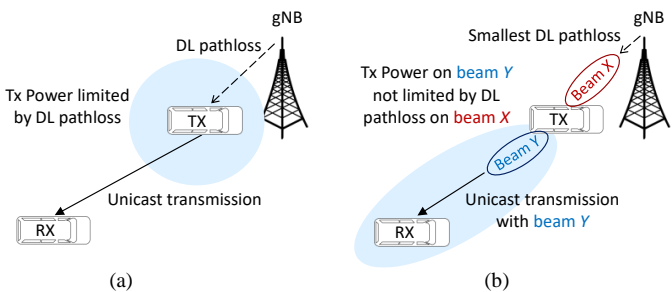

Fig. 34. Sidelink power control based on DL pathloss. (a) No beam-based power control/transmission. (b) Beam-based power control/transmission.

### B. Relative positioning via sidelink measurements

A study item on scenarios and requirements of in coverage, partial coverage, and out of coverage positioning use cases has been approved for Rel. 17 [112]. The objectives of the study are to identify the positioning use cases and requirements for V2X and public safety, as well as to identify potential deployment and operation scenarios for UEs in network coverage, partial network coverage, and out of network coverage. For instance, V2X use cases like platooning and coordinated maneuvers may impose requirements on relative positioning of nearby vehicles. In an overtaking maneuver, for example, a collision with a nearby vehicle can be avoided if the overtaking vehicle knows the relative position of the nearby vehicle as shown in Fig. 35. In such cases, relative positioning based on SL measurements (i.e., SL positioning or ranging) may be beneficial, as it can be supported for UEs independent of the network coverage.

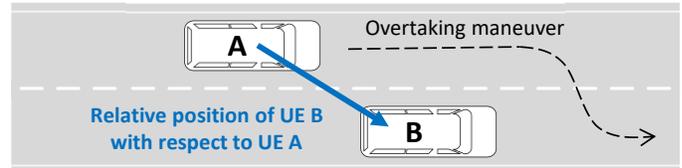

Fig. 35. Relative positioning by an overtaking vehicle (UE A) of a nearby vehicle (UE B).

Relative positioning can also be obtained with network-based positioning. However, network-based positioning requires UEs to be in network coverage and it may incur increased latency and signalling. In fact, the network first needs to obtain the absolute position of two UEs in order to derive their relative position, based on the difference between the two absolute positions. The relative position should then be shared with at least one of the UEs. In addition, deriving the relative position based on absolute positions can lead to imposing more stringent requirements for absolute positioning. This is the case as the uncertainties (i.e., positioning errors) in the absolute position of each of the two UEs add up when obtaining the relative positioning (e.g., based on the difference between the absolute positions of the two UEs shown in Fig. 36).

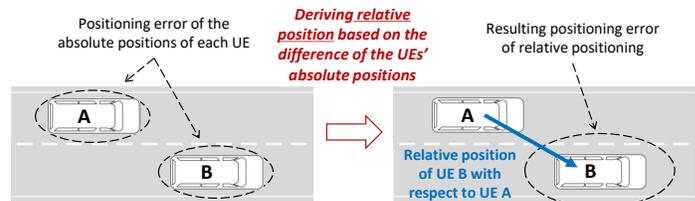

Fig. 36. Deriving relative positioning based on absolute positions of two UEs.

As some coordinated maneuvers in V2X only require avoiding a collision with a nearby vehicle, i.e., via relative positioning, the absolute position of a nearby vehicle is not really necessary. Relative positioning can be performed based on SL measurements, i.e., via SL positioning or ranging. Sidelink positioning can potentially benefit from the smaller SL pathloss between UEs (compared with the DL pathloss), the support for out of coverage, and the lower SL latency and signaling (compared with the Uu link) [113]. The SL measurements can be based on time and angular measurements

---

[58]In contrast to NR V2X, the uplink power control in NR is closed loop, i.e., based on TPC commands from the gNB. However, prior to receiving any TPC commands, the uplink transmit power is based on the DL pathloss.

[114], similar to the measurements used for network-based positioning in Rel. 16. For V2X, the large dimensions of the vehicles can be exploited for placing distributed antennas or arrays for performing the SL measurements [114]. In addition, SL positioning can be used to complement or enhance any existing positioning schemes, e.g., based on network-based positioning, sensors, or GNSS, which may be limited due to obstructions or other factors. For this purpose, techniques for relative positioning based on SL measurements are of interest.

### C. Enhancements to resource allocation

A new work item on NR Sidelink enhancement has been agreed for Rel. 17 [110]. One of the objectives of this Rel. 17 work item is to enhance Rel. 16 NR V2X resource allocation mode 2. Some planned enhancements focus on power saving and on improving KPIs such as reliability and latency. All enhancements must be able to coexist in the same resource pool (i.e., co-channel coexistence) with Rel. 16 NR V2X.

Mode 1 and mode 2 in Rel. 16 NR V2X have been designed for UEs such as vehicles or RSUs, which do not have strong power limitations. These limitations are present in other types of UEs (e.g., smartphones) that are used by VRUs such as pedestrians. With the current Rel. 16 mode 2 specifications, a UE requires long sensing intervals that severely impact the battery consumption. An enhancement adopted for study under Rel. 17 is a variant of mode 2 that would reduce power consumption by using partial sensing [110]. Partial sensing was already considered in Rel. 14 for a variant of LTE V2X mode 4, where a UE only senses a subset of subframes contained in the sensing interval [23], [115].

Another SL enhancement adopted for Rel. 17 is the support of inter-UE coordination [110]. Inter-UE coordination was analysed under Rel. 16 where it was referred to as mode 2(b) [116], [62], but it was not standardized. With inter-UE coordination, an UE can assist other UE(s) in their resource selection process. There are two types of inter-UE coordination: type A and type B [116]. Under type A, an *assisting* UE (UE$_1$) restricts the resources that can be used by an *assisted* UE (UE$_2$). Under type B, UE$_1$ sends recommendations on which resources should be selected by UE$_2$. UE$_2$ decides whether or not to follow these recommendations. Contents of recommendations or restrictions have not been decided by 3GPP yet, but the recommendations could contain, for example, the available resources detected by the assisting UE.

For Rel. 17, inter-UE coordination type B is considered [110]. An important benefit from inter-UE coordination is the reduction of the hidden terminal problem as illustrated in Fig. 37. Fig. 37 illustrates a scenario where UE B is under the transmission range of UE A and UE C. UE B can then receive transmissions and detect the resource reservation of UE A and UE C. However, this is not the case between UE A and UE C (or viceversa) that experience the hidden terminal problem. In this case, UE A and UE C can end up selecting the same resources for their transmissions. In Fig. 37, UE B has reserved the resource consisting of sub-channel $SC_2$ at slot $S_2$. UE C has reserved resources consisting of sub-channel $SC_1$ at slot $S_2$ and sub-channel $SC_2$ at slot $S_4$. The resource reservation of UE B is detected by UE A and by UE C. However, reservations of UE C are only detected by UE B and not by UE A due to the hidden terminal problem. In slot $S_1$, UE A generates a TB. Let's suppose that the selection window of UE A starts at $S_2$ and ends at $S_5$. Since UE A has not detected the reservations of UE C, UE A would detect as available the resources reserved by UE C if there is no inter-UE coordination (Fig. 37(a)). In this case, UE A can select resources reserved by UE C with the consequent risk of packet collisions. Fig. 37(b) shows that inter-UE coordination reduces the hidden terminal problem. In this scenario, UE B acts as an assisting UE and sends a recommendation (type B) to UE A for its resource selection. As a result, the resource selection of UE A can consider the status of the resources detected by UE B (which include the reservations of UE C) and reduce the probability of collisions caused by the hidden terminal problem.

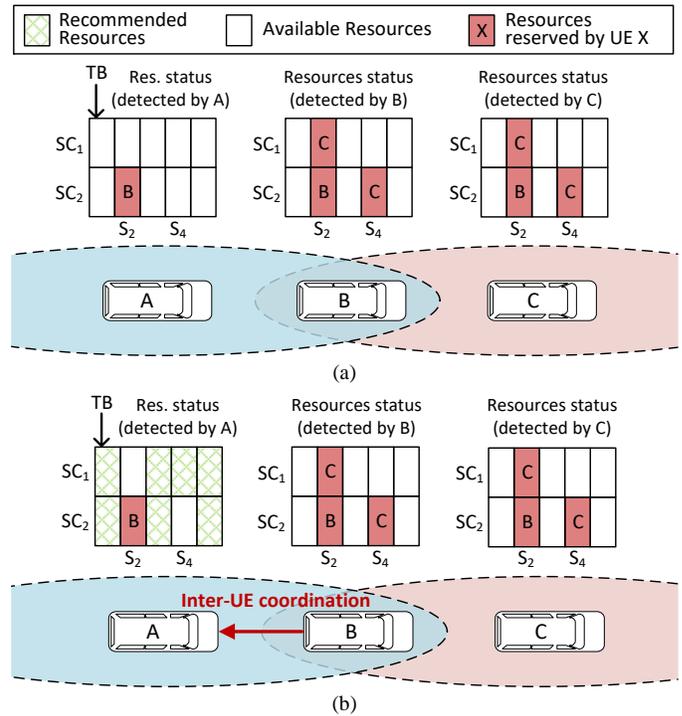

Fig. 37. Benefits of inter-UE coordination to mitigate the hidden terminal problem. (a) Without inter-UE coordination. (b) With inter-UE coordination.

It is also interesting to highlight other enhancements that were analysed for NR V2X mode 2 during Rel. 16, but that were not adopted in Rel. 16 or Rel. 17. These enhancements could be of interest in future releases (Rel. 18 and beyond). One such enhancement is the possibility for a UE to schedule resources of other UEs [117]; it was referred to as mode 2(d) during the standardization process [62]. This enhancement requires establishing a group of member UEs, where a member UE acts as a scheduling UE[59] to schedule the transmissions of all the member UEs in the group. Mode 2(d) requires member UEs (or at least the scheduling UE) to operate in network coverage.

---

[59]Different versions of the TR 38.885 refers to this UE as scheduling UE [118] or UE-A [62].

With mode 2(d), the scheduling UE acts as an intermediary between the gNB and other member UEs in the management of the SL radio resources. Different options were proposed to select the scheduling UE [118]. For example, the scheduling UE could be selected by the gNB or by the group members.

Another enhancement proposed is to perform the first transmission of a TB using NR V2X mode 2 and its retransmissions using NR V2X mode 1 in order to increase the reliability of NR V2X communications [120].

Another idea proposed for mode 2 is the concept of standalone reservation [121] where the selection window is divided into two windows W1 and W2. A UE that wants to transmit a TB selects one resource in each window, and announces in the resource reserved in W1 the resource it has selected in W2 to transmit the TB.

Another enhancement analysed (but finally not adopted) for mode 2 during Rel. 16 is the use of Time-Frequency Resource Patterns (TFRPs) [122]. Each TFRP contains a set of (pre-)configured resources in the resource pool. Each TFRP is adapted to meet specific QoS requirements, for example, by adjusting the number of resources in the TFRP, its periodicity and the number of resources reserved for retransmissions. TFRPs are repeated with certain periodicity. The concept of TFRP is associated with the resource allocation proposal referred to as mode 2(c) during the study item phase of Rel. 16 [62]. A TFRP can be assigned to a UE or it can be shared by a group of UEs. Several proposals were made for selecting the TFRP and avoid collisions between UEs. For example, a UE can select the TFRP using sensing information [123] or using information about the relative position of other UEs [122].

### D. UE Relaying

3GPP has been developing the concept of UE relaying since Rel. 13. The initial scope focused mainly on public safety scenarios in order to extend the coverage range. First specifications referred to UE relaying as ProSe (Proximity Service) and UE-to-network relaying [81]. This was later extended in Rel. 14 and 15 for use cases where power-limited devices (e.g., wearables) use other UEs (e.g., smartphones) as relays to connect to the network [124]. Rel. 17 includes a work item on UE relaying with the objective to achieve maximum commonality between commercial (i.e., smartphones), critical (i.e., public safety) and V2X use cases [125]. There are several motivations to consider UE relaying for V2X communications in Rel. 17: extension of the network coverage and the SL transmission range, higher data rates, enhanced reliability, power saving, and spectral efficiency enhancements. UE relaying will be designed under Rel. 17 to operate in different network and spectrum scenarios. It shall be able to operate in network coverage, partial coverage, and out network coverage (see Fig. 38[60]). It should also be able to operate using licensed, unlicensed, and ITS SL spectrum. It is expected that UE relaying will cover both UE-to-network relaying and UE-to-UE relaying (see Fig. 38), where relays forward the communication between a UE and the infrastructure or between two UEs, respectively. The work will primarily focus on utilizing a single relay, which is referred to in 3GPP as single-hop NR SL-based relay [125]. However, forward compatibility for multi-hop relay support in a future release will be taken into account. The technical report entitled 'Study on NR sidelink relay' (3GPP TR 38.836) will collect the progresses related to UE relaying that will include (among others): relay discovery and selection procedures, service continuity during path switching between direct Uu connection and a connection via a UE-to-Network relay, and extending Rel-16 QoS management to UE-to-UE or UE-to-Network relaying.

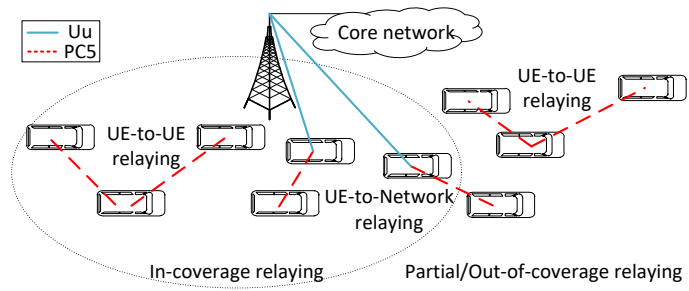

Fig. 38. UE relaying scenarios.

### E. Uu multicast communications

Advanced eV2X use cases such as sensors monitoring, software updates, or update/download of high definition (HD) maps [13] could benefit from Uu multicasting or multicast broadcast services (MBS) to deliver the same information to different groups of UEs. In contrast to a broadcast transmission, a gNB sends content to a specific group of UEs in a multicast transmission. Multicast transmissions can support feedback to improve reliability. For MBS, the group of UEs can be determined based on UEs with a subscription to certain services or UEs which have indicated their interest in the content. MBS allows a more efficient use of the resources, since the same information can be sent to a group of UEs with fewer resources than in unicast transmissions. For instance, a vehicle may require real-time HD map updates. These updates may impact the navigation route of several vehicles in the same region. Thus, the updates can be delivered to a group of UEs in a multicast manner and avoid independent unicast transmissions to each UE over multiple resources. The group of UEs in a MBS transmission can be identified based on their location, but also on other factors like their heading. In fact, UEs near a traffic jam are interested in different content depending whether they are heading towards the traffic jam or in the opposite direction.

Besides V2X applications, there are other relevant applications and uses cases that can benefit from multicast and broadcast services (e.g., public safety and mission critical, IPTV, software updates, group communications and IoT) to improve user experience and service latency while efficiently utilizing the radio resources. Since Rel. 15 and Rel. 16 do not support MBS in NR, a new work item to support MBS has been agreed in Rel. 17 [126]. The objectives of this work item include the specification of RAN basic functions for MBS by providing a group scheduling mechanism to allow UEs to receive MBS, as well as the support for simultaneous operation with unicast reception. Within the work item, the support for basic mobility and dynamic change between multicast and unicast delivery with service continuity for a given UE should be specified. Necessary changes for addressing improved reliability of MBS, e.g., via feedback in the uplink, should also be specified. As a gNB may consist of a gNB-Control Unit (CU) and one or more gNB-Distributed Units (DUs), any required changes on the 5G-RAN architecture and interfaces between the gNB-CU and gNB-DUs should be defined assuming that the coordination functions are handled by the gNB-CU. The support for dynamic control (and its management) of the MBS transmission area within one gNB-DU should also be studied.

---

[60] Although UEs are represented in the figure by vehicles, the UE relaying scenario envisioned in Rel. 17 might also include smartphones, wearables, etc.

The work item on NR Multicast and Broadcast Services is limited to current Rel. 15 numerologies, physical channels and signals. Furthermore, lower priority will be given in this work item to any enhancements related to FR2. Although a flexible resource allocation between unicast services and MBS should be possible, allocating all the resources for MBS is not a mandatory requirement for this work item. The impact on UE implementation and UE complexity should be minimized to enable the deployment of the broadcast/multicast feature.

## XII. CONCLUSIONS

This paper has presented an in-depth tutorial of the 3GPP Rel. 16 NR V2X standard, the first V2X standard based on the 5G NR air interface. This standard focuses mainly on the sidelink aspects that were not developed in Rel. 15 where the 5G NR air interface was first introduced. The 5G NR V2X standard has been developed to complement the capabilities of LTE V2X and be able to support more advanced eV2X use cases, including those related to connected and automated driving, where 5G NR is expected to play an important role. This tutorial will help the community to better understand the new standard and its advanced functionalities as a first step towards a thorough evaluation and further enhancement of its performance and capabilities. To this aim, the paper has also presented the evaluation methodology and the system and link level simulations that were defined in 3GPP for the development of Rel. 16. These can be used as a basis for the evaluation of 5G NR V2X and its future enhancements. Indeed, Rel. 16 has made significant contributions by developing the first V2X standard and the first NR SL standard based on the 5G NR air interface. However, improvements are possible, and some have already been identified as study items in Rel. 17. In this context, this paper also discusses some possible future enhancements related to beamforming, sidelink positioning, resource allocation (including the coexistence of different communication modes), relaying, and multicast communications. This discussion should help the research community to identify potential topics and contributions that could have a strong impact on Rel. 17 and beyond.

## ACKNOWLEDGMENT

We would like to thank Konstantinos Manolakis and Lu Lei and for their useful inputs on the structure of the paper, the physical layer section, and the QoS framework section. We would also like to thank Matthew Webb for his helpful comments and suggestions.

This work was supported in part by the Ministerio de Ciencia e Innovación (MCI), AEI and FEDER funds through the project TEC2017-88612-R, and the Ministerio de Universidades (FPU18/00691, IJC2018-036862-I).

The authors would also like to thank and provide full copyright acknowledgment to the following copyright holders:
For Table IX in this paper (Table 5.4.4-1 in [51]):
"© 2020. 3GPP™ TSs and RSs are the property of ARIB, ATIS, CCSA, ETSI, TSDSI, TTA and TTC who jointly own the copyright in them. They are subject to further modifications and are therefore provided to you "as is" for information purposes only. Further use is strictly prohibited."
For Figs. 31 and 32 in this paper (Figure A-1 and A-2 in [103]):
"© 2019. 3GPP™ TSs and RSs are the property of ARIB, ATIS, CCSA, ETSI, TSDSI, TTA and TTC who jointly own the copyright in them. They are subject to further modifications and are therefore provided to you "as is" for information purposes only. Further use is strictly prohibited."

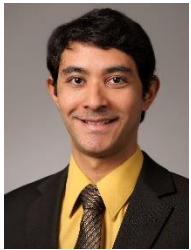

**Mario H. Castañeda Garcia** received the B.Sc. degree in electrical engineering from the Universidad Nacional Autónoma de Honduras, Tegucigalpa, Honduras, in 2001, and the M.Sc. degree in communications engineering and the Dr.-Ing. degree from the Technische Universität München (TUM), Munich, Germany, in 2004 and 2014, respectively. From 2005 to 2014, he was a research and teaching assistant with the Institute for Circuit Theory and Signal Processing, TUM, where he was involved in several industry and government projects in the areas of signal processing for wireless communication systems and for satellite-based positioning. In 2014, he joined the Munich Research Center, Huawei Technologies Düsseldorf GmbH, Munich, where he has been involved in several EU funded research projects as well as in 5G research and standardization activities. He was the recipient of the Best Overall Paper Award at IEEE VTC Fall 2018. His research interests comprise multi-antenna technologies, millimeter-wave systems, V2X communication and wireless positioning.

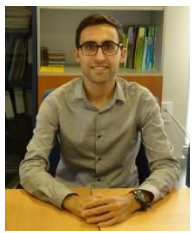

**Alejandro Molina-Galan** received a Telecommunications Engineering Degree in 2016 and a Master's Degree in Telecommunications Engineering in 2018, both from the Universidad Miguel Hernandez (UMH) de Elche, Spain. He received the Best Student award in the Telecommunications Engineering Degree and the Master's Degree by UMH. He joined the UWICORE research laboratory in October 2018, and he was awarded in September 2019 a highly competitive PhD fellowship by the Spanish government. He is currently pursuing his PhD on 5G and beyond V2X networks for connected and automated vehicles.

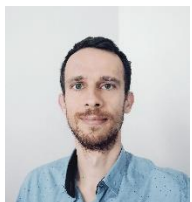

**Mate Boban** received the Diploma degree in informatics from the University of Zagreb, Croatia, and the Ph.D. degree in electrical and computer engineering from Carnegie Mellon University, Pittsburgh, PA, USA, in 2004 and 2012, respectively. He is a Principal Research Engineer with Huawei Munich Research Center, Germany. Before Huawei, he was with NEC Labs Europe and Apple. He is an alumni of the Fulbright Scholar program. He has co-chaired several IEEE workshops and conferences and has been involved in European Union funded projects (5G-CAR, DRIVE-C2X, and TEAM) as a Work Package Leader and editor of deliverables. He is actively involved in key industry and standardization bodies: 3GPP, 5GAA, and ETSI. His current research interests include resource allocation, machine learning applied to wireless communication systems (in particular, V2X and IIoT), and channel modeling. He coauthored three papers that received the Best Paper Award, at IEEE VTC Spring 2014, IEEE VNC 2014, and EuCAP 2019.

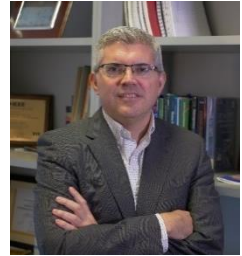

**Javier Gozalvez** received an electronics engineering degree from the Engineering School ENSEIRB (Bordeaux, France), and a PhD in mobile communications from the University of Strathclyde, Glasgow, U.K. Since October 2002, he is with the Universidad Miguel Hernández de Elche (UMH), Spain, where he is currently Full Professor and Director of the UWICORE laboratory. At UWICORE, he leads research activities in the areas of vehicular networks, 5G and Beyond, and industrial wireless networks. He has received several awards at international and national conferences, the best research paper award from the Journal of Network and Computer Applications (Elsevier) in 2014, the Runner-up prize for the "Juan López de Peñalver" award of the Royal Academy of Engineering in Spain that recognizes the most notable Spanish engineers aged below 40, and the IEEE VTS Outstanding Service Award in 2019. He is an elected member to the Board of Governors of the IEEE Vehicular Technology Society (IEEE VTS) and was the 2016 and 2017 President of the IEEE VTS. He has served as IEEE Distinguished Speaker and IEEE Distinguished Lecturer for the IEEE VTS. He is the Editor in Chief of the IEEE Vehicular Technology Magazine, and the Founder and General Co-Chair for the IEEE Connected and Automated Vehicles Symposium. He was the General Co-Chair for the IEEE VTC-Spring 2015 conference in Glasgow (UK), ACM VANET 2013, ACM VANET 2012 and ISWCS 2006, and TPC Co-Chair for 2011 IEEE VTC-Fall and 2009 IEEE VTC-Spring. He is a scientific evaluator and reviewer for the European Commission and various national scientific and technical agencies.

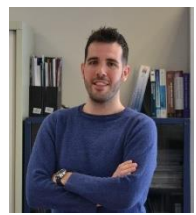

**Baldomero Coll-Perales** received his M.Sc. and Ph.D. degrees in Telecommunications Engineering from the Universidad Miguel Hernandez (UMH) de Elche, Spain. He is currently a Juan de la Cierva Research Fellow at the UWICORE laboratory (UMH). He has hold visiting research positions at Hyundai Motor Europe Technical Center (Germany), Italian National Research Council (Italy), WINLAB-Rutgers University (USA) and Institute of Telecommunications-King's College London (UK). His research interests lie in the field of advanced mobile and wireless communication systems, including the design of device-centric technologies for future wireless Beyond 5G network and connected and automated vehicles. He serves/has served as Associate Editor for the International Journal of Sensor Networks and Springer's Telecommunication Systems. He has served as Track Co-Chair for IEEE VTC-Fall 2018, and as member of TPC in over 35 international conferences.

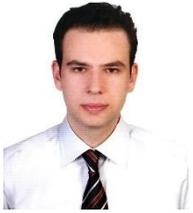

**Taylan Şahin** received his B.S. degree in electrical and electronics engineering from the Middle East Technical University, Ankara, Turkey, in 2014, and his M.Sc. degree in communications engineering from the Technical University of Munich, Germany, in 2016. Since February 2017, he has been a doctoral candidate at the Technical University of Berlin, Germany, and worked as a student in Huawei Germany Research Center, Munich, where he also conducted his master's work prior to that. Since December 2020, he is a radio access research specialist at Nokia Bell Labs, Munich. His research interests include radio resource management aspects of radio access networks, and V2X communications, targeting the next generations of wireless mobile networks.

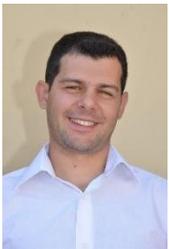

**Apostolos Kousaridas** received his Ph.D. from the Department of Informatics & Telecommunications at the University of Athens. Athens. He holds B.Sc. degree in Informatics and M.Sc. degree in Information Systems from the Department of Informatics at Athens University of Economics and Business. Currently, he is a principal research engineer of the Huawei Research Center in Munich, contributing to the design of 5G and beyond 5G communication systems, by generating patents and standardization contributions to 3GPP. He has disseminated over 50 publications and he is also serving as delegate to the 5G Automotive Association (5GAA) and vice-chair of the 5G-PPP Automotive WG. He has participated in several impacting large and medium-scale European R&D projects, having technical and management roles. His research interests include vehicular communications, wireless networks and artificial intelligence.